\documentclass[12pt]{article}
\usepackage[a4paper, margin=2.5cm]{geometry}

\usepackage{graphicx}
\usepackage{newtxtext}
\usepackage{newtxmath}
\usepackage{natbib}
\usepackage{caption}
\usepackage{hyperref}
\usepackage{float}
\hypersetup{
    colorlinks = true,
    urlcolor   = blue,
    citecolor  = black,
}

\newcommand{\RomanNumeralCaps}[1]
\linenumbers
\usepackage{subcaption}
\title{Non-linear evolution of the horizontal shear instability in stratified rotating fluids under the complete Coriolis acceleration.}

\author{
  C. Moisset\textsuperscript{1},
  P. Billant\textsuperscript{2},
  J. Park\textsuperscript{3},
  S. Mathis\textsuperscript{1}
}

\begin{document}
\maketitle

\footnotetext[1]{Université Paris-Saclay, Université Paris Cité, CEA, CNRS, 91191 Gif-sur-Yvette, France}
\footnotetext[2]{LadHyX, Ecole Polytechnique / CNRS, Palaiseau, France}
\footnotetext[3]{Centre for Fluid and Complex Systems, Coventry University, Priory Street, Coventry CV15FB, UK}

\begin{abstract}

This paper investigates the non-linear dynamics of horizontal shear instability in an incompressible, stratified and rotating fluid in the non-traditional $f$-plane, i.e. with the full Coriolis acceleration, using direct numerical simulations. The study is restricted to two-dimensional horizontal perturbations. It is therefore independent of the vertical (traditional) Coriolis parameter. However, the flow has three velocity components due to the horizontal (non-traditional) Coriolis parameter. Three different scenarios of non-linear evolution of the shear instability are identified, depending on the non-dimensional Brunt-V\"ais\"al\"a frequency $N$ and the non-dimensional non-traditional Coriolis parameter $\tilde{f}$ (non-dimensionalized by the maximum shear), in the range $\tilde{f}<N$ for fixed Reynolds and Schmidt numbers $Re=2000$, $Sc=1$. When the stratification is strong $N\gg 1$, the shear instability generates stable Kelvin-Helmholtz billows like in the traditional limit $\tilde{f}=0$. Furthermore, when $N\gg1$, the governing equations for any $\tilde{f}$ can be transformed into those for $\tilde{f}=0$. This enables us to directly predict the characteristics of the flow depending on $\tilde{f}$ and $N$. When $N$ is around unity and $\tilde{f}$ is above a threshold, the primary Kelvin-Helmholtz vortex is destabilized by secondary instabilities but it remains coherent. For weaker stratification, $N\leqslant0.5$ and $\tilde{f}$ large enough, secondary instabilities develop vigorously and destroy the primary vortex into small-scales turbulence. Concomitantly, the enstrophy rises to high values by stretching/tilting as in fully three-dimensional flows. A local analysis of the flow prior to the onset of secondary instabilities reveals that the Fj\o rtoft necessary condition for instability is satisfied, suggesting that they correspond to shear instabilities.
\end{abstract}

\section{Introduction}
Rotating and stratified fluids are ubiquitous in stellar interiors, as well as planetary atmospheres and oceans. To predict their structure and evolution, one of the main challenges is to understand how momentum, heat and chemical elements are redistributed and mixed by various transport mechanisms (e.g. large scale circulations, waves or hydrodynamical instabilities) and small-scale processes in presence of stratification and rotation \citep[e.g.][]{Vallis2006, Kippenhahn2013, AertsMathisRogers2019}. In this context, hydrodynamic instabilities have been thoroughly studied because of their prime importance in transport processes \citep[see][]{Zahn1984, Rieutord2006, SmythCarpenter2019}. Their non-linear evolutions produce efficient momentum and heat fluxes and strongly modify the initial structure of the flows.\\

Among various instabilities, shear instability is most frequently encountered as it only necessitates a velocity shear to develop \citep[][]{Peltier2003, Carpenteretal2012, Caulfield2021}, a configuration which is generic in geo- and astrophysical flows. In planetary atmospheres and oceans, shear is present in shear layers, jets, wakes and boundary layers flows \citep[][]{Johannessen1996, Flamentetal2001}. In stellar interiors, in particular in stably stratified radiative regions, the differential rotation in the radial and latitudinal directions is associated with large-scale sheared motions. The resulting instabilities are thought to control the turbulent transport and mixing \citep[][]{Zahn1992, ChaboyerZahn1992, MathisZahn2004, Mathisetal2018}.\\

For two-dimensional, homogeneous and inviscid parallel flows, the Rayleigh and Fj\o rtoft necessary conditions for shear instability are well-known: the former states that the velocity profile must have an inflection point whereas the latter requires that it should correspond to a shear maximum \citep[][]{SmythCarpenter2019}. Furthermore, for non-rotating and non-stratified shear flows, Squire's theorem ensures that the most unstable mode is two-dimensional. As reviewed below, previous works have studied how a stable stratification, a planetary rotation, viscous effects and heat or mass diffusion modify these classical results \citep[e.g.][]{Deloncleetal2007, AroboneSarkar2012, Parketal2021}. The following discussion will be restricted to the configuration in which the flow and shear are along the horizontal whereas the stratification or rotation component is along the vertical. In this case, only the three-dimensional stability of the flow is modified by stratification or rotation whereas the two-dimensional stability (i.e. in the horizontal plane) is unaffected.\\

The impact of a vertical stable stratification on the horizontal shear instability has been studied by \cite{Deloncleetal2007}. They have demonstrated that the fastest growing mode remains two-dimensional even if Squire's theorem no longer holds in a stratified fluid. However, the range of unstable vertical wave numbers is broadened as the stratification increases, meaning that three-dimensional perturbations tend to be more unstable in stratified fluids than in homogeneous fluids.\\

In the presence of a background rotation about the vertical axis, the most unstable mode of the shear instability remains also two-dimensional but the band of unstable vertical wavenumbers shrinks as the rotation increases \citep[][]{Yanaseetal1993, AroboneSarkar2012}. The inertial instability, which is of different nature and is three-dimensional, can also exist in an intermediate range of anticyclonic background rotation when the Pedley criterion is satisfied \citep[][]{Pedley1969}. This criterion is the analogue for parallel shear flows of the generalized Rayleigh criterion for the centrifugal instability of vortices. \cite{AroboneSarkar2012} have also considered the combined effects of rotation and stable stratification. Their effects on the shear instability for non-zero vertical wavenumbers are globally the same as when each effect is considered separately.\\

Recently, \cite{Parketal2021} have considered the linear stability of a horizontally sheared flow, the hyperbolic tangent velocity profile, in stratified rotating fluids in presence of the non-traditional Coriolis force. This force, which comes from the horizontal component of background rotation, is generally neglected when studying geo- and astro-physical flows. Such ``traditional'' approximation is legitimate if the vertical velocity is small compared to the horizontal velocity or if the flow is sufficiently away from the equator \citep[][]{Gerkemaetal2008}. However, several works have demonstrated that small-scale processes can be affected by non-traditional effects: gravito-inertial waves \citep[][]{GerkemaShrira2005} and their coupling with convectively induced turbulence \citep[][]{MathisNeinerTranMihn2014}, the dynamics of vortices \citep[][]{ToghraeiBillant2022, ToghraeiBillant2025}, or the inertial instability \citep[][]{Zeitlin2018, Parketal2021, ParkMathis2025}.\\ 

In presence of the non-traditional Coriolis force, \cite{Parketal2021} have shown that the horizontal shear instability can remain two-dimensional, if it is initially so, but with three velocity components, i.e. with a non-zero vertical velocity field due to the non-traditional Coriolis force. As a consequence, the stability problem depends on the non-dimensional Brunt-V\"ais\"al\"a frequency $N$ characterising the stratification and on the non-dimensional, non-traditional Coriolis parameter $\tilde{f}$, both non-dimensionalized using the maximum shear. In the inviscid and non-diffusive limits, they have shown that the maximum growth rate decreases with $N$ and increases with $\tilde{f}$. However, in presence of strong thermal diffusion, which suppresses the effect of stratification, the classical results for the two-dimensional horizontal shear instability are recovered \citep[][]{Parketal2021}.\\

It is well-known that the non-linear evolution of the shear instability leads to the formation of periodic Kelvin-Helmholtz billows. In two-dimensional homogeneous fluids, these vortices remain coherent and are essentially stable except for the pairing instability \citep[][]{Patnaiketal1976, CorcosSherman1976, KlaassenPeltier1985a, KlaassenPeltier1985, KlaassenPeltier1989}. Nowadays, such classical configuration serves even as a benchmark for testing numerical codes \citep[][]{Lecoanetetal2016}. In contrast, in two-dimensional stably stratified vertically sheared layers, secondary instabilities of the primary Kelvin-Helmholtz vortices and of the braids in between can occur \citep[][]{Staquet1995, KlaassenPeltier1991, MashayekPeltier2012, MashayekPeltier2012b}. These secondary instabilities are shear-driven or buoyancy-driven. The latter one is due to the vorticity intensification made possible by baroclinic effects. A similar effect occurs in two-dimensional, incompressible, variable density shear flows \citep[][]{ReinaudJolyChassaing2000, FontaneJoly2008}. In addition, many studies have looked at the non-linear three-dimensional evolution   of the Kelvin-Helmholtz vortices in various configurations such as homogeneous fluids \citep[][]{PierrehumbertWidnall1982}, stably stratified with vertical shear \citep[][]{MashayekPeltier2011, MashayekPeltier2012, MashayekPeltier2012b} or stratified and/or rotating with horizontal shear \citep[][]{SmythPeltier1994, BasakSarkar2006, AroboneSarkar2012, AroboneSarkar2013, Copeetal2020}. A wealth of different three-dimensional instabilities and non-linear behaviours can occur but, since the present paper is restricted to a two-dimensional configuration, they will not be discussed further.\\

The present work develops upon \cite{Parketal2021} by studying the non-linear evolution of the horizontal shear instability of the hyperbolic tangent velocity profile in the presence of stratification and non-traditional rotation. As justified later in section \S \ref{Sect2}, the disturbances will be imposed to depend only on the horizontal coordinates, i.e. they will be two-dimensional, but with three velocity components. This configuration will allow us to concentrate only on the dynamics of the shear instability. Regarding the non-linear evolution of inertial instability, we refer to the recent work by \cite{Barkeretal2019, Barkeretal2020} and \cite{Dymottetal2023} on the analogous Goldreich-Schubert-Fricke (GSF) instability, which not only is enabled when the Solberg-Hoiland criteria for stability are violated but also holds in the presence of viscous and diffusive processes.\\

The outline of the paper is as follows. In section \S \ref{Sect2}, the problem and numerical methods are presented. Section \S \ref{nonlin_evol} first presents typical non-linear evolutions as the non-traditional Coriolis parameter and the Brunt-V\"ais\"al\"a frequency are varied. Sections \S \ref{ener_analyse} and \S \ref{enstro_analyse} describe the effects of the stratification and horizontal rotation component on the evolutions of the kinetic and potential energies, as well as the enstrophy. In \S \ref{sect_secondary_inst}, we investigate the origin of the secondary instabilities that are observed for weak and moderate stratification. Finally, in \S \ref{strong_strat}, the strongly stratified limit is investigated in detail. It is shown that the governing equations in this limit can be transformed back into the traditional equations thanks to a change of variables. This allows us to derive scaling laws for the energies and they are tested against the results of the numerical simulations. Section \S \ref{Sect_conclusion} draws the conclusions and discusses the results in the context of stellar or planetary atmospheric/oceanic dynamics, before giving ideas for further research. 
\section{Methods}\label{Sect2}
\subsection{Formulation of the problem}

\subsubsection{Governing Equations}
\begin{figure}
        \centering
        \begin{subfigure}[b]{0.45\textwidth}
        \centering
         \includegraphics[width=0.55\textwidth]{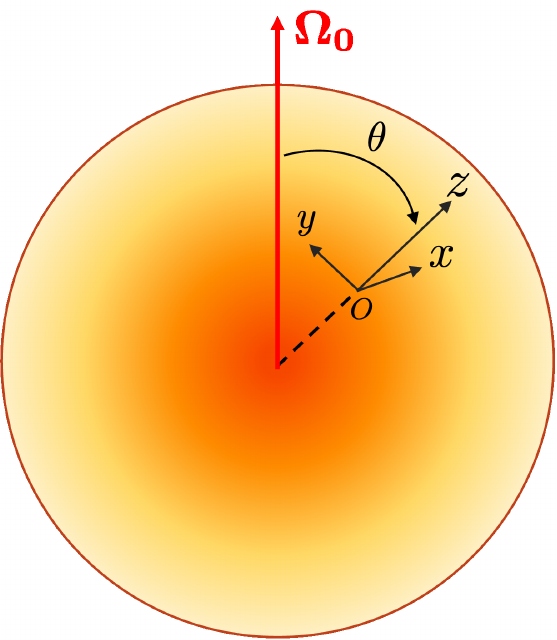}
        \caption{ }
        \end{subfigure}
        \begin{subfigure}[b]{0.45\textwidth}
        \centering
         \includegraphics[width=0.68\linewidth, trim=0cm 0cm 0cm 0cm, clip]{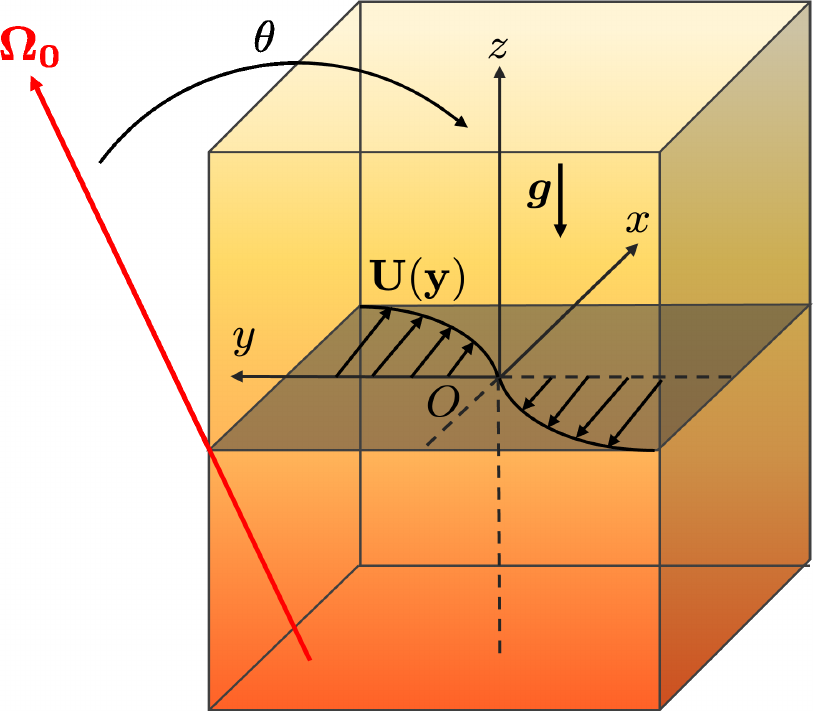}
        \caption{ }
        \label{hyperbolic tangent}
        \end{subfigure}
        \captionsetup{width=\linewidth, justification=justified, format=plain}
        \caption{(a) Sketch of the local Cartesian frame at the colatitude $\theta$ in a sphere rotating at rate $\Omega_{0}$. (b) Sketch of the horizontal shear flow in a local Cartesian frame. The colour gradient from yellow to orange illustrates the increasing density as $z$ decreases.}
         \label{contexte}
\end{figure}

We consider the incompressible Navier-Stokes equations under the Boussinesq and f-plane approximations: 
\begin{subequations}
\begin{align}
    \displaystyle \partial_t \boldsymbol{u}+(\boldsymbol{u}\cdot\boldsymbol{\nabla})\boldsymbol{u}+\boldsymbol{f_{0}}\times\boldsymbol{u} &=-\frac{1}{\rho_0}\boldsymbol{\nabla} p+b\boldsymbol{e_{z}}+\nu_{0}\boldsymbol{\nabla^2}\boldsymbol{u}, \label{systeme_vectoriel_a} \\
    \displaystyle \partial_t b+\boldsymbol{u}\cdot\boldsymbol{\nabla}b+N_{0}^{2}w&=D_{0}\boldsymbol{\nabla^2}b, \label{systeme_vectoriel_b} \\
    \displaystyle \boldsymbol{\nabla}\cdot\boldsymbol{u}&=0, \label{systeme_vectoriel_c}
\end{align}
\label{systeme_vectoriel}
\end{subequations}
where $\boldsymbol{u}=(u, v, w)$ is the velocity field in Cartesian coordinates $(x,y,z)$, with unit vectors $(\boldsymbol{e_x},\boldsymbol{e_y}, \boldsymbol{e_z})$ for the zonal, meridional and vertical directions, respectively (figure \ref{contexte}). $\boldsymbol{f_{0}}=(0,\tilde{f_{0}},f_{0})$ is the complete Coriolis vector with $f_{0}=2\Omega_{0}\cos\theta$ and $\tilde{f_0}=2\Omega_{0}\sin\theta$ the traditional and non-traditional Coriolis components, respectively, where $\Omega_{0}$ is the angular velocity and $\theta$ is the colatitude (see figure \ref{contexte}). $p$ is the pressure, $b=-g\hat{\rho}/\rho_0$ is the buoyancy perturbation, where $g$ is the gravity, and the density has been decomposed as $\rho=\rho_0+\overline{\rho}(z)+\hat{\rho}(x,y,z,t)$ with $\rho_0$ a reference density, $\overline{\rho}$ the mean density profile and $\hat{\rho}$ the perturbation density. $\nu_{0}$ is the kinematic viscosity, $\displaystyle N_{0}=\sqrt{-g\partial_{z}\overline{\rho}/\rho_{0}}$ is the Brunt-Väisälä frequency and $D_{0}$ is the diffusivity of the stratifying agent, which could represent temperature or salinity.
\subsubsection{Base state}\label{presentation_base_state}
We consider as base velocity a horizontal shear flow with the classical hyperbolic tangent profile (figure \ref{hyperbolic tangent}) \citep[][]{Michalke1964}
\begin{equation}
    \boldsymbol{U}=(U(y),0,0) \quad \textrm{with} \quad
    U(y)=U_0\,\textrm{tanh}\left(\frac{y}{L_0}-y_{c}\right),
    \label{profil_vitesse_base}
\end{equation}
where $U_0$ and $L_0$ are the characteristic velocity and length, respectively and $y_{c}$ is the center of the shear flow. Such a flow is a solution of (\ref{systeme_vectoriel}) in the inviscid and non-diffusive limits in presence of the base buoyancy field \citep[][]{Parketal2021}:
\begin{equation}
	B=-\tilde{f}_{0}U,
    \label{flottabilite_de_base}
\end{equation}
This relation derives from the vertical momentum equation and could be interpreted as  the thermal wind balance in presence of the full Coriolis acceleration.

\subsubsection{Governing equations for the perturbations}
The base state (\ref{profil_vitesse_base}-\ref{flottabilite_de_base}) is subjected to perturbations denoted by a hat:
\begin{equation}
    (u,v,w,p,b)=(U,0,0,0,B)+(\hat{u},\hat{v},\hat{w},\hat{p},\hat{b}).
    \label{base_state_perturbation_decomposition}
\end{equation}
By introducing this decomposition into (\ref{systeme_vectoriel}) and by neglecting the viscous and mass diffusions of the base state, we obtain the governing equations for the perturbations:
\begin{subequations}
\begin{align}
    \displaystyle \partial_t \hat{u}+U\partial_x\hat{u}+\hat{v}\partial_yU+\boldsymbol{\hat{u}}\cdot\boldsymbol{\nabla}\hat{u}-f_{0}\hat{v}+\tilde{f_{0}}\hat{w}&=-\frac{1}{\rho_{0}}\partial_x \hat{p}+\nu_{0}\boldsymbol{\nabla^2}\hat{u},
    \label{systeme_brut_equations_a}\\
    \displaystyle \partial_t \hat{v}+U\partial_x\hat{v}+\boldsymbol{\hat{u}}\cdot\boldsymbol{\nabla}\hat{v}+f_{0}\hat{u}&=-\frac{1}{\rho_{0}}\partial_y \hat{p}+\nu_{0}\boldsymbol{\nabla^2}\hat{v},
    \label{systeme_brut_equations_b}\\
    \displaystyle \partial_t \hat{w}+U\partial_x\hat{w}+\boldsymbol{\hat{u}}\cdot\boldsymbol{\nabla}\hat{w}-\tilde{f_{0}}\hat{u}&=-\frac{1}{\rho_{0}}\partial_{z}\hat{p}+\hat{b}+\nu_{0}\boldsymbol{\nabla^2}\hat{w}, 
    \label{systeme_brut_equations_c}\\
    \displaystyle \partial_t \hat{b}+U\partial_x\hat{b}+\hat{v}\partial_{y}B+\boldsymbol{\hat{u}}\cdot\boldsymbol{\nabla}\hat{b}+N_{0}^2\hat{w}&=D_{0}\boldsymbol{\nabla^2}\hat{b},
    \label{systeme_brut_equations_d}\\
    \boldsymbol{\nabla}\cdot\boldsymbol{\hat{u}}&=0.
    \label{systeme_brut_equations_e}
\end{align}
\label{systeme_brut_equations}
\end{subequations}
\noindent From (\ref{systeme_brut_equations}), it can be seen that if the perturbations are independent of the vertical coordinate initially, they will remain two-dimensional for any time but with three velocity components due to the non-traditional Coriolis force \citep[][]{ToghraeiBillant2022}. Hence, we can restrict the following study to two-dimensional perturbations depending only on the horizontal coordinates. This implies that the divergence equation (\ref{systeme_brut_equations_e}) can be automatically satisfied by introducing a streamfunction $\hat{\psi}$ such as $\hat{u}=-\partial_y \hat{\psi}$, $v=\partial_x \hat{\psi}$. As a consequence, the traditional Coriolis force can be written as $(-f_{0}\hat{v}, +f_{0}\hat{u},0)=-f_{0}\boldsymbol{\nabla}\hat{\psi}$ and therefore removed from (\ref{systeme_brut_equations}) by redefining the pressure $\hat{p}\rightarrow\hat{p}-\rho_{0}f_{0}\hat{\psi}$. Thus, (\ref{systeme_brut_equations}) becomes formally independent of $f_{0}$ even if it is non-zero.\\
Finally, (\ref{systeme_brut_equations}) is non-dimensionalized by using $L_{0}$ and $L_{0}/U_{0}$ as length and time units. It becomes:
\begin{subequations}
\begin{align}
    \displaystyle \partial_t \hat{u}+U\partial_x\hat{u}+\hat{v}\partial_yU+\boldsymbol{\hat{u}}\cdot\boldsymbol{\nabla}\hat{u}+\tilde{f}\hat{w}&=-\partial_x \hat{p}+\frac{1}{Re}\boldsymbol{\nabla^2}\hat{u},
    \label{systeme_adim_equations_a}\\
    \displaystyle \partial_t \hat{v}+U\partial_x\hat{v}+\boldsymbol{\hat{u}}\cdot\boldsymbol{\nabla}\hat{v}&=-\partial_y \hat{p}+\frac{1}{Re}\boldsymbol{\nabla^2}\hat{v},
    \label{systeme_adim_equations_b}\\
    \displaystyle \partial_t \hat{w}+U\partial_x\hat{w}+\boldsymbol{\hat{u}}\cdot\boldsymbol{\nabla}\hat{w}-\tilde{f}\hat{u}&=\hat{b}+\frac{1}{Re}\boldsymbol{\nabla^2}\hat{w}, 
    \label{systeme_adim_equations_c}\\
    \displaystyle \partial_t \hat{b}+U\partial_x\hat{b}-\tilde{f}\hat{v}\partial_yU+\boldsymbol{\hat{u}}\cdot\boldsymbol{\nabla}\hat{b}+N^2\hat{w}&=\frac{1}{Re Sc}\boldsymbol{\nabla^2}\hat{b},
    \label{systeme_adim_equations_d}\\
    \partial_x \hat{u}+\partial_y \hat{v}&=0,
    \label{systeme_adim_equations_e}
\end{align}
\label{systeme_adim_equations}
\end{subequations}
\noindent where
\begin{equation}
	N=\frac{N_{0}L_{0}}{U_{0}}, \quad \tilde{f}=\frac{\tilde{f_{0}}L_{0}}{U_{0}}, \quad Re=\frac{U_{0}L_{0}}{\nu_{0}}, \quad Sc=\frac{\nu_{0}}{D_{0}},
	\label{nb_adim}
\end{equation}
\noindent are the non-dimensional Brunt-Väisälä frequency, the non-dimensional non-traditional Coriolis parameter, the Reynolds number and the Schmidt number, respectively.
\subsection{Numerical methods}
\begin{table}
  \begin{center}
\def~{\hphantom{0}}
  \begin{tabular}{lccccccccc}
      Reference of the simulation  & $\mathit{N}$ & $\mathit{\tilde{f}}$  & $k_{x\textrm{m}}$ &  $L_x$ & $L_y$ & $N_x\times N_y$ & $dt$ & $k_{\rm{max}}$ & $k_{\eta}$\\[3pt]
       ``Traditional''   &$ 5$ & $0$  & $0.45$ & $13.96$ & $18$ &  $512\:\textrm{x}\:512$ & $0.005$ & $89.36$ & $14.85$\\
       ``Traditional-like''   & $2$ & $1.5$ & $0.52$ & $11.97$ & $36$ &  $512\:\textrm{x}\:1024$ & $0.005$ & $89.36$ & $15.23$\\
       ``Mixed''   & $1$ & $0.5$ &  $0.45$ & $13.96$ & $36$ &  $512\:\textrm{x}\:1024$ & $0.005$ & $89.36$ & $17.86$\\
       ``Non-traditional''   & $0.5$ & $0.5$  & $0.3$ & $20.94$ & $72$ &  $1024\:\textrm{x}\:2048$ & $0.002$ & $153.6$ & $38.93$     
  \end{tabular}
  \captionsetup{width=1.\linewidth, justification=justified, format=plain}
  \caption{Summary of the physical and numerical parameters of typical simulations. $N$ is the non-dimensional Brunt-Väisälä frequency, $\tilde{f}$ the non-dimensional non-traditional parameter, $k_{x\textrm{m}}$ the most amplified wave number, $L_x$ and $L_{y}$ the sizes of the domain with $L_x=2\pi/k_{x\textrm{m}}$, $N_x\times N_y$ the number of collocation points in the $x$ and $y$ directions, $dt$ the time step, $k_{\rm{max}}$ the maximum wave number and $k_{\eta}$ the Kolmogorov wavenumber. The Reynolds and Schmidt numbers are fixed to $Re=2000$ and $Sc=1$ throughout the paper.}
  \label{tab_ref_simu}
  \end{center}
\end{table}
The equations (\ref{systeme_adim_equations}) are integrated by means of a Fourier pseudo-spectral method in a doubly periodic domain \citep[][]{Deloncleetal2008, Deloncleetal2014}. The time integration is carried out by a fourth-order Runge-Kutta scheme except for the viscous and diffusive terms that are integrated exactly. The $2/3$ rule for de-aliasing is used (i.e. the upper third of the modes in each direction is removed). The velocity field of the perturbation is initialised by a divergence-free random noise of amplitude 0.001 whereas the buoyancy $\hat{b}$ is initially set to zero. The form of the initial perturbation has no particular effect on the results. The random noise serves simply as a seed for the instability development.\\ 

\noindent The Reynolds number and the Schmidt number will be fixed to $Re=2000$ and $Sc=1$ throughout the paper. This value of $Re$ has been chosen so as to have weak viscous and diffusion effects with a reasonable computational cost. The effects of varying $Re$ and $Sc$ will be studied in a future paper.\\

\noindent The size $L_{x}$ of the domain (table \ref{tab_ref_simu}) is set to the most amplified wavelength for each set of parameters $(\tilde{f}, N, Re, Sc)$. This prevents the occurrence of the pairing instability which leads to the merging of neighbouring Kelvin-Helmholtz vortices \citep[][]{HoHuerre1984, KlaassenPeltier1989}, allowing us to focus on the non-linear evolution of the shear instability. However, in appendix \ref{appB}, we present some simulations with $L_{x}$ equal to twice the dominant wavelength in which the pairing instability occurs. The size $L_{y}$ is chosen sufficiently large to minimise the effect of the periodic boundary conditions and the shear flow is centered in the domain $y_{c}=L_{y}/2$. The resolution before de-aliasing, $(N_{x}, N_{y})$, is given in table \ref{tab_ref_simu} for four typical sets of parameters. It ranges from 512$\,\times\,$512 to 1024$\,\times\,$2048 depending on the size of the domain and whether or not there is a transition to small-scale turbulence. The time step has been chosen so as to satisfy the Courant-Friedrichs-Lewy condition for each case (table \ref{tab_ref_simu}).\\

\noindent When there is no transition to turbulence, (simulations denoted ``Traditional'', ``Traditional-like'', and ``Mixed'' in table \ref{tab_ref_simu}), the accuracy of the simulations have been checked by running control simulations with a double resolution. In order to have the same initial conditions regardless of the resolution, the original simulations (with resolution $(N_{x}, N_{y})$) and the control simulations (with resolution $(2N_{x}, 2N_{y})$) have been initialised by the most unstable eigenmode with a small amplitude instead of a random noise. The total velocity has been observed to differ by less than 1\% at each common point between the two simulations. When there is a transition to turbulence (simulations denoted ``Non-traditional'' in table \ref{tab_ref_simu}), the resolution has been fixed so as to have $k_{\rm{max}}/k_{\eta}\geqslant1$ where $k_{\rm{max}}$ is the highest wavenumber and $k_{\eta}=(\overline{\epsilon}/\nu_{0}^3)^{1/4}$ the Kolmogorov wavenumber, where $\overline{\epsilon}$ is the maximum value over time of the dissipation rate of kinetic energy averaged in space $\displaystyle\overline{\epsilon}=\frac{1}{L_{x}L_{y}Re}\int\int(\nabla\hat{u})^{2}dxdy$  \citep[][]{deBruynKopsRiley1998}.\\

\noindent Similarly, the effect of the periodic boundary conditions in the $y$ direction has been checked by increasing the size $L_{y}$ keeping the same resolution $dy=L_{y}/N_{y}$ in simulations initialised by the most unstable eigenmode. When there is no transition to turbulence, the evolution of the total kinetic energy of the perturbations has been found to differ by less than 2\% when the size is doubled. For the simulation ``Non-traditional'' in table \ref{tab_ref_simu} for which turbulence arises, a maximum instantaneous difference of 5\% has been observed before t$=$100 when $L_{y}$ is increased from 72 to 144. However, for $t>100$, the difference increases up to 15\% due to the high sensitivity of turbulence on initial conditions.\\

\noindent In general, the simulations require a larger resolution and a larger size $L_{y}$ as $\tilde{f}$ is increased for a given buoyancy frequency, especially for low values of $N$. In order to keep computational costs reasonable, the highest value of $\tilde{f}$ that has been achieved for each $N$ for $Re=2000$ is $\tilde{f}=N$.\\ 

\noindent As mentioned before, the viscous and mass diffusions of the base state are neglected in (\ref{systeme_adim_equations}). This approximation is tested in appendix \ref{appA}, for two simulations of table \ref{tab_ref_simu}, against simulations where the base state is allowed to decay. It is shown that simulations that take into account the base state dissipations are very similar, both qualitatively and quantitatively, to those presented in the following since the Reynolds number is large $Re=2000$.

\subsection{Linear stability analysis}\label{subsect3}
\noindent Prior to performing a non-linear simulation for a given set of parameters, the most unstable eigenmode has been determined. To do so, the equations (\ref{systeme_adim_equations}) without non-linear terms have been integrated over a long time in several simulations for different horizontal sizes $L_{x}$, i.e. different streamwise wavenumbers $k_{x}=2\pi/L_{x}$. The perturbation then tends toward the most unstable eigenmode in each simulation. Provided that the instability is non-oscillatory, its growth rate can be retrieved from:
\begin{equation}
	\sigma=\lim_{t\rightarrow\infty}\frac{1}{2}\frac{d\ln(\hat{K})}{dt},
	\label{sigma_}
\end{equation}
where $\hat{K}$ is the perturbations kinetic energy integrated over the domain. The horizontal size $L_{x}$ in the non-linear simulation has been then fixed to $L_{x}=2\pi/k_{x\textrm{m}}$, where $k_{x\textrm{m}}$ is the wavenumber yielding the maximum growth rate, $\rm{max}(\sigma)$.
\begin{figure}
        \centering
         \begin{subfigure}[b]{0.48\textwidth}
        \centering
        \includegraphics[width=\textwidth]{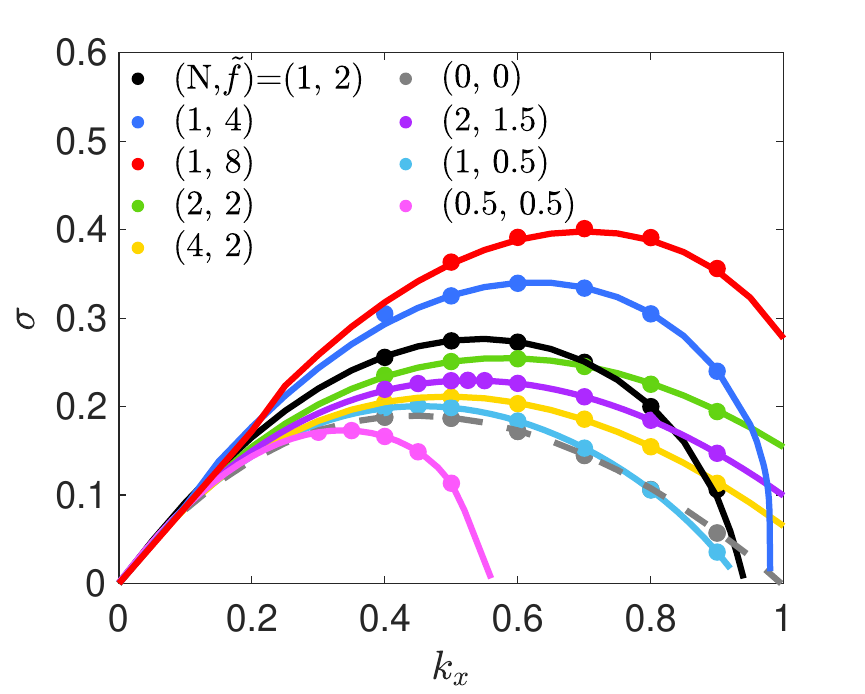}
        \caption{ }
        \label{linear_growthrate_Park_et_al}
        \end{subfigure}
        \begin{subfigure}[b]{0.48\textwidth}
        \centering
        \includegraphics[width=\textwidth]{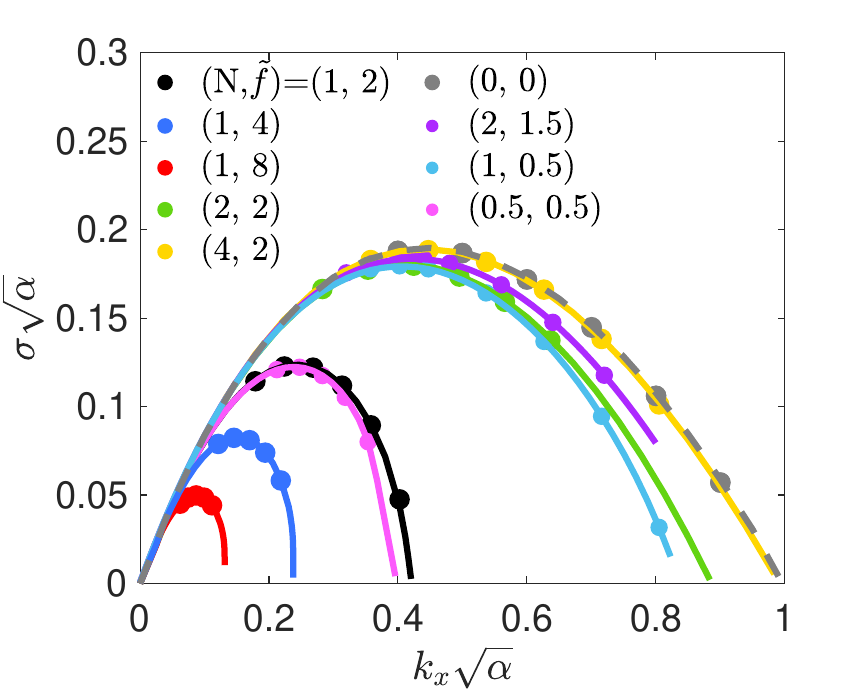}
        \caption{ }
        \label{scaled_Park_et_al}
        \end{subfigure}
        \captionsetup{width=\linewidth, justification=justified, format=plain}
        \caption{(a) Growth rate $\sigma$ as a function of the wavenumber $k_{x}$ for different parameters $(N, \tilde{f})$, coming from \cite{Parketal2021} and from the cases listed in table \ref{tab_ref_simu}. The lines represent the results of \cite{Parketal2021} in the inviscid limit, whereas the symbols show the present results for $Re=2000$ and $Sc=1$. (b) Rescaled growth rate $\sigma\sqrt{\alpha}$ as a function of the rescaled wavenumber $k_{x}\sqrt{\alpha}$, where $\alpha=N^{2}/(\tilde{f}^{2}+N^{2})$.}
        \label{linear_growthrate}
\end{figure}
Figure \ref{linear_growthrate_Park_et_al} shows the results of the linear stability analysis for the set of parameters $(N, \tilde{f})$ studied by \cite{Parketal2021} by means of a pseudospectral Chebyshev method in the inviscid and non-diffusive limits. The results for the parameters listed in table \ref{tab_ref_simu} are also displayed. The growth rates obtained by the present method (symbols) are in good agreement with those of \cite{Parketal2021} (solid lines) for all the wavenumbers $k_{x}$ investigated, even if the Reynolds number is here $Re=2000$. This validates our numerical method and indicates that viscous and diffusive effects have a very weak impact on the stability of the tangent hyperbolic profile for $Re=2000$ and $Sc=1$.\\

\noindent Since only the most unstable mode is computed for a given $k_{x}$, i.e. for a given size $L_{x}$ of the domain, our method does not allow us to compute the growth rate of too low wavenumber $k_{x}$ because, then, the mode is dominated by the one for 2$k_{x}$ if it has a larger growth rate. However, this limitation is not a problem here since we are interested only by the most amplified wavenumber for each set of parameters $(N, \tilde{f})$.\\

\noindent As discussed by \cite{Parketal2021}, figure \ref{linear_growthrate_Park_et_al} shows that the non-traditional Coriolis force is destabilizing when $N\geqslant1$ since the maximum growth rate $\sigma_{\textrm{m}}$ increases with $\tilde{f}$ for a given value of $N$ and is always larger than the one in the traditional limit $\tilde{f}=0$: $\sigma_{\textrm{m}}=0.1891$. In contrast, the maximum growth rate decreases as $N$ increases for a given value of $\tilde{f}$. It is also noteworthy that the most amplified wavenumber $k_{x\textrm{m}}$ tends to increase with $\tilde{f}$ and to decrease with $N$. When $N\gg$1, it is always larger than the value $k_{x\textrm{m}}=0.44$ in the traditional limit (figure \ref{linear_growthrate_Park_et_al}). In contrast, for $(N, \tilde f)=(0.5,0.5)$, the maximum growth rate and most amplified wavenumber are both lower than for $\tilde f=0$. Figure \ref{scaled_Park_et_al} will be discussed in \S \ref{strong_strat}.

\section{Non-linear dynamics} \label{Sect4}
\subsection{Exploration of the parameter space}\label{nonlin_evol}
\begin{figure}
	\centering
	\includegraphics[width=0.75\linewidth, trim=0cm 0cm 0cm 0cm, clip]{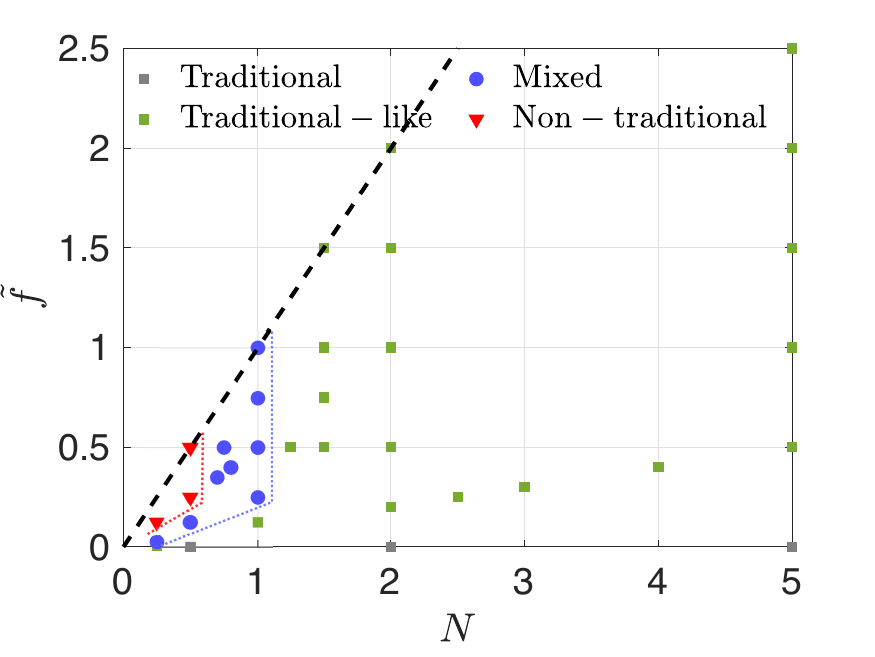}
	 \captionsetup{width=\linewidth, justification=justified, format=plain}
	\caption{Map of the simulations in the parameter space $(N,\tilde{f})$ for $Re=2000$ and $Sc=1$. Grey, green, blue and red symbols represent the simulations where ``Traditional'', ``Traditional-like'', ``Mixed'' and ``Non-traditional'' behaviours have been observed, respectively. The black dashed line corresponds to the line $N=\tilde{f}$ delimiting the explored region in the present paper. The blue and red dotted lines indicate approximately the limits of the ``mixed'' domain.}
	\label{espace_parametres}
\end{figure}
Many non-linear simulations have been performed for different values of $N$ and $\tilde{f}$ in the ranges $0\leqslant N\leqslant5$, and $0\leqslant\tilde{f}\leqslant\textrm{min}(N, 2.5)$. Only positive values of $\tilde{f}$ have been considered since (\ref{systeme_adim_equations}) is invariant under the changes $\tilde{f}, \hat{w}, \hat{b} \rightarrow -\tilde{f}, -\hat{w}, -\hat{b}$. As mentioned before, the horizontal size $L_{x}$ has been considered as $L_{x}=2\pi/k_{xm}$, where $k_{xm}$ is the wavelength of the most unstable mode determined by linear stability analysis for each set of parameters $(N,\tilde{f})$.\\
Figure \ref{espace_parametres} summarizes all the simulations in the parameter space $(N, \tilde{f})$. Four distinct non-linear evolutions have been observed as indicated by the different symbols in figure \ref{espace_parametres}. The limits between the different domains are indicated approximately by red and blue dotted lines.
\begin{figure}
        \begin{subfigure}[b]{0.495\textwidth}
        \includegraphics[width=\linewidth, trim=1.6cm 0cm 2.2cm 0.5cm, clip]{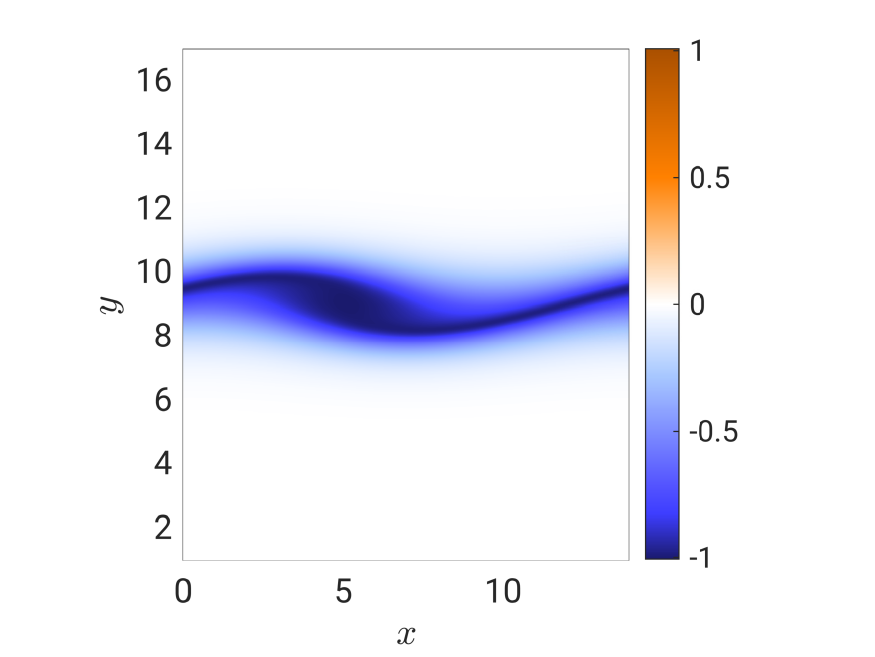}
        \caption{ }
        \label{fig_trad_a}
        \end{subfigure}
        \begin{subfigure}[b]{0.495\textwidth}
        \includegraphics[width=\linewidth, trim=1.6cm 0cm 2.2cm 0.5cm, clip]{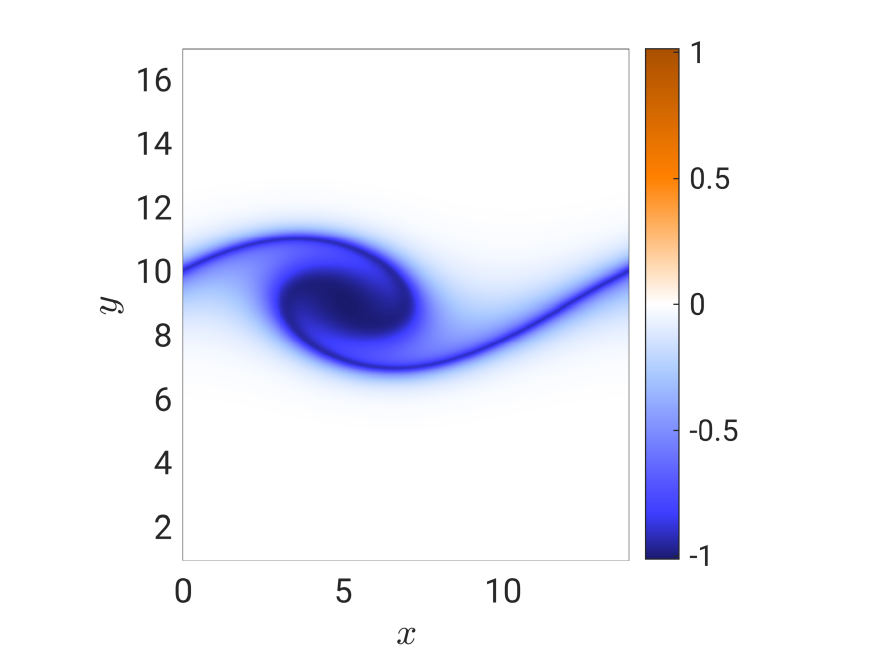}
        \caption{ }
        \label{fig_trad_b}
        \end{subfigure}
        \begin{subfigure}[b]{0.495\textwidth}
        \includegraphics[width=\linewidth, trim=1.6cm 0cm 2.2cm 0.5cm, clip]{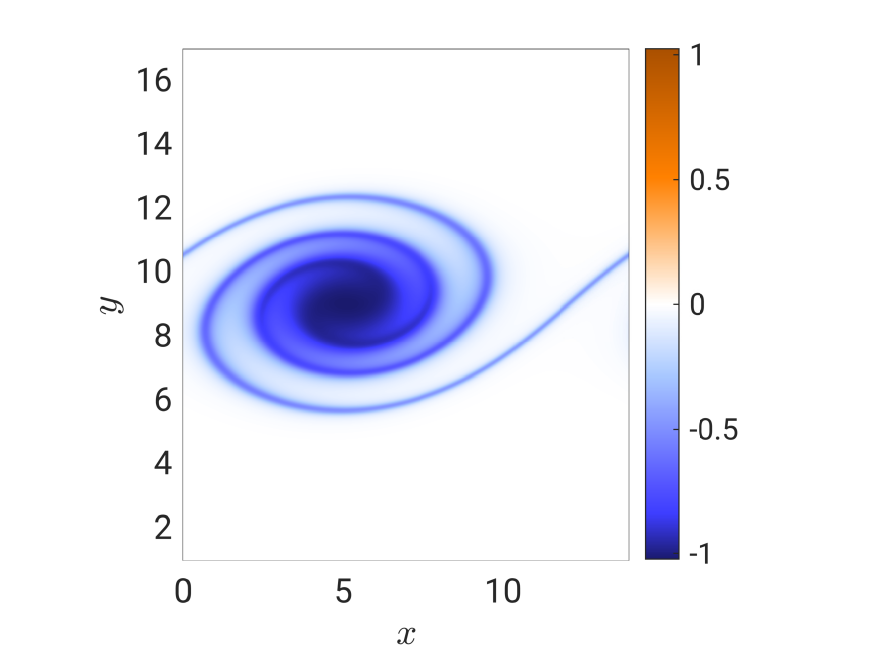}
        \caption{ }
        \label{fig_trad_c}
        \end{subfigure}
        \begin{subfigure}[b]{0.495\textwidth}
        \includegraphics[width=\linewidth, trim=1.6cm 0cm 2.2cm 0.5cm, clip]{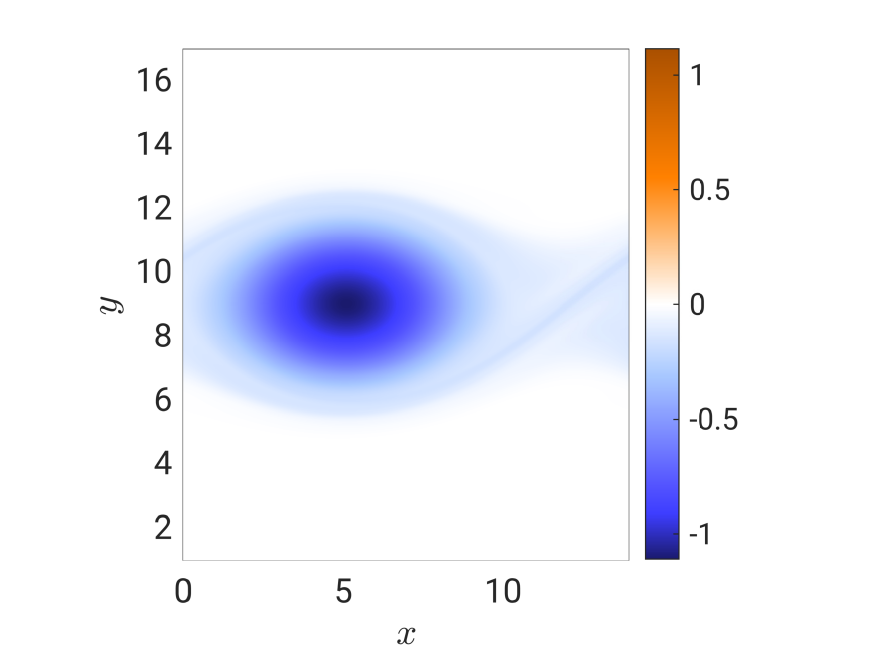}
        \caption{ }
        \label{fig_trad_d}
        \end{subfigure}
        \captionsetup{width=1.\linewidth, justification=justified, format=plain}
        \caption{Total vertical vorticity ($\xi=\partial_x v-\partial_y u$) at, (a), $t=60$, (b), $t=67$, (c), $t=78$, (d), $t=200$ for $\tilde{f}=0$, $Re=2000$ and $Sc=1$ (Traditional evolution).}
        \label{Quatre_instants_Cas_trad}
\end{figure}
\begin{figure}
        \begin{subfigure}[b]{0.49\textwidth}
        \includegraphics[width=\linewidth, trim=1.6cm 0cm 2.2cm 0.5cm, clip]{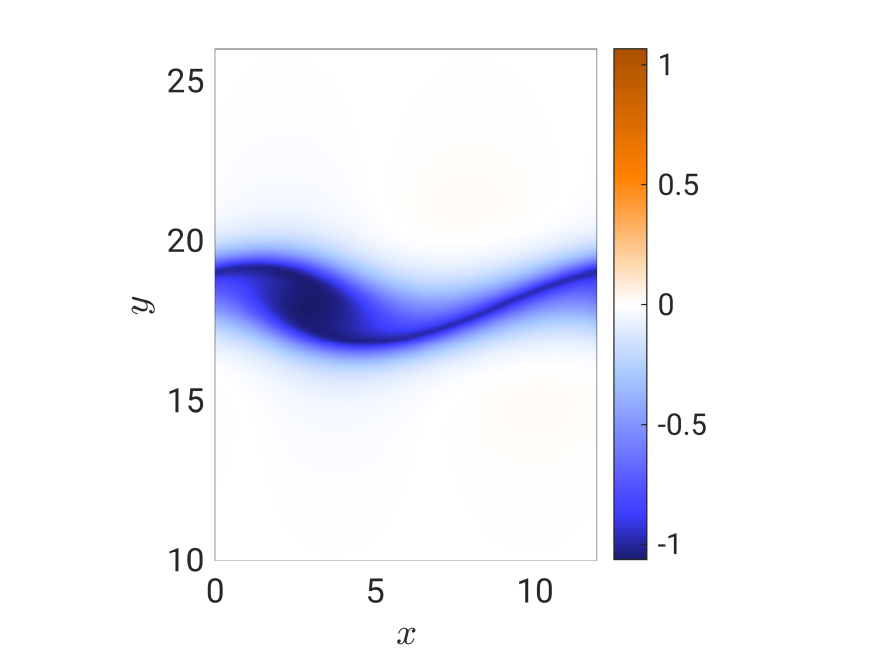}
        \caption{ }
        \label{fig_tradi-like_a}
        \end{subfigure}
        \begin{subfigure}[b]{0.49\textwidth}
        \includegraphics[width=\linewidth, trim=1.6cm 0cm 2.2cm 0.5cm, clip]{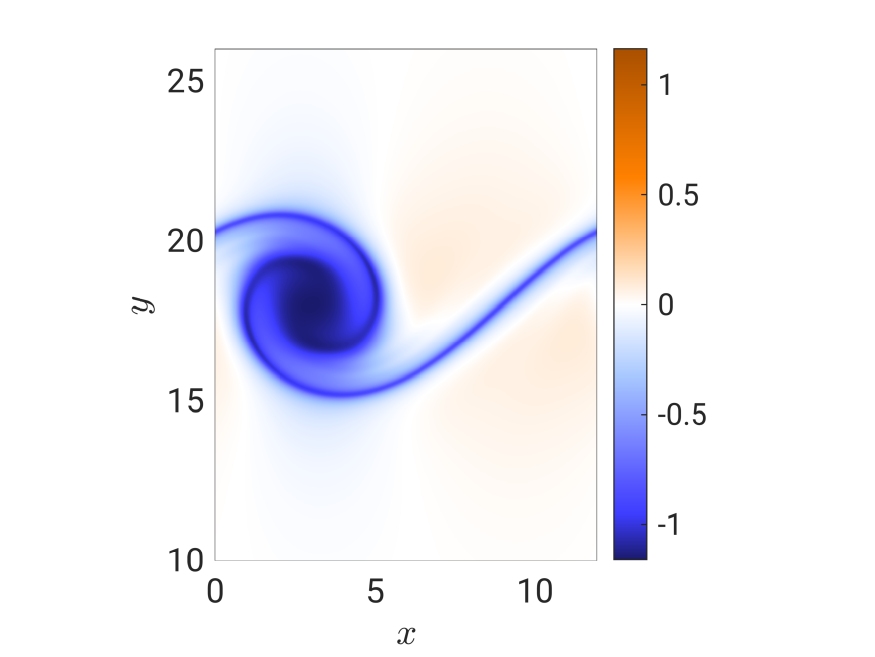}
        \caption{ }
         \label{fig_tradi-like_b}
        \end{subfigure}
        \begin{subfigure}[b]{0.49\textwidth}
        \includegraphics[width=\linewidth, trim=1.6cm 0cm 2.2cm 0.5cm, clip]{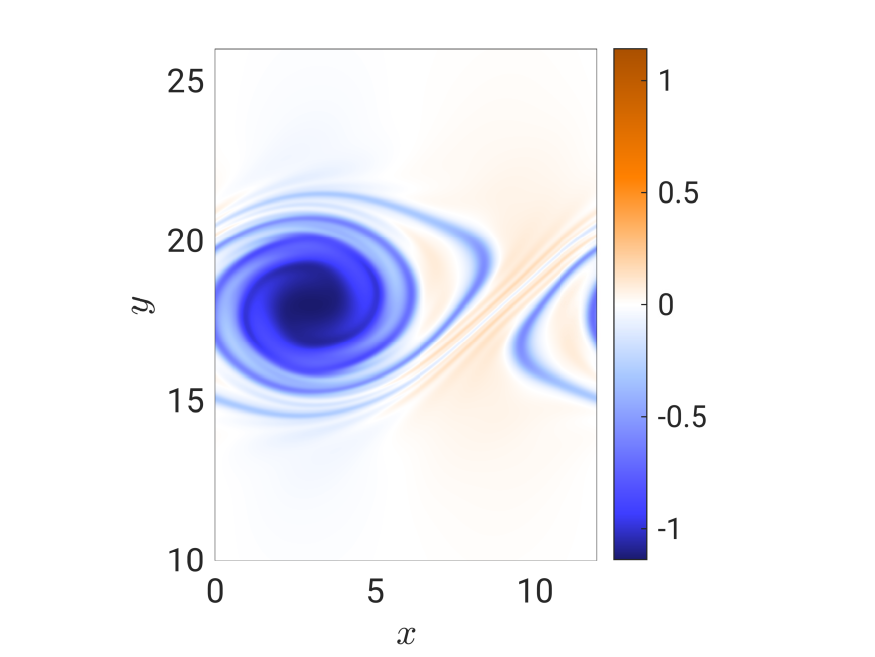}
        \caption{ }
         \label{fig_tradi-like_c}
        \end{subfigure}
        \begin{subfigure}[b]{0.49\textwidth}
        \includegraphics[width=\linewidth, trim=1.6cm 0cm 2.2cm 0.5cm, clip]{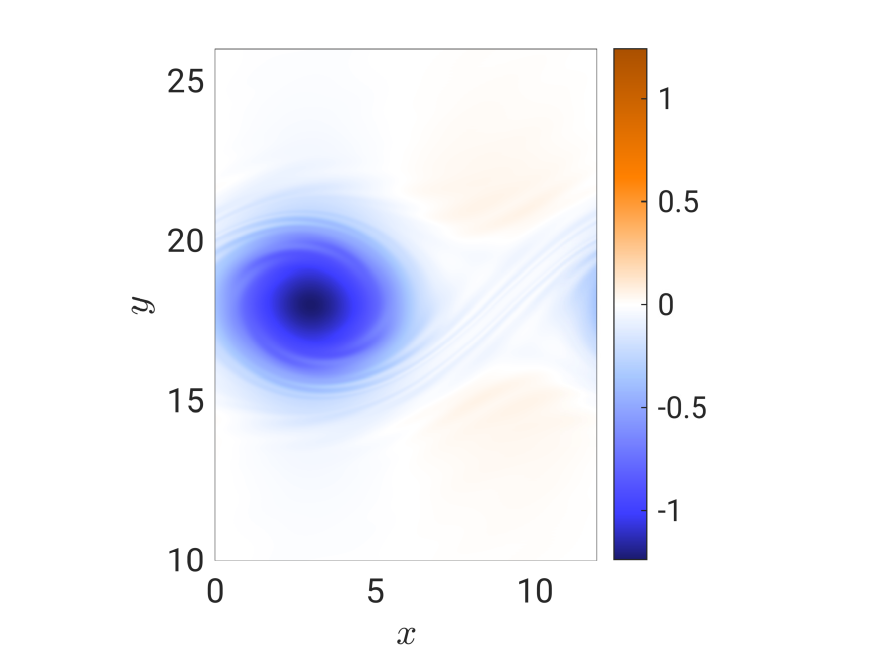}
        \caption{ }
         \label{fig_tradi-like_d}
        \end{subfigure}
        \captionsetup{width=1.\linewidth, justification=justified, format=plain}
        \caption{Same as figure \ref{Quatre_instants_Cas_trad} but for $(N, \tilde{f})=(2, 1.5)$ and (a), $t=51$, (b), $t=58$, (c), $t=70$, (d), $t=200$ (traditional-like evolution).The $y$-axis has been cropped compared to the original computational domain.}
        \label{Quatre_instants_Cas_N_grand}
\end{figure}
\begin{figure}
        \begin{subfigure}[b]{0.49\textwidth}
        \includegraphics[width=\linewidth, trim=1.6cm 0cm 2.2cm 0.5cm, clip]{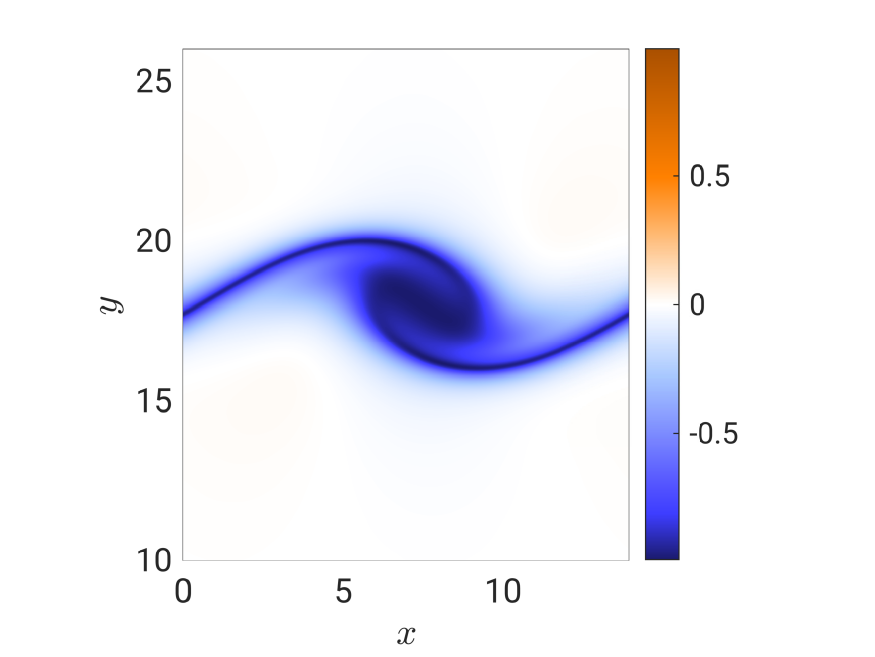}
        \caption{ }
        \label{fig_localSI_a}
        \end{subfigure}
        \begin{subfigure}[b]{0.49\textwidth}
        \includegraphics[width=\linewidth, trim=1.6cm 0cm 2.2cm 0.5cm, clip]{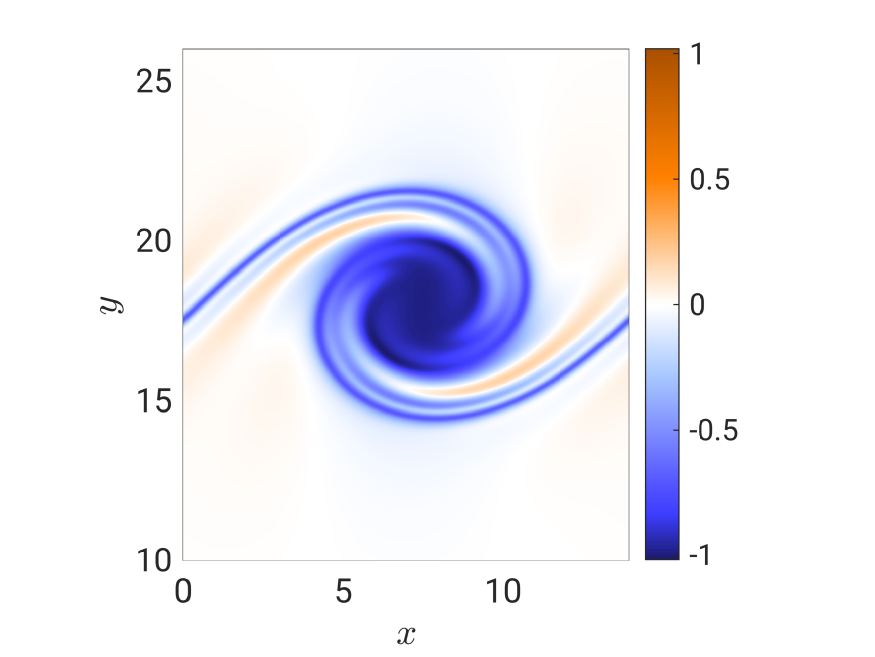}
        \caption{ }
        \label{fig_localSI_b}
        \end{subfigure}
        \begin{subfigure}[b]{0.49\textwidth}
        \includegraphics[width=\linewidth, trim=1.6cm 0cm 2.2cm 0.5cm, clip]{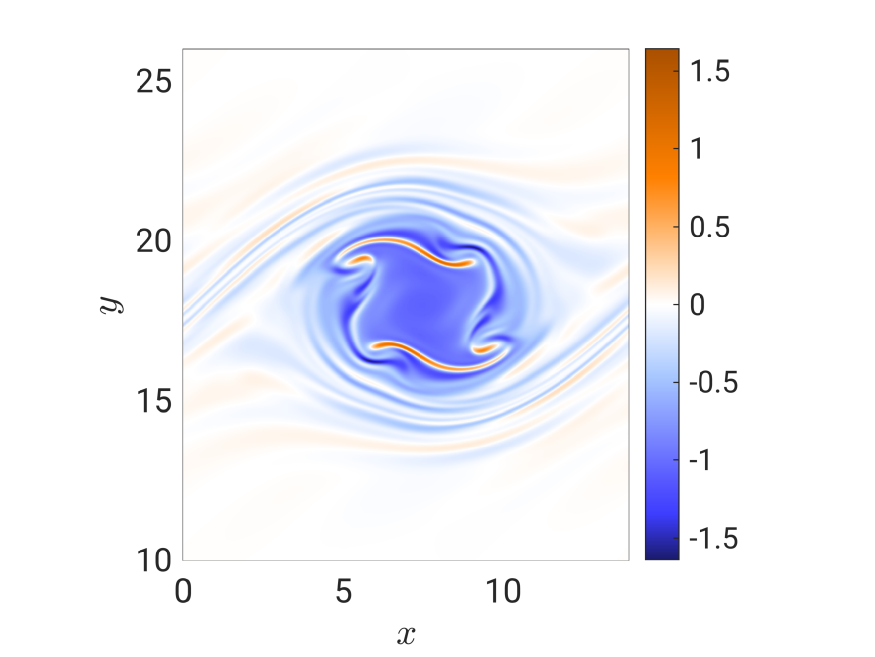}
        \caption{ }
        \label{fig_localSI_c}
        \end{subfigure}
        \begin{subfigure}[b]{0.49\textwidth}
        \includegraphics[width=\linewidth, trim=1.6cm 0cm 2.2cm 0.5cm, clip]{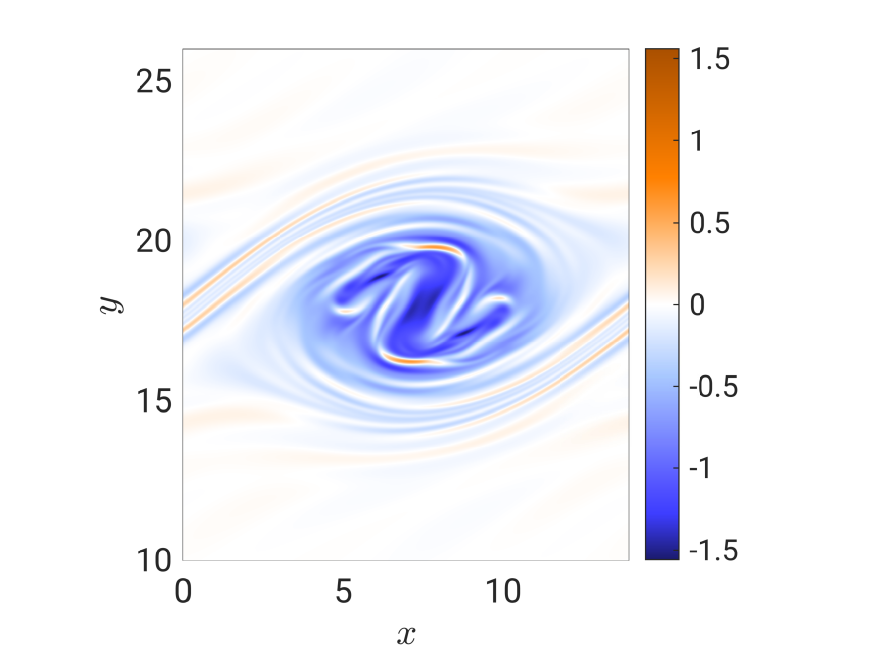}
        \caption{ }
        \label{fig_localSI_d}
        \end{subfigure}
        \captionsetup{width=1.\linewidth, justification=justified, format=plain}
        \caption{Same as figure \ref{Quatre_instants_Cas_trad} but for $(N, \tilde{f})=(1, 0.5)$ and (a), $t=65$, (b), $t=73$, (c), $t=145$, (d), $t=200$ (mixed evolution).The $y$-axis has been cropped compared to the original computational domain.}
        \label{Quatre_instants_Cas_mixte}
\end{figure}
\begin{figure}
     \begin{subfigure}[b]{0.49\textwidth}
     \includegraphics[width=\linewidth, trim=1.5cm 0cm 2.1cm 0.5cm, clip]{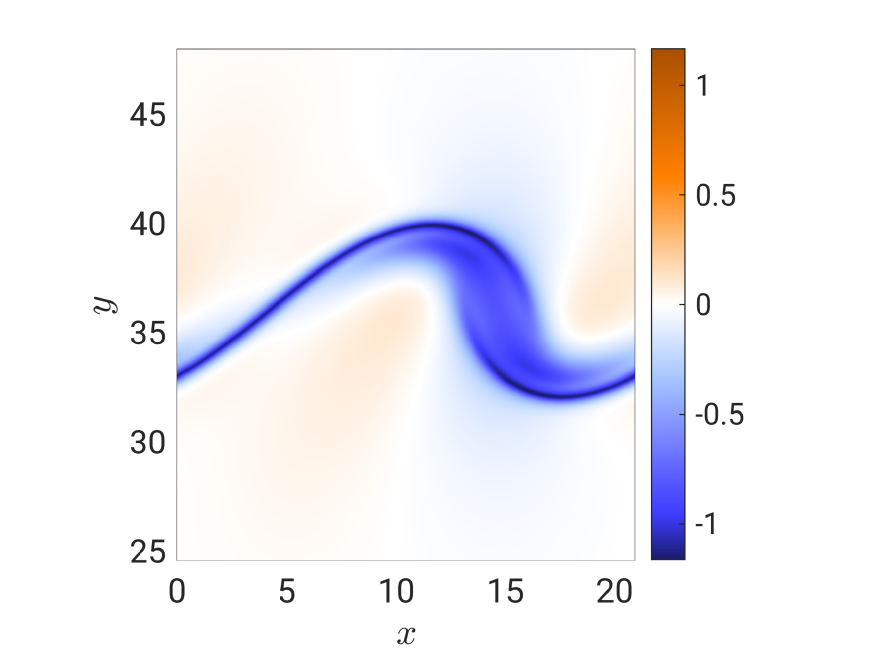}
     \caption{ }
     \label{fig_stringSI_a}
     \end{subfigure}
     \begin{subfigure}[b]{0.49\textwidth}
     \includegraphics[width=\linewidth, trim=1.5cm 0cm 2.1cm 0.5cm, clip]{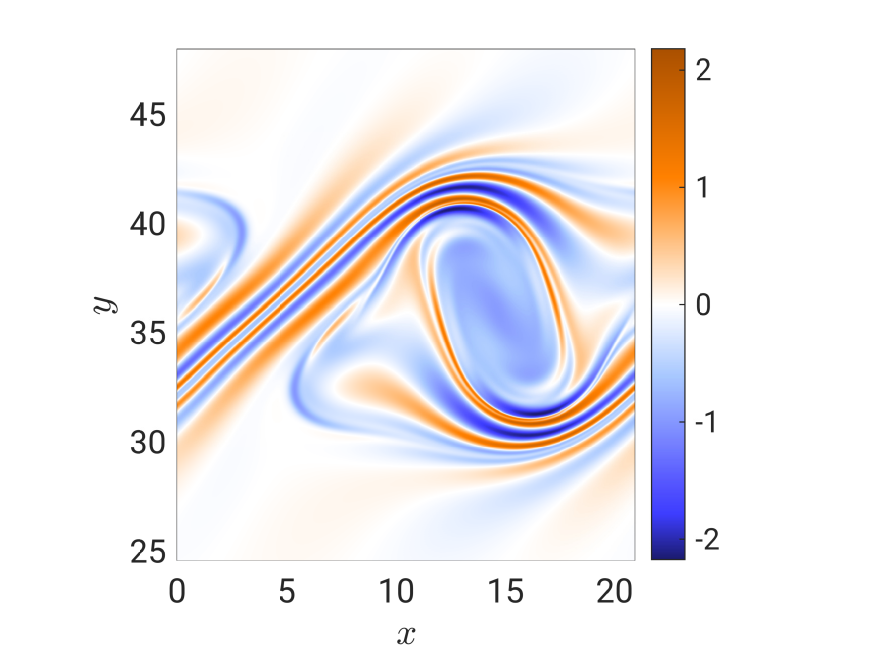}
     \caption{ }
     \label{fig_stringSI_b}
     \end{subfigure}
     \begin{subfigure}[b]{0.49\textwidth}
     \includegraphics[width=\linewidth, trim=1.5cm 0cm 2.1cm 0.5cm, clip]{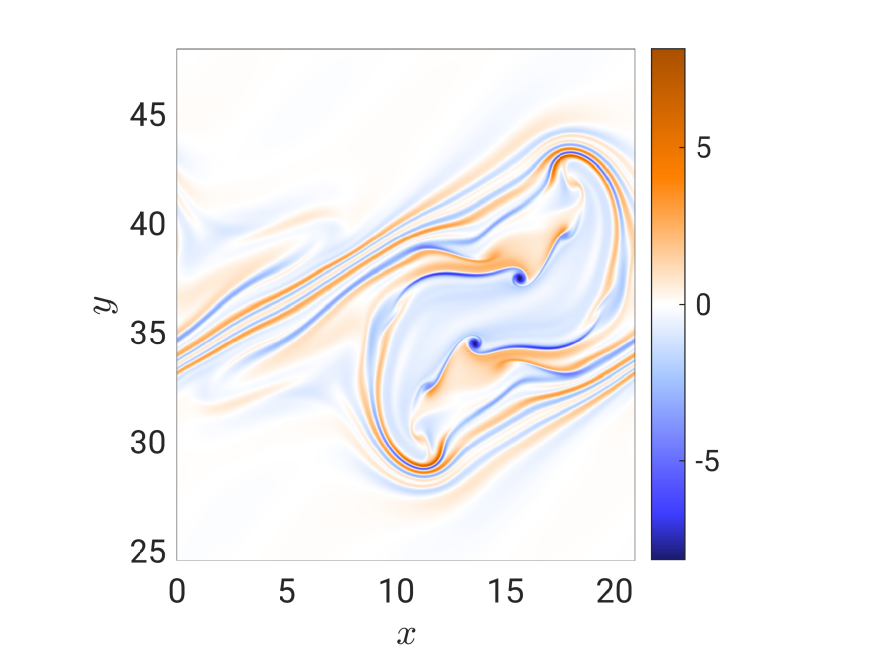}
     \caption{ }
     \label{fig_stringSI_c}
     \end{subfigure}
     \begin{subfigure}[b]{0.49\textwidth}
     \includegraphics[width=\linewidth, trim=1.5cm 0cm 2.1cm 0.5cm, clip]{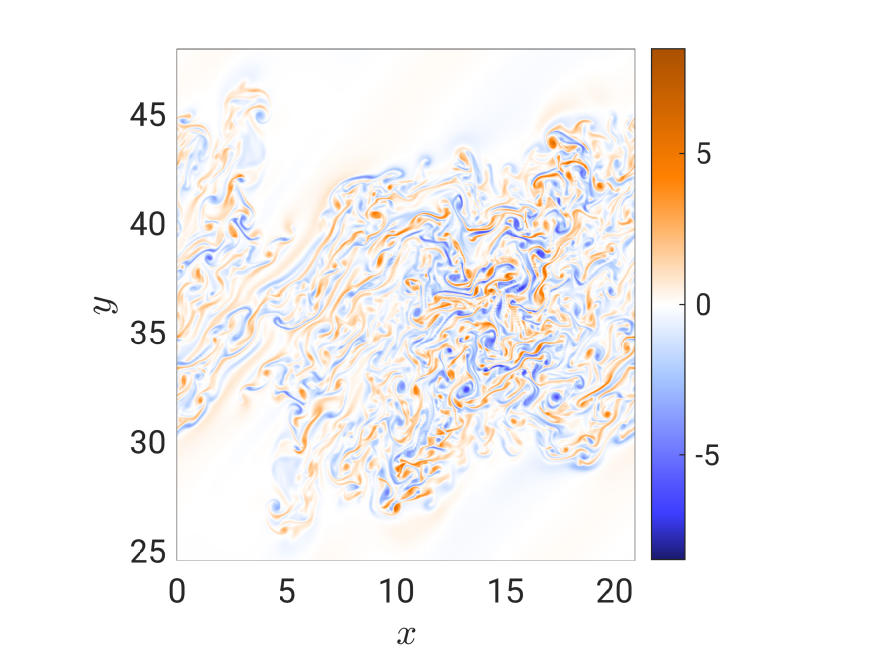}
     \caption{ }
     \label{fig_stringSI_d}
     \end{subfigure}
     \captionsetup{width=\linewidth, justification=justified, format=plain}
    \caption{Same as figure \ref{Quatre_instants_Cas_trad} but for $(N, \tilde{f})=(0.5, 0.5)$ and (a), $t=89$, (b), $t=110$, (c), $t=119$, (d), $t=146$ (non-traditional evolution). The $y$-axis has been cropped compared to the original computational domain.}
    \label{Quatre_instants_Cas_N_petit}
\end{figure}

First, the traditional evolution (figure \ref{Quatre_instants_Cas_trad}) is observed for $\tilde{f}=0$ and any value of $N$ since the governing equations for the horizontal velocity (\ref{systeme_brut_equations_a}, \ref{systeme_brut_equations_b}, \ref{systeme_brut_equations_e}) are decoupled from those for $\hat{w}$, $\hat{b}$ (\ref{systeme_brut_equations_c}, \ref{systeme_brut_equations_d}) when $\tilde{f}=0$. As seen in figure \ref{Quatre_instants_Cas_trad}, the total vertical vorticity $\xi=\partial_x v-\partial_y u$ evolves toward a quasi-steady vortex. No pairing instability occurs since the streamwise length of the domain $L_{x}$ accommodates only one wavelength. As already mentioned, appendix \ref{appB} describes the consequences of allowing two wavelengths to develop within a double-sized domain for the parameters $(N,\tilde{f})=(0.5, 0.5)$.\\
When the stratification is sufficiently strong $N\gtrsim1$ for any value of $\tilde{f}$ or when $N\lesssim 1$ for sufficiently low values of $\tilde{f}$ (green symbols in figure \ref{espace_parametres}), the evolution of the total vertical vorticity (figure \ref{Quatre_instants_Cas_N_grand}) is very similar to the one observed in the traditional limit. However, we can notice the generation of positive vorticity in the (yellowish) region surrounding the vorticity braid. The negative vorticity also goes slightly below the initial minimum value $-1$. This contrasts with the traditional case where the vertical vorticity is conserved in the inviscid limit and remains in the range $-1\leqslant\xi\leqslant0$ for $Re=2000$ (figure \ref{Quatre_instants_Cas_trad}). However, the instability generates a stable and coherent vortex like in figure \ref{Quatre_instants_Cas_trad}. These cases will be called ``traditional-like'' since we will see in the following that some differences exist compared to the traditional scenario depending on $N$ and $\tilde{f}$.\\
For moderate stratification $0.7\lesssim N\lesssim 1$ and sufficiently large values of $\tilde{f}$ or for $N \lesssim 0.5$ and intermediates values of $\tilde{f}$ (blue symbols in figure \ref{espace_parametres}), the total vertical vorticity first evolves like in the traditional case (figure \ref{fig_localSI_a} and \ref{fig_localSI_b}) but, subsequently, small secondary vortices develop within the core of the primary Kelvin-Helmholtz vortex (figure \ref{fig_localSI_c} and \ref{fig_localSI_d}). Nevertheless, it keeps its integrity. We can also notice that thin positive vorticity layers are generated and entrained inside or at the periphery of the Kelvin-Helmholtz vortex  (figure \ref{fig_localSI_b}, \ref{fig_localSI_c}, \ref{fig_localSI_d}). In addition, the minimum of the vertical vorticity decreases significantly below the initial value $-1$. Such evolution will be termed as ``mixed'' in the following.\\
Finally, when the stratification is weak $(N\leqslant0.5)$ and $\tilde{f}$ larger than a critical value depending on $N$ ($\tilde{f}\geqslant$ 0.25 for $N=0.5$ and $\tilde{f}\geqslant$ 0.125 for $N=0.25$, red symbols in figure \ref{espace_parametres}), the non-linear evolution of the shear instability differs widely from the traditional case (figure \ref{Quatre_instants_Cas_N_petit}). Thin layers with large positive and negative values of vertical vorticity are generated (figure \ref{fig_stringSI_b}) and, subsequently, stretched and amplified (figure \ref{fig_stringSI_c}) leading to an abrupt transition to small-scale turbulence (figure \ref{fig_stringSI_d}). This evolution will be called ``non-traditional'' in the following. It is also noteworthy that the size in the $y$ direction of the developing Kelvin-Helmholtz billow is around 12, i.e. twice the one for the traditional case (figure \ref{fig_stringSI_b} compared to figure \ref{fig_trad_c}). This is the main reason why the size of the domain $L_{y}$ has to be increased for small and moderate values of $N$ as $\tilde{f}$ is increased (table \ref{tab_ref_simu}). We can also notice that the shear instability appears slightly later (figure \ref{fig_stringSI_a}) since its growth rate is somewhat smaller than in the traditional limit (figure \ref{linear_growthrate_Park_et_al}). Although this regime is represented by only three points in the parameter space (figure \ref{espace_parametres}), several preliminary simulations with a lower resolution for other parameter values in this region are fully consistent with the existence of this regime.\\

\noindent The origin of these different behaviors of the vertical vorticity field is the vertical velocity and buoyancy. Indeed, while they are zero in the traditional limit, we see that the non-traditional Coriolis acceleration forces the vertical momentum equation (\ref{systeme_adim_equations_c}) as soon as the streamwise velocity perturbation $\hat u$ is non-zero. Similarly, the advection of the base buoyancy $B$ by the cross-stream velocity $\hat v$ forces the buoyancy equation (\ref{systeme_adim_equations_d}). In turn, the non-traditional Coriolis acceleration due to the vertical velocity forces the horizontal momentum equation (\ref{systeme_adim_equations_a}). Equivalently, looking at the vertical vorticity equation:
\begin{equation}
	\frac{D \xi}{Dt}=\partial_{t}\hat{\xi}+U\partial_{x}\hat{\xi}-\hat{v}\partial_{y}^{2}U+\boldsymbol{\hat{u}}\cdot\boldsymbol{\nabla}\hat{\xi}=\tilde{f}\partial_{y}\hat{w}+\frac{1}{Re}\boldsymbol{\nabla^{2}}\hat{\xi},
	\label{Equation_evolution_vorticity}
\end{equation}
where $D/Dt$ is the lagrangian derivative, $\xi$ is the total vertical vorticity and $\hat{\xi}=\partial_{x}\hat{v}-\partial_{y}\hat{u}$ is the vertical vorticity perturbation, we see that the right-hand side contains a tilting term of the horizontal background vorticity by the vertical velocity: $\tilde{f}\partial_{y}\hat{w}$.\\

\begin{figure}
  \centering
  \begin{subfigure}[b]{0.32\textwidth}
     \includegraphics[width=\linewidth, trim=1.5cm 0cm 2.1cm 0.5cm, clip]{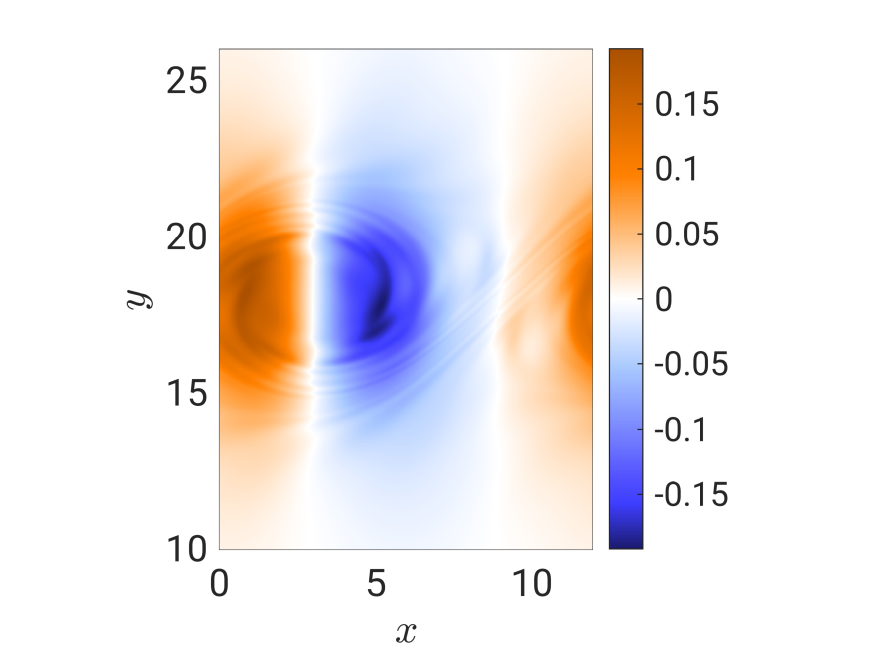}
     \caption{ }
     \label{fig_vert_velocity_a}
     \end{subfigure}
     \begin{subfigure}[b]{0.32\textwidth}
     \includegraphics[width=\linewidth, trim=1.5cm 0cm 2.1cm 0.5cm, clip]{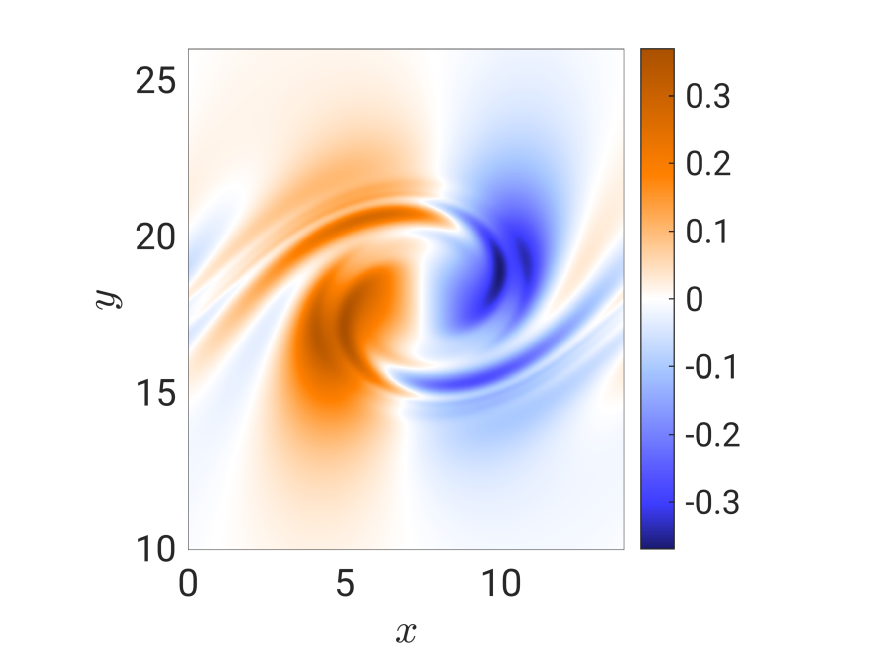}
     \caption{ }
     \label{fig_vert_velocity_b}
     \end{subfigure}
     \begin{subfigure}[b]{0.32\textwidth}
     \includegraphics[width=\linewidth, trim=1.5cm 0cm 2.1cm 0.5cm, clip]{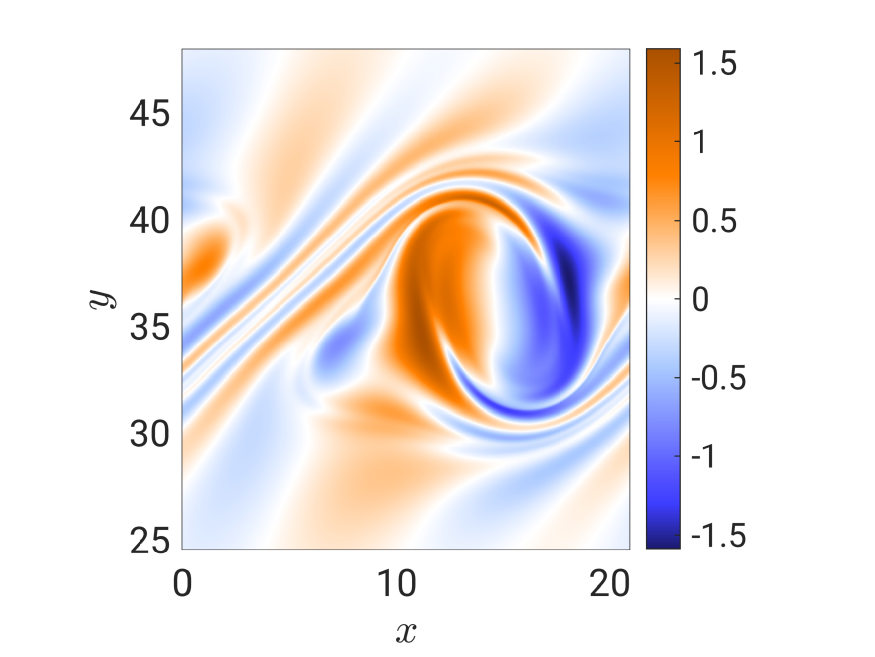}
     \caption{ }
     \label{fig_vert_velocity_c}
     \end{subfigure}
     \captionsetup{width=\linewidth, justification=justified, format=plain} 
   \caption{Vertical velocity for (a) $(N,\tilde{f})=(2,1.5)$, $t=70$, (b) $(N,\tilde{f})=(1,0.5)$, $t=73$, (c) $(N,\tilde{f})=(0.5,0.5)$, $t=110$ for $Re=2000$ and $Sc=1$}   
    \label{fig:u_z} 
 \end{figure}

\noindent Figure \ref{fig:u_z} shows the vertical velocity at one instant for the traditional-like, mixed and non-traditional simulations. The instant displayed corresponds to the ones in figures \ref{fig_tradi-like_c}, \ref{fig_localSI_b}, \ref{fig_stringSI_b}, i.e. at the beginning of the non-linear saturation regime and prior to the onset of secondary instabilities when they develop. In the three cases, the vertical velocity field has a global dipolar structure with positive/negative values on the left/right of the core of the primary vortex. Hence, the tilting term $\tilde{f}\partial_{y}\hat{w}$ of (\ref{Equation_evolution_vorticity}) is negative in the transverse band encompassing the vortex core and positive outside of it. This explains the appearance of vorticity in the regions surrounding the core and the braid (reddish and blueish regions in figures \ref{Quatre_instants_Cas_N_grand}, \ref{Quatre_instants_Cas_mixte}, \ref{Quatre_instants_Cas_N_petit}) In addition, thin bands of enhanced vertical velocity are superimposed on the dipolar structure of the vertical velocity, especially when $N \le 1$. However, the main difference between the three cases is the amplitude of the vertical velocity: while its maximum is around 0.15 in figure \ref{fig_vert_velocity_a}, it is five and ten times larger in figures \ref{fig_vert_velocity_b} and \ref{fig_vert_velocity_c}, i.e. of the same order as the horizontal velocity. It is therefore not surprising that the vertical vorticity dynamics can be more or less modified compared to the traditional case due to the tilting term $\tilde{f} \partial_y \hat w$ in the right-hand side of (\ref{Equation_evolution_vorticity}).\\

\begin{figure}
  \centering
       \begin{subfigure}[b]{0.47\textwidth}
     \includegraphics[width=\linewidth, trim=0cm 0cm 0cm 0cm, clip]{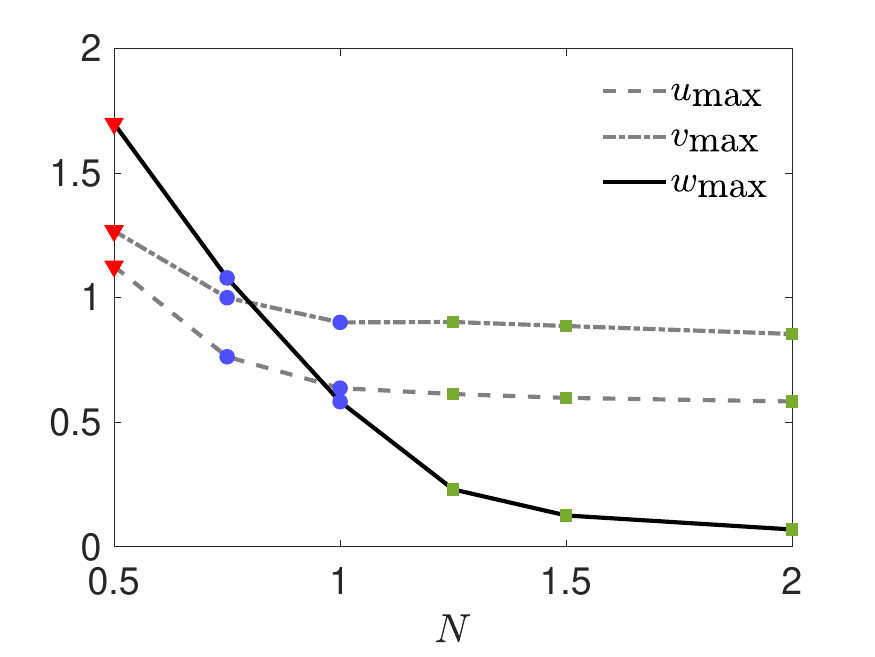}
     \caption{ }
     \label{fig_max_w_a}
     \end{subfigure} 
     \begin{subfigure}[b]{0.47\textwidth}
     \includegraphics[width=\linewidth, trim=0cm 0cm 0cm 0cm, clip]{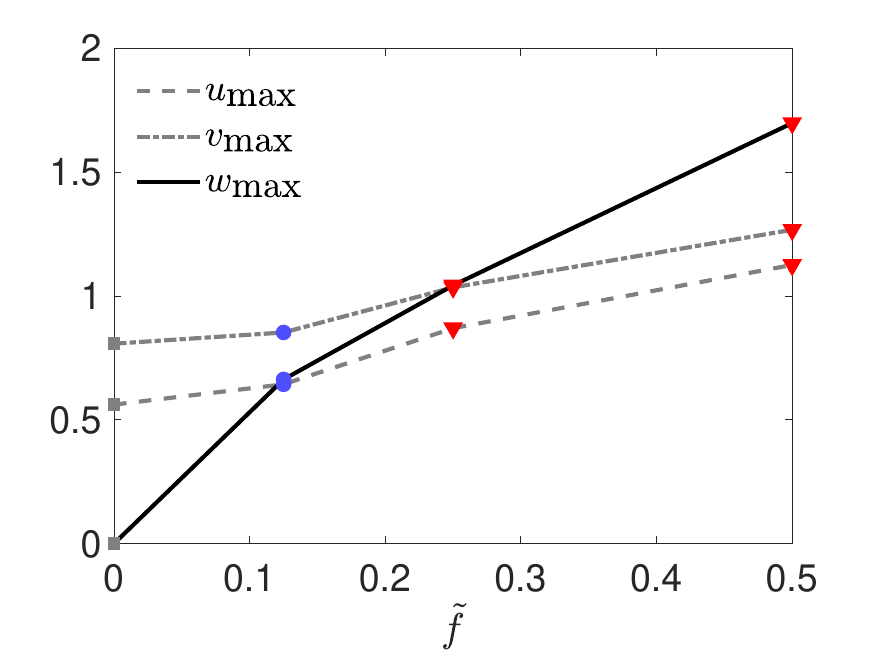}
     \caption{ }
     \label{fig_max_w_b}
     \end{subfigure} 
     \captionsetup{width=\linewidth, justification=justified, format=plain} 
   \caption{Maximum of the velocity component $w$ (solid line), $u$ (dashed line), $v$ (dashed dotted line) observed at the beginning of the saturated regime and before the onset of secondary instabilities as a function of (a), $N$ for $\tilde{f}=0.5$ and (b), $\tilde{f}$ for $N=0.5$. The symbols indicate the different non-linear behaviors as in figure \ref{espace_parametres}.}   
    \label{fig:maxu_z} 
 \end{figure}

\noindent Figure \ref{fig:maxu_z} summarizes the maxima of the three velocity components observed at the times prior to the onset of secondary instabilities (when they occur), for varying $N$ for $\tilde{f}=0.5$ (figure \ref{fig_max_w_a}) and for varying $\tilde{f}$ for $N=0.5$ (figure \ref{fig_max_w_b}). We see that the maximum of the vertical velocity increases when $N$ decreases for a fixed $\tilde{f}$ or when $\tilde{f}$ increases for a constant value of $N$. In contrast, the maxima of the horizontal velocities ($u_{\textrm{max}}$ and $v_{\textrm{max}}$) vary much less and remain always of order unity. The non-traditional and traditional-like evolutions are observed when the maximum vertical velocity is typically larger and much smaller than the maxima of the horizontal velocity components, respectively. The mixed evolution occurs when the maximum vertical velocity is in the intermediate range between these two cases.\\

\noindent The approximate boundaries between the different domains in the parameter space depend both on $N$ and $\tilde{f}$ (red and blue dotted lines in figure \ref{espace_parametres}). Although there are only few points in some regions, they seem however to occur for fixed values of the non-dimensional buoyancy frequency $N$: $N \simeq 0.5$ and $N \simeq 1$, for sufficiently large $\tilde{f}$. This feature is reminiscent of the dynamics of a vertical vortex under the complete Coriolis acceleration in a stratified fluid \citep[][]{ToghraeiBillant2022}. A strong vertical velocity field is generated when the buoyancy frequency is smaller than the maximum angular velocity of the vortex. This field develops near the critical radius where the angular velocity of the vortex is equal to the buoyancy frequency. In turn, the vertical velocity generates an anomaly of vertical vorticity which may trigger a shear instability above a critical non-traditional Coriolis parameter which depends on the Reynolds number. \\

\noindent In the present case, when the non-traditional parameter is weak, the vortex generated by the Kelvin-Helmholtz instability has a maximum absolute angular velocity approximately equal to 1/2 since the minimum vertical vorticity is around -1. Therefore, if the same mechanism as in \cite{ToghraeiBillant2022} is at play, we could expect that strong vertical velocities may develop via critical layers when the non-dimensional Brunt-Väisälä frequency is less than $1/2$. 
Since this critical value should probably be considered only as an order of magnitude, it could explain why the transition between the regimes in figure \ref{espace_parametres} seem to occur for fixed values of the non-dimensional buoyancy frequency $N$. In addition, the thin bands of enhanced vertical velocity seen in figure \ref{fig:u_z} could be a signature of the presence of critical layers. However, it is difficult to make more quantitative comparisons since, here, the dynamics is more intricate and complicated than in \cite{ToghraeiBillant2022}. In particular, the Kelvin-Helmholtz vortex and the vertical velocity field develop concomitantly and there is an instantaneous mutual interaction between them, while the vortex is assumed to pre-exist in the study of \cite{ToghraeiBillant2022}.\\

\subsection{Energy analysis}\label{ener_analyse}
We now study the influence of the parameters $N$ and $\tilde{f}$ on the evolution of the kinetic and potential energies of the perturbations, defined as:  
\begin{equation}
    \hat{K}=\frac{1}{2L_{x}}\int_0^{L_x}\int_0^{L_y}(\hat{u}^2+\hat{v}^2+\hat{w}^2)dy dx, \quad \hat{P}=\frac{1}{2L_{x}}\frac{1}{N^2}\int_0^{L_x}\int_0^{L_y}\hat{b}^2\,dy dx,
\label{Expression_Ec_et_Ep}
\end{equation}
\noindent respectively. The surface integrals in (\ref{Expression_Ec_et_Ep}) are normalized by $L_{x}$, i.e. these energies are per unit wavelength in order to be able to compare simulations for different wavelengths. In contrast, they are not normalized by $L_y$ because the perturbations decay in the $y$ direction. In other words, the energies are $x$-averaged and $y$-integrated.\\
The kinetic energy can be further decomposed into horizontal and vertical components:
\begin{equation}
    \hat{K}_h=\frac{1}{2L_{x}}\int_0^{L_x}\int_0^{L_y}(\hat{u}^2+\hat{v}^2)dy dx, \quad  \hat{K}_v=\frac{1}{2L_{x}}\int_0^{L_x}\int_0^{L_y}\hat{w}^2\,dy dx.
    \label{hat_Kh_}
\end{equation}

\begin{figure}
\centering
        \begin{subfigure}[b]{0.49\textwidth}
        \includegraphics[width=\linewidth, trim=0cm 0.2cm 0.9cm 0.8cm, clip]{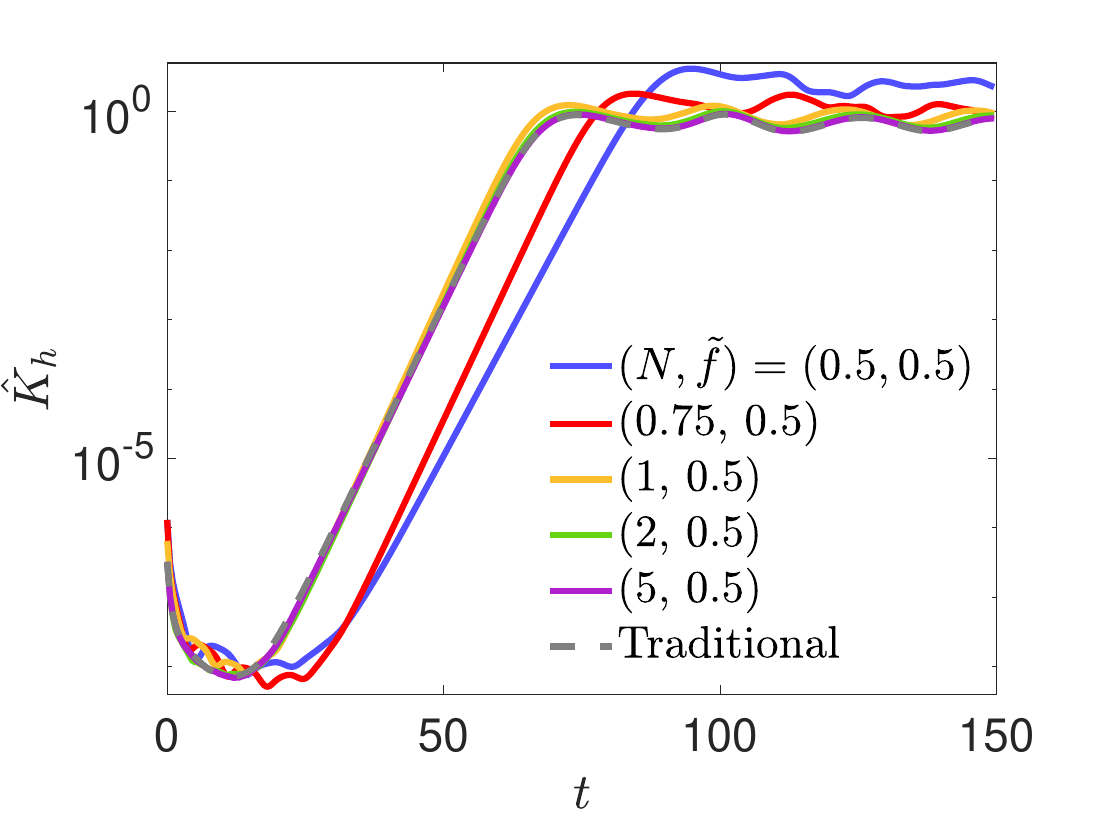}
        \caption{ }
        \label{Ech_ftilde_fixe}
        \end{subfigure}
         \begin{subfigure}[b]{0.49\textwidth}
        \includegraphics[width=\linewidth, trim=0cm 0.2cm 0.9cm 0.5cm, clip]{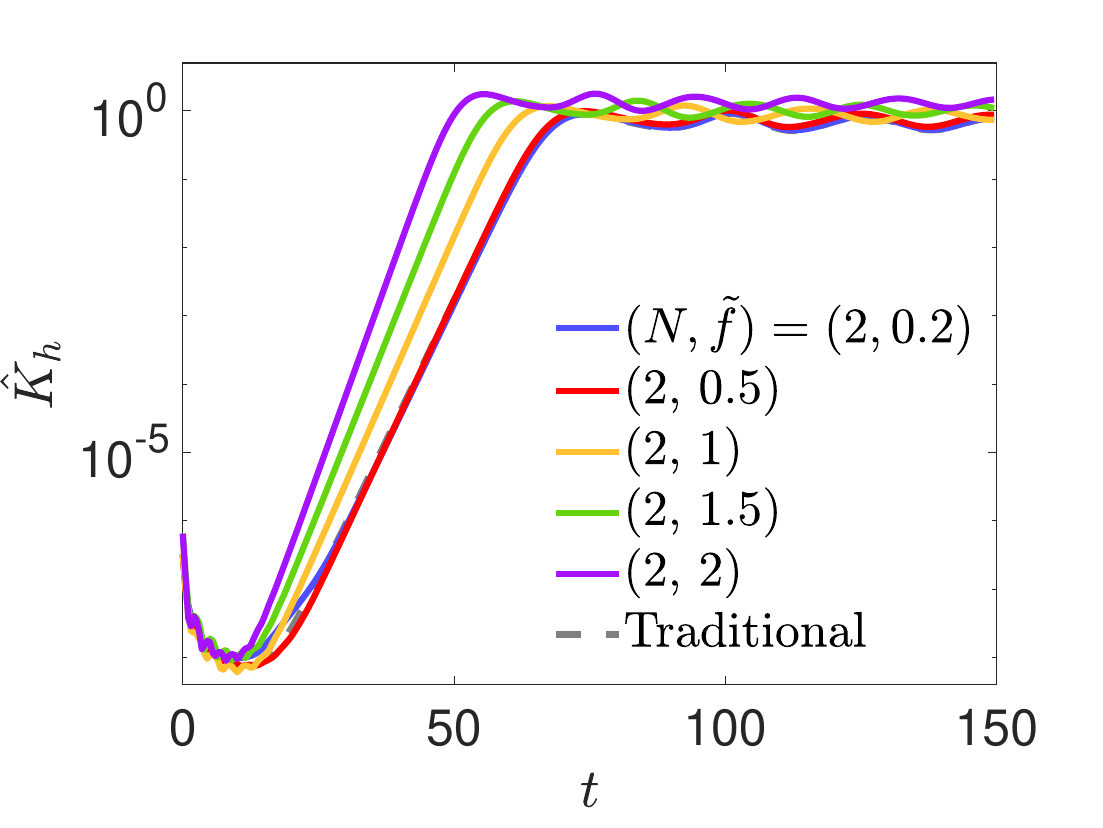}
        \caption{}
        \label{Ech_N_fixe}
        \end{subfigure}
        \begin{subfigure}[b]{0.49\textwidth}
        \includegraphics[width=\linewidth, trim=0cm 0.2cm 0.9cm 0.8cm, clip]{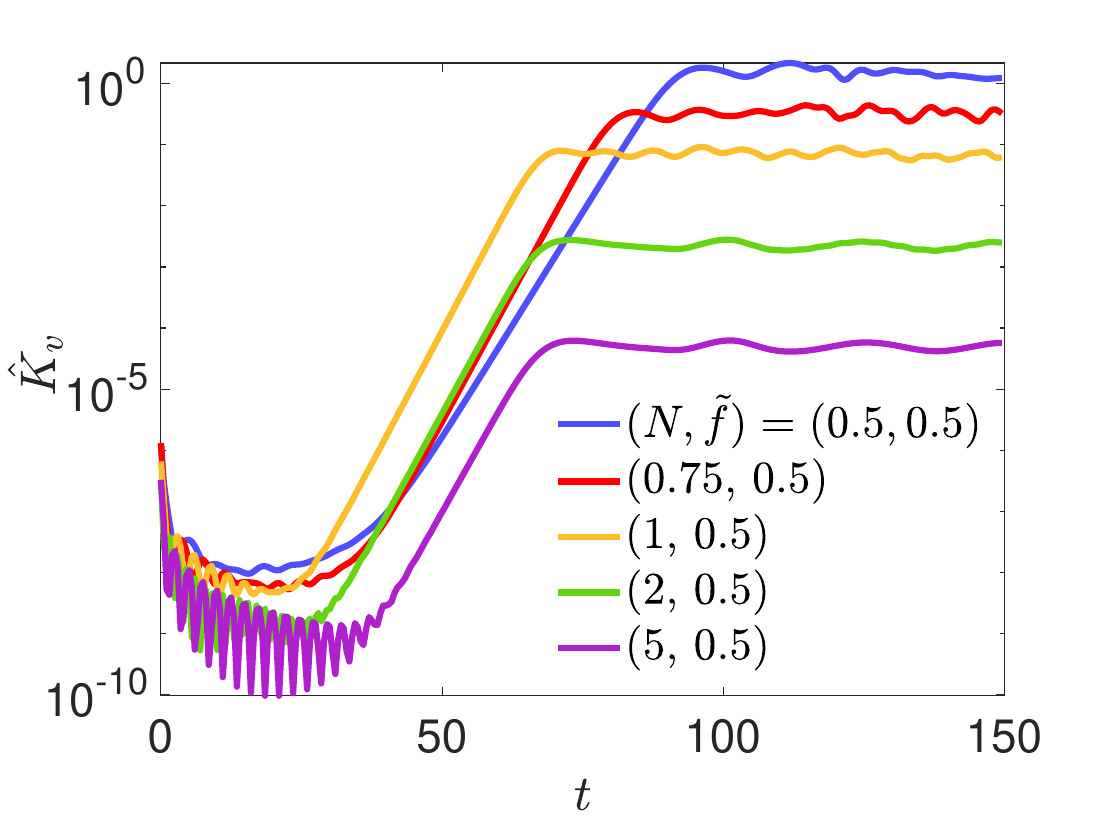}
        \caption{ }
        \label{Ecv_ftilde_fixe}
        \end{subfigure}
        \begin{subfigure}[b]{0.49\textwidth}
        \includegraphics[width=\linewidth, trim=0cm 0.2cm 0.9cm 0.5cm, clip]{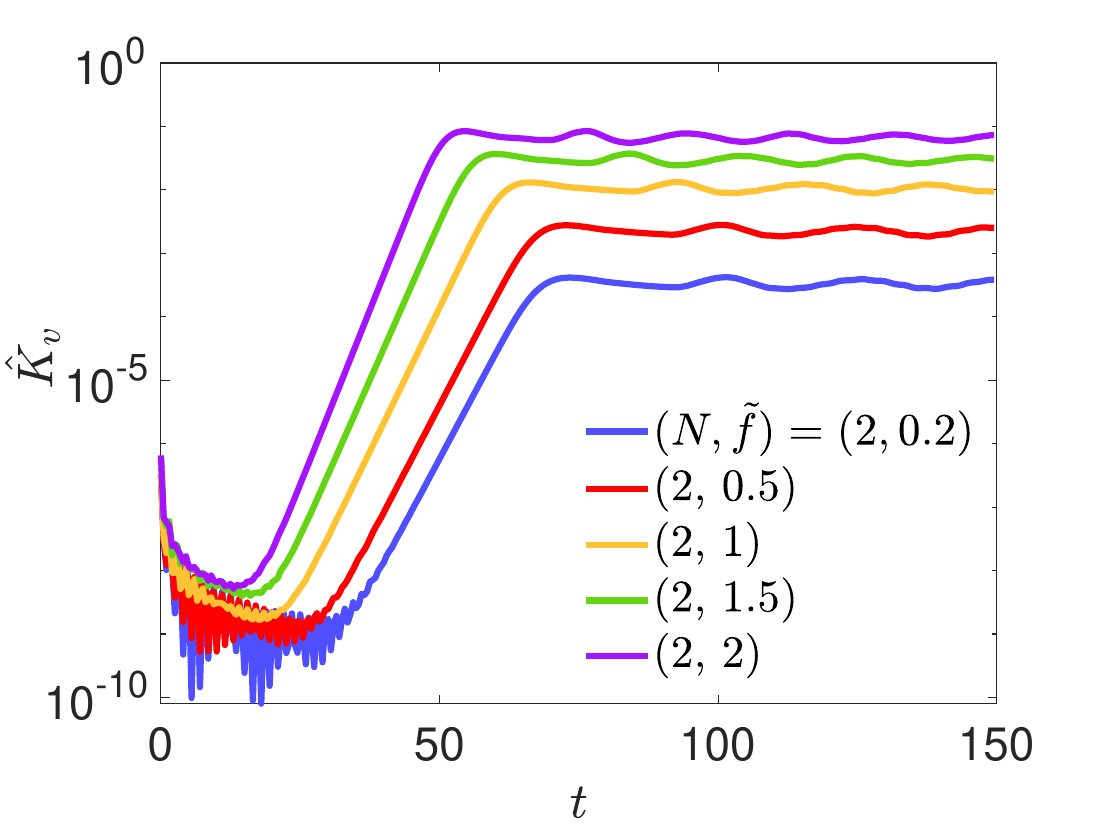}
        \caption{ }
        \label{Ecv_N_fixe}
        \end{subfigure}
        \begin{subfigure}[b]{0.49\textwidth}
        \includegraphics[width=\linewidth, trim=0cm 0.2cm 0.9cm 0.8cm, clip]{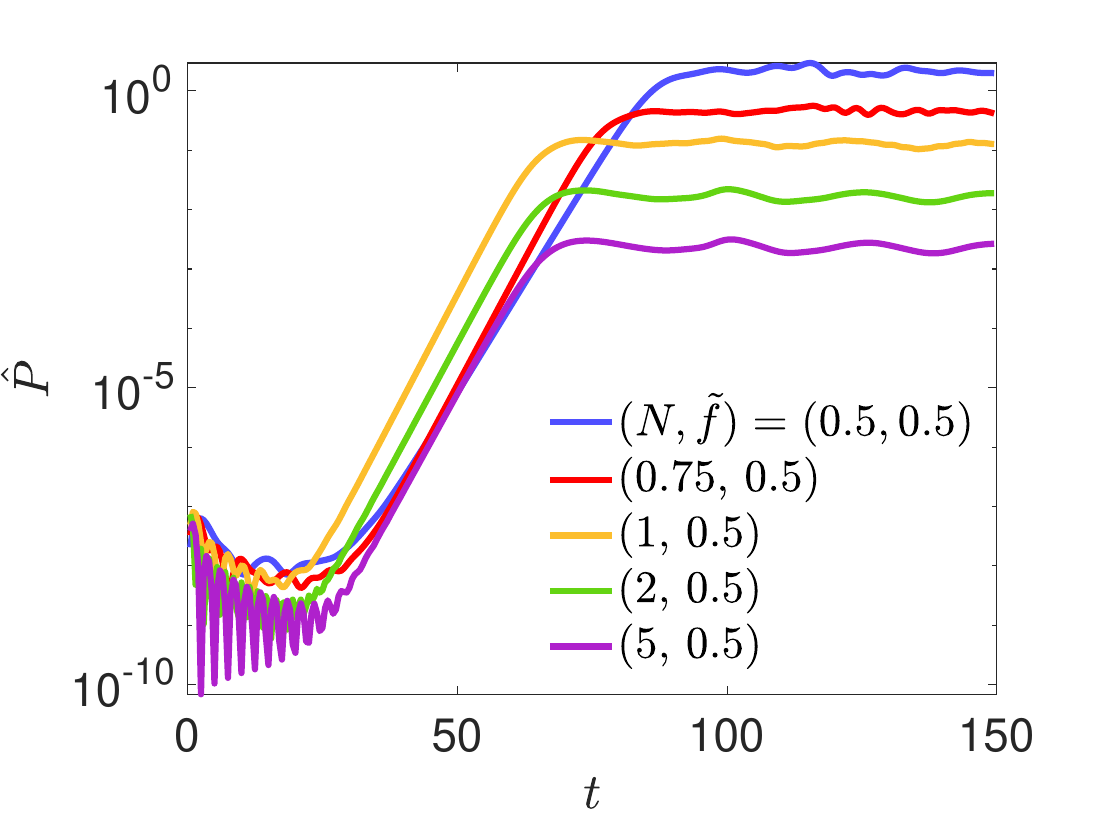}
        \caption{ }
        \label{Ep_ftilde_fixe}
        \end{subfigure}
        \begin{subfigure}[b]{0.49\textwidth}
        \includegraphics[width=\linewidth, trim=0cm 0.2cm 0.9cm 0.5cm, clip]{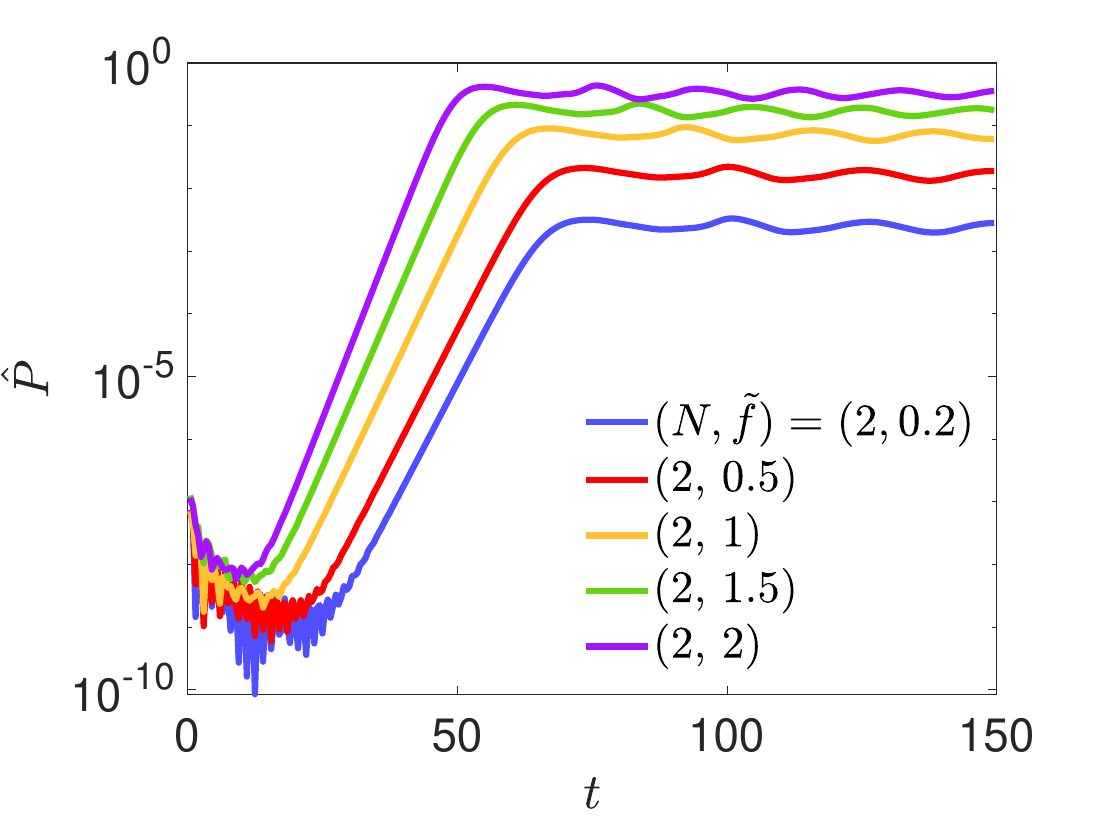}
        \caption{ }
        \label{Ep_N_fixe}
        \end{subfigure}
        \captionsetup{width=1.\linewidth, justification=justified, format=plain}
        \caption{Evolutions of the (a, b) horizontal kinetic energy, (c, d) vertical kinetic energy and, (e, f), potential energy of the perturbations for different stratifications $N$ and non-traditional parameter $\tilde{f}$ values for $Re=2000$, $Sc=1$. (a, c, e): $\tilde{f}=0.5$ and $N$ variable. (b, d, f): $N=2$ and $\tilde{f}$ variable.}
        \label{Ec_Ep_ftilde_et_N_fixe}
\end{figure}

\noindent The left panels of figure \ref{Ec_Ep_ftilde_et_N_fixe} shows the evolution of the horizontal and vertical kinetic energy and potential energy of the perturbations, for varying stratification $N$ and a non-traditional parameter fixed to $\tilde{f}=0.5$. They all start at small values corresponding to the amplitude of the white noise and, after a transient period, increase exponentially before non-linear saturation. However, the level of saturation oscillates slightly but regularly, especially when $N\geq1$. As discussed by \cite{KlaassenPeltier1985a} and \cite{KlaassenPeltier1991}, the oscillation is linked to a nutation of the Kelvin-Helmholtz vortex, during which an exchange of energy between the mean-flow and the perturbations occurs. The four typical non-linear evolutions described in the previous section are present in figure \ref{Ech_ftilde_fixe}, \ref{Ecv_ftilde_fixe}, \ref{Ep_ftilde_fixe}, namely: traditional (figure \ref{Quatre_instants_Cas_trad}), traditional-like for $N=2$ and $N=5$ (figure \ref{Quatre_instants_Cas_N_grand}), mixed for $N=1$ and $N=0.75$ (figure \ref{Quatre_instants_Cas_mixte}) and non-traditional for $N=0.5$ (figure \ref{Quatre_instants_Cas_N_petit}). As seen in figure \ref{Ech_ftilde_fixe}, the level of saturation of the horizontal kinetic energy decreases as $N$ increases and tend, when $N\geqslant 1$, to the traditional case (dashed line). When $N$ increases, the levels of saturation of the vertical kinetic energy and potential energy also decrease but in a much more pronounced manner than the horizontal kinetic energy. They are identically zero in the traditional limit (hence, not plotted), very low for $N=5$ while for $N=0.5$, they are comparable to the horizontal kinetic energy. In section \ref{strong_strat}, we will further investigate what govern the levels of saturation of these different energies.\\

In turn, the right panels of figure \ref{Ec_Ep_ftilde_et_N_fixe} show the evolution of the energies for a fixed stratification $N=2$ and varying $\tilde{f}$. In this case, all the simulations shown exhibit a traditional-like behaviour. The growth rate and level of saturation of the horizontal kinetic energy slightly increases with $\tilde{f}$ but remains close to the one in the traditional limit (figure \ref{Ech_N_fixe}). In contrast, the amplitudes of saturation of the vertical kinetic energy and potential energy (figure \ref{Ecv_N_fixe}, \ref{Ep_N_fixe}) increase strongly with $\tilde{f}$ but they remain below the level of the horizontal kinetic energy for $\tilde{f}\leqslant N$. Analytical expressions for the energies in terms of $N$ and $\tilde{f}$ will be derived in \S\ref{strong_strat} in the limit of strong stratification. In summary, the stratification and non-traditional effects have a competing influence on the non-linear dynamics of the shear instability: when the stratification is increased keeping $\tilde{f}$ constant, the evolutions of the energies tends to the traditional ones while, when the non-traditional parameter is increased for fixed $N$, they depart more and more from the traditional case.

\subsection{Enstrophy analysis}\label{enstro_analyse}
The evolution of the total enstrophy,
\begin{equation}
Z=\frac{1}{2L_{x}}\int_0^{L_x}\int_0^{L_y}\boldsymbol{\omega}^{2}dy dx,
\label{expression_enstrophy}
\end{equation}
where $\boldsymbol{\omega}$ is the total vorticity $\boldsymbol{\omega}=\boldsymbol{\omega_{h}}+\xi\boldsymbol{e_{z}}$, with $\boldsymbol{\omega_{h}}=\partial_{y}w\boldsymbol{e_{x}}-\partial_{x}w\boldsymbol{e_{y}}$ the horizontal vorticity, has been studied for the four simulations presented in section \ref{nonlin_evol}. Like for the energy, the total enstrophy can be further decomposed into horizontal and vertical components:
\begin{equation}
Z_{h}=\frac{1}{2L_{x}}\int_0^{L_x}\int_0^{L_y}\omega_{h}^{2}dy dx, \quad Z_{v}=\frac{1}{2L_{x}}\int_0^{L_x}\int_0^{L_y}\xi^{2}dy dx.
\label{enstrophy_pert}
\end{equation}
Equivalent quantities denoted with a hat $\hat{Z}$, $\hat{Z}_{h}$, $\hat{Z}_{v}$ can be defined for the vorticity of the perturbation $\boldsymbol{\hat{\omega}}$.\\

\begin{figure}
\centering
        \begin{subfigure}[b]{0.48\textwidth}
        \includegraphics[width=\linewidth, trim=0cm 0cm 0cm 0cm, clip]{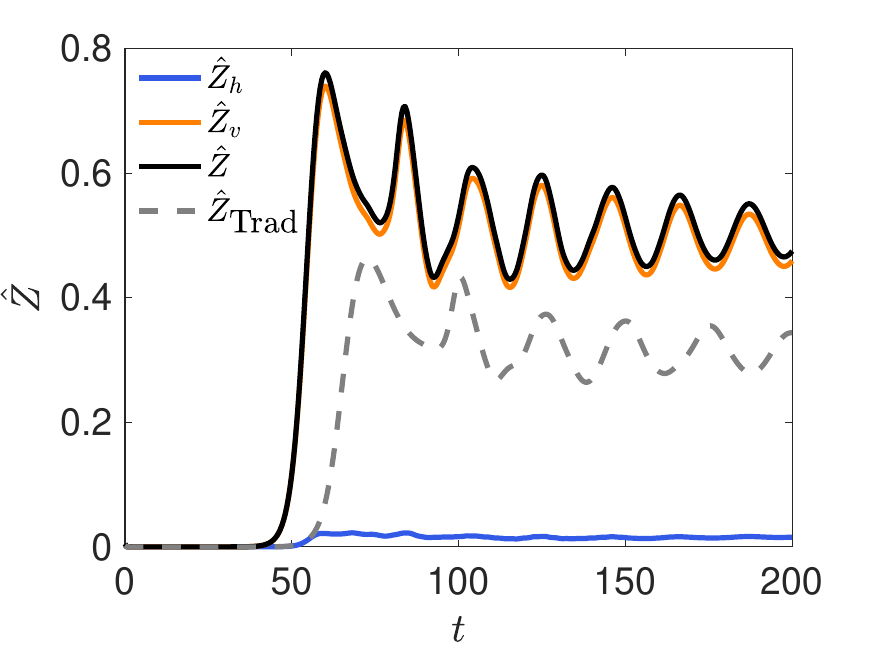}
        \caption{}
        \label{Enst_N5_ftilde0_et_1_a}
        \end{subfigure}
        \begin{subfigure}[b]{0.48\textwidth}
        \includegraphics[width=\linewidth, trim=0cm 0cm 0cm 0cm, clip]{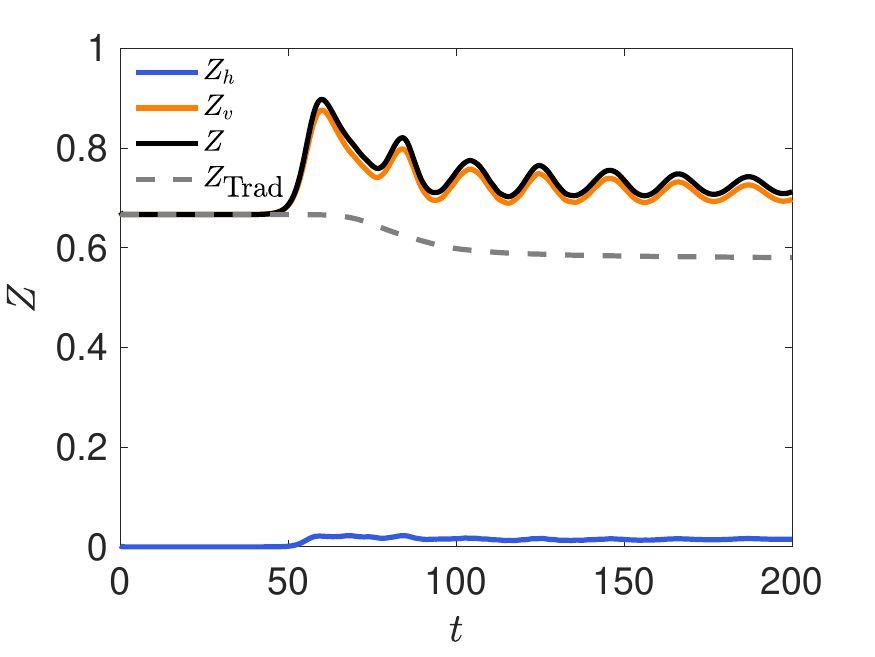}
        \caption{}
        \label{Enst_N5_ftilde0_et_1_b}
        \end{subfigure}
        \captionsetup{width=1.\linewidth, justification=justified, format=plain}
        \caption{(a) Evolution of the horizontal (blue line), vertical (green line) and total (black line) enstrophy of the perturbations for $(N, \tilde{f})=(2, 1.5)$. The gray dashed line corresponds to the total enstrophy of the perturbation in the traditional limit. (b) Same as (a) but for the total enstrophies $Z_{h}$, $Z_{v}$ and $Z$.}
        \label{Enst_N5_ftilde0_et_1}
\end{figure}
\begin{figure}
\centering
        \begin{subfigure}[b]{0.48\textwidth}
        \includegraphics[width=\linewidth, trim=0cm 0cm 0cm 0cm, clip]{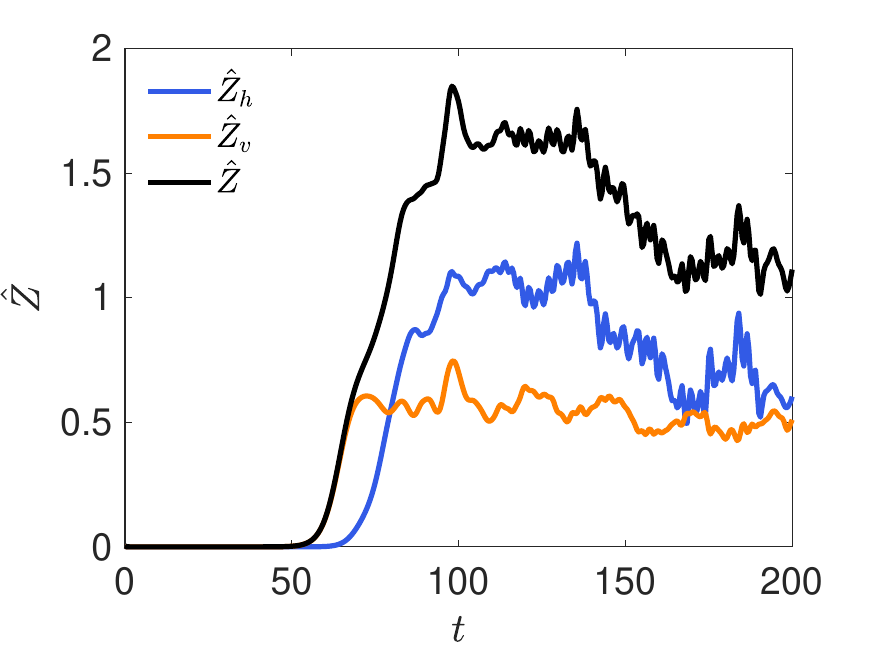}
        \caption{}
        \label{Enst_N1_ftilde05_a}
        \end{subfigure}
        \begin{subfigure}[b]{0.48\textwidth}
        \includegraphics[width=\linewidth, trim=0cm 0cm 0cm 0cm, clip]{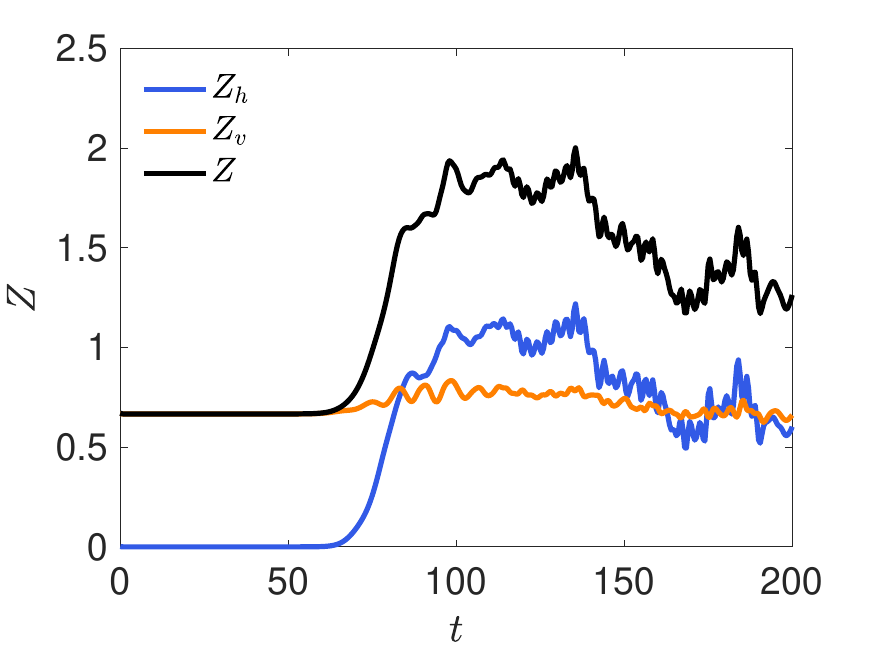}
        \caption{}
        \label{Enst_N1_ftilde05_b}
        \end{subfigure}
        \captionsetup{width=1.\linewidth, justification=justified, format=plain}
        \caption{(a) Evolution of the horizontal (blue line), vertical (green line) and total (black line) enstrophy of the perturbations for $(N, \tilde{f})=(1, 0.5)$. (b) Same as (a) but for the total enstrophies $Z_{h}$, $Z_{v}$ and $Z$.}
        \label{Enst_N1_ftilde05}
\end{figure}
\begin{figure}
\centering
        \begin{subfigure}[b]{0.48\textwidth}
        \includegraphics[width=\linewidth, trim=0cm 0cm 0cm 0cm, clip]{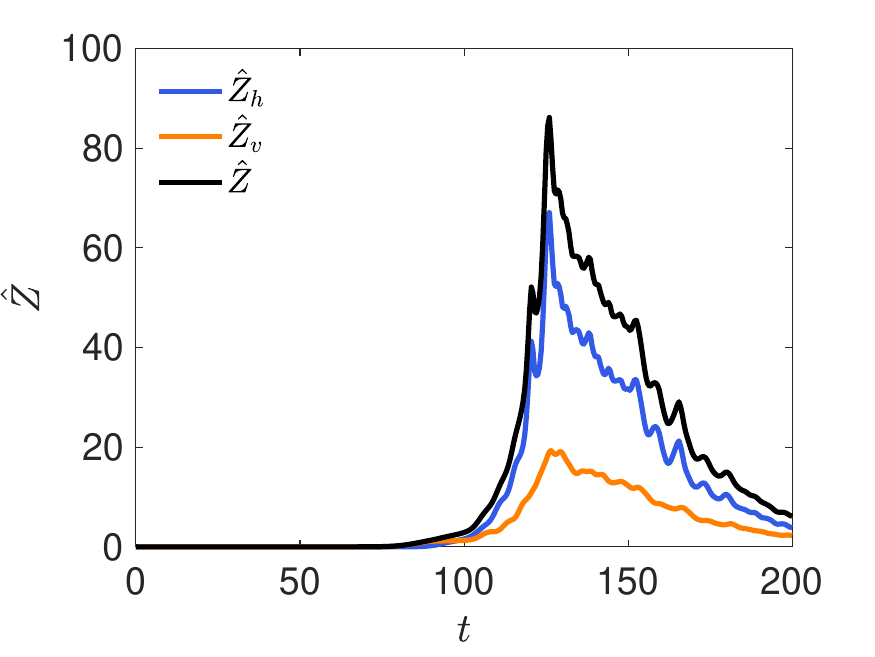}
        \caption{}
        \label{Enst_N05_ftilde05_a}
        \end{subfigure}
        \begin{subfigure}[b]{0.48\textwidth}
        \includegraphics[width=\linewidth, trim=0cm 0cm 0cm 0cm, clip]{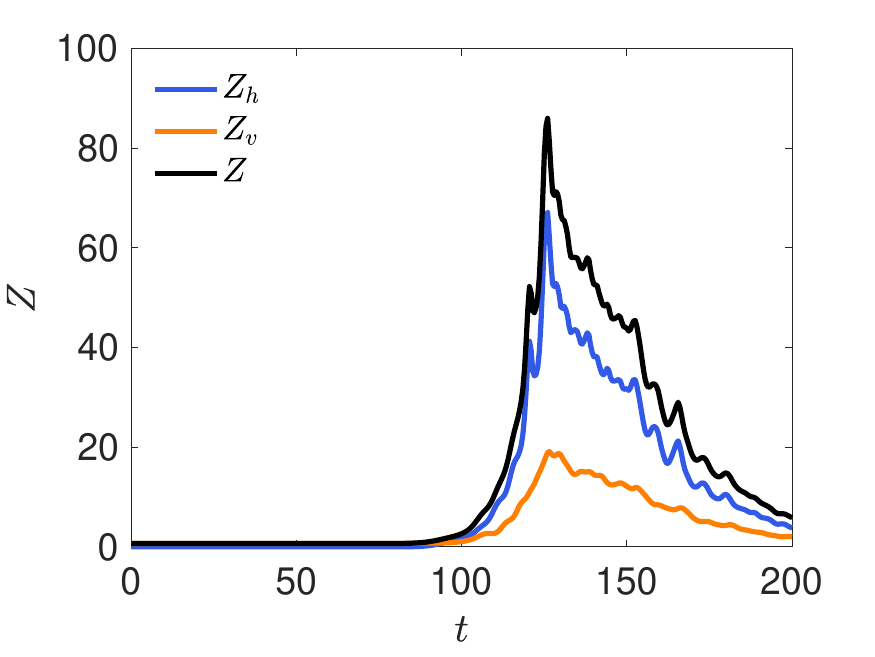}
        \caption{}
        \label{Enst_N05_ftilde05_b}
        \end{subfigure}
        \captionsetup{width=1.\linewidth, justification=justified, format=plain}
        \caption{Same as figure \ref{Enst_N1_ftilde05} but for $(N, \tilde{f})=(0.5, 0.5)$.}
        \label{Enst_N05_ftilde05}
\end{figure}
\noindent Figure \ref{Enst_N5_ftilde0_et_1} shows the evolutions of $\hat{Z}_{h}$, $\hat{Z}_{v}$, $\hat{Z}$ and of the total enstrophy for the simulation in the traditional limit (figure \ref{Quatre_instants_Cas_trad}) and for the traditional-like simulation for $(N, \tilde{f})=(2, 1.5)$ (figure \ref{Quatre_instants_Cas_N_grand}). In the traditional limit, only the total enstrophy is shown (dashed line) since the horizontal enstrophy remains zero. As seen in figure \ref{Enst_N5_ftilde0_et_1_a}, the enstrophy of the perturbations for $\tilde{f}=0$ rises exponentially and then tends to saturate but with some oscillations superimposed as observed for the kinetic energy. The total enstrophy $Z$ (dashed line in figure \ref{Enst_N5_ftilde0_et_1_b}) decreases monotonically as expected since the flow is purely two-dimensional for $\tilde{f}=0$. The total decay of $Z$ is however relatively small (note the vertical scale). We remind that the dissipation of the base flow is not taken into account.\\
For the traditional-like simulation (solid lines in figure \ref{Enst_N5_ftilde0_et_1}), the horizontal enstrophy remains always small since the vertical velocity is small. The evolution of $\hat{Z}$ is very similar to the one in the traditional limit but with a higher value as the maximum and minimum vertical vorticity are no longer 0 and -1 but are outside this range. For this reason, the total enstrophy $Z$ increases somewhat before oscillating (figure \ref{Enst_N5_ftilde0_et_1_b}). This is possible since the vorticity is no longer conserved in the inviscid limit for non-zero $\tilde{f}$.\\

Figure \ref{Enst_N1_ftilde05} shows the evolution of the enstrophy for the mixed case for $(N, \tilde{f})=(1, 0.5)$ (figure \ref{Quatre_instants_Cas_mixte}). The vertical enstrophy of the perturbation $\hat{Z}_{v}$ saturates at approximately the same level as in figure \ref{Enst_N5_ftilde0_et_1}. When $\hat{Z}_{v}$ begins to saturate, the horizontal enstrophy of the perturbation $\hat{Z}_{h}$ starts to grow and becomes dominant afterwards. Thereby, the total enstrophy (figure \ref{Enst_N1_ftilde05_b}) increases significantly instead of decaying. The total vertical enstrophy remains approximately constant. It is worth pointing out that the increase of the enstrophies of the perturbation occurs during the linear and non-linear stages of the primary Kelvin-Helmholtz instability while the secondary instabilities develop only later $t\geqslant120$ (small wiggles are then visible on the curves in figure \ref{Enst_N1_ftilde05}).\\

For the non-traditional case for $(N, \tilde{f})=(0.5, 0.5)$ (figure \ref{Enst_N05_ftilde05}), there is a dramatic increase of the horizontal enstrophy of the perturbation up to $\hat{Z}_{h}\sim 70$ and then it eventually decays. The vertical enstrophy $\hat{Z}_{v}$ increases also but in a less pronounced manner. It is interesting to note that both components of enstrophy begin to increase when the non-linear regime sets in at $t\sim 90$ as in figure \ref{Enst_N1_ftilde05}, but the main increase actually occurs later at $t\sim 120$ when the secondary instabilities develop and the flow becomes turbulent. A high peak of enstrophy has been also observed by \cite{Staquet1995} when secondary instabilities develop in a stratified vertically sheared flow. The influence of the Reynolds number and Schmidt number on the level of this enstrophy maximum will be investigated in a future paper.\\

\noindent To interpret the evolution of the enstrophy, we consider the equation for the vertical vorticity (\ref{Equation_evolution_vorticity}) and the one for the horizontal vorticity of the perturbations $\boldsymbol{\hat{\omega}_{h}}$, which can be derived from (\ref{systeme_adim_equations}): 
\begin{equation}
	    \displaystyle \partial_t \boldsymbol{\hat{\omega}_{h}}+U\partial_x\boldsymbol{\hat{\omega}_{h}}+(\boldsymbol{\hat{u}}\cdot\boldsymbol{\nabla})\boldsymbol{\hat{\omega}_{h}}=\tilde{f}\partial_{y}\boldsymbol{\hat{u}}+(\boldsymbol{\hat{\omega}_{h}}\cdot\boldsymbol{\nabla})(U\boldsymbol{e_{x}}+\boldsymbol{\hat{u}})+\boldsymbol{\nabla}\times(\hat{b}\boldsymbol{e_{z}})+\frac{1}{Re}\boldsymbol{\nabla^2}\boldsymbol{\hat{\omega}_{h}}. 
	    \label{equation_horizontal_vorticity}
\end{equation}
It is worth noticing that the right-hand side of (\ref{equation_horizontal_vorticity}) contains the following effects: first, the tilting and stretching of the horizontal background vorticity $\tilde{f}$, second, the tilting and stretching of the horizontal vorticity of the perturbation by the base flow and the velocity of the perturbations, third, the baroclinic torque and, finally, the viscous effects.\\

The equations (\ref{Equation_evolution_vorticity}, \ref{equation_horizontal_vorticity}) can be combined to give an equation for the evolution of the enstrophy of the perturbation:
\begin{subequations}
\begin{align}
    \displaystyle \frac{D\hat{Z}}{Dt}&=\mathcal{S}+\mathcal{I}+\mathcal{B}+\mathcal{D},
	\label{bilan_enstro}\\
\noindent \textrm{where:} \quad\quad &\\
    \displaystyle \mathcal{S}&=\frac{1}{L_{x}}\int_{0}^{L_{x}}\int_{0}^{L_{y}}\boldsymbol{\hat{\omega}}\cdot\left[(\boldsymbol{\hat{\omega}}+\tilde{f}\boldsymbol{e_{y}})\cdot\boldsymbol{\nabla}\boldsymbol{\hat{u}}\right]\,dydx,
    \label{terme_bilan_enstro_S}\\
    \displaystyle \mathcal{I}&=\frac{1}{L_{x}}\int_{0}^{L_{x}}\int_{0}^{L_{y}}\boldsymbol{\hat{\omega}}\cdot\left[(\boldsymbol{\hat{\omega}}\cdot\boldsymbol{\nabla}U\boldsymbol{e_{x}})+(\boldsymbol{\omega_{b}}\cdot\boldsymbol{\nabla})\boldsymbol{\hat{u}}-(\boldsymbol{\hat{u}}\cdot\boldsymbol{\nabla})\boldsymbol{\omega_{b}})\right]\,dydx,
    \label{terme_bilan_enstro_I}\\
        \displaystyle \mathcal{B}&=\frac{1}{L_{x}}\int_{0}^{L_{x}}\int_{0}^{L_{y}}\boldsymbol{\hat{\omega}}\cdot\boldsymbol{\nabla}\times(\hat{b}\boldsymbol{e_{z}})\,dydx,
    \label{terme_bilan_enstro_B}\\
    \displaystyle \mathcal{D}&=-\frac{1}{L_{x}}\frac{1}{Re}\int_{0}^{L_{x}}\int_{0}^{L_{y}}(\boldsymbol{\nabla}\boldsymbol{\hat{\omega}})^{2}\,dydx,
    \label{terme_bilan_enstro_D}
\end{align}
\label{termes_bilan_enstro}
\end{subequations}
\noindent are respectively stretching/tilting effect, interaction with the base flow, baroclinic and viscous effects. $\boldsymbol{\omega_{b}}=-\partial_{y}U\,\boldsymbol{e_{z}}$ is the vorticity of the base flow.\\

\begin{figure}
        \centering
        \begin{subfigure}[b]{0.495\textwidth}
        \includegraphics[width=\linewidth, trim=0cm 0cm 0cm 0.1cm, clip]{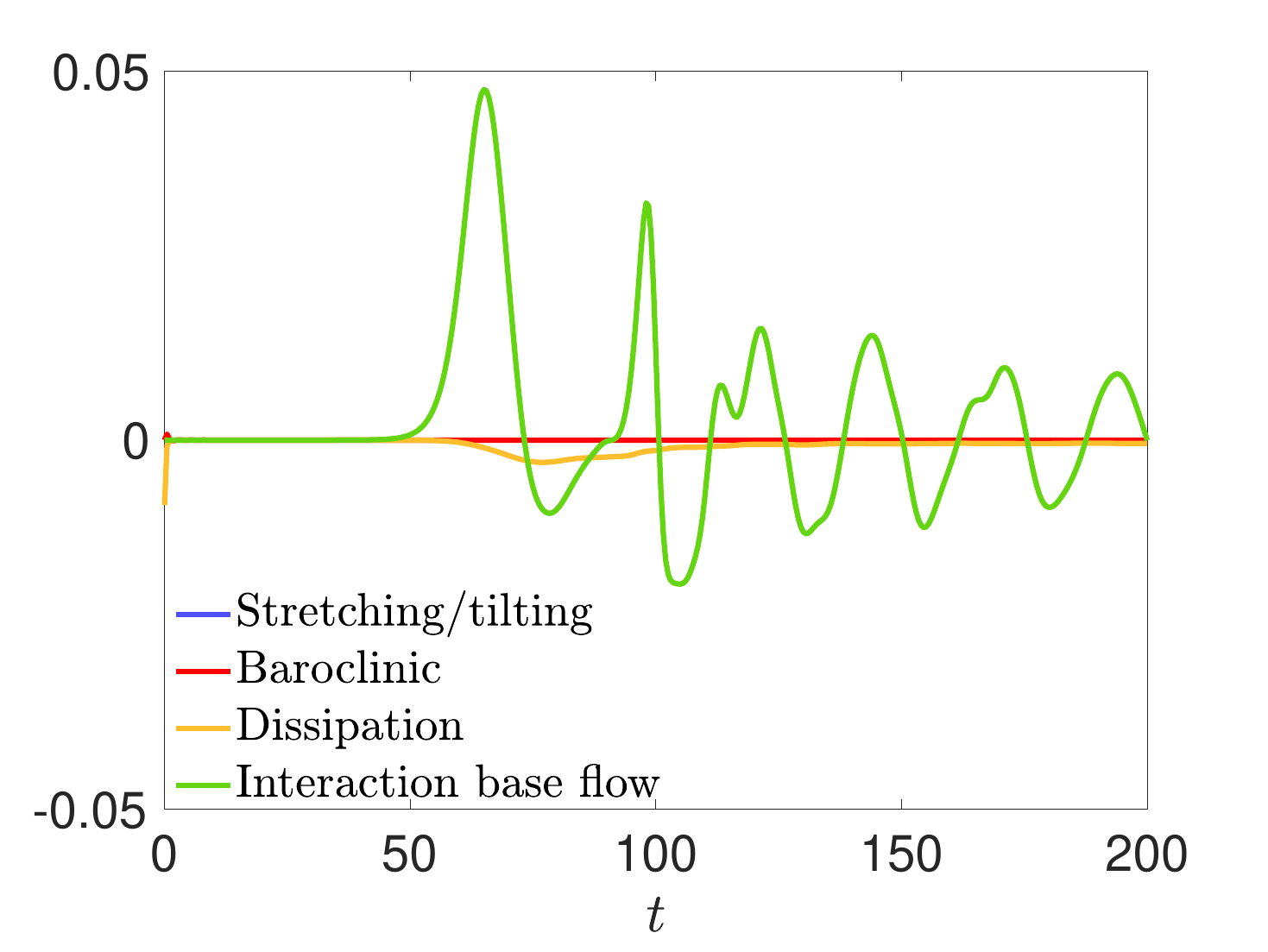}
        \caption{}
        \label{Bilan_enstro_N5_ftilde0_a}
        \end{subfigure}
        \begin{subfigure}[b]{0.495\textwidth}
        \includegraphics[width=\linewidth, trim=0cm 0cm 0cm 0.1cm, clip]{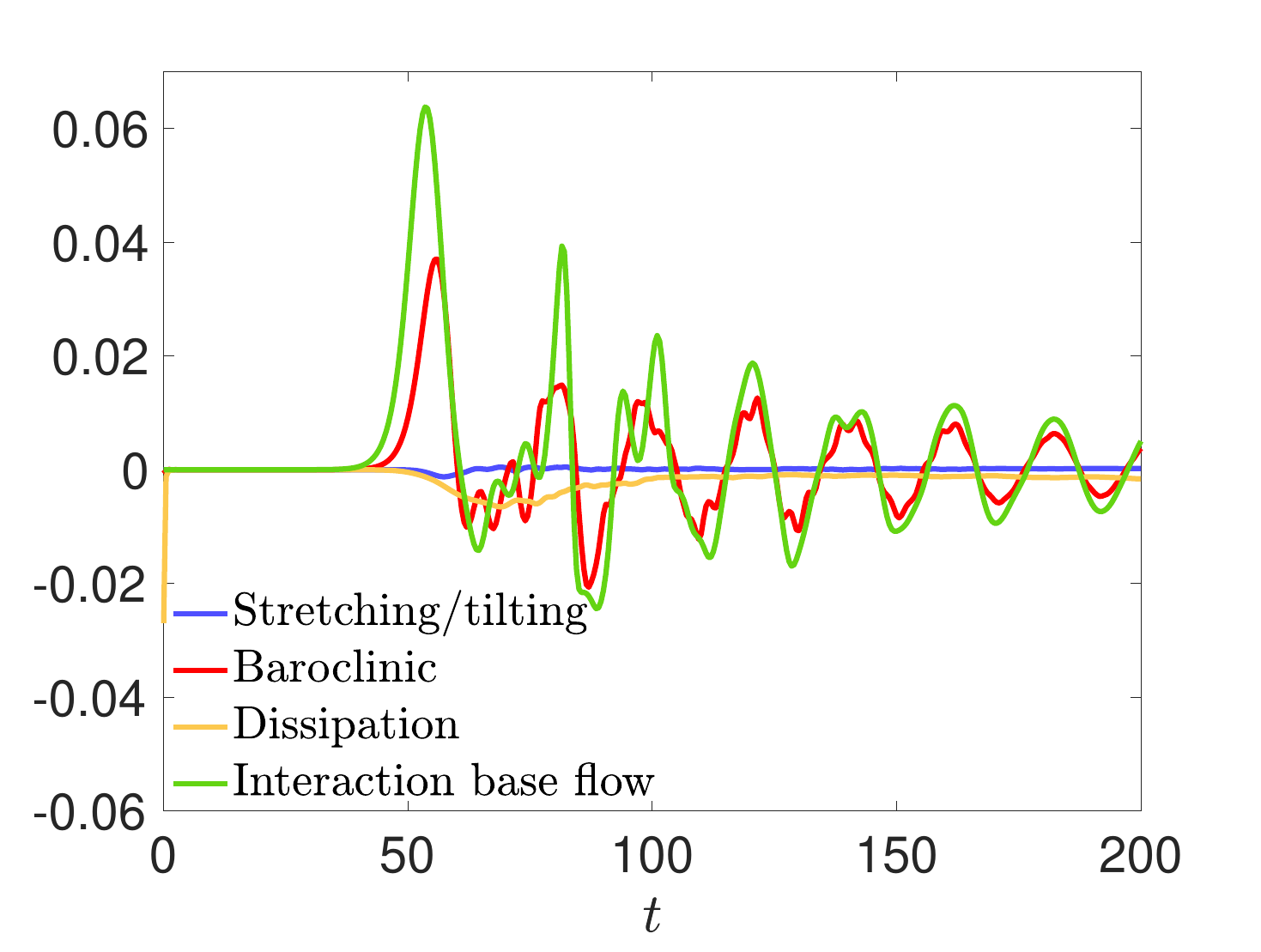}
        \caption{}
        \label{Bilan_enstro_N5_ftilde1_b}
        \end{subfigure}
        \begin{subfigure}[b]{0.495\textwidth}
        \includegraphics[width=\linewidth, trim=0cm 0.2cm 0cm 0cm, clip]{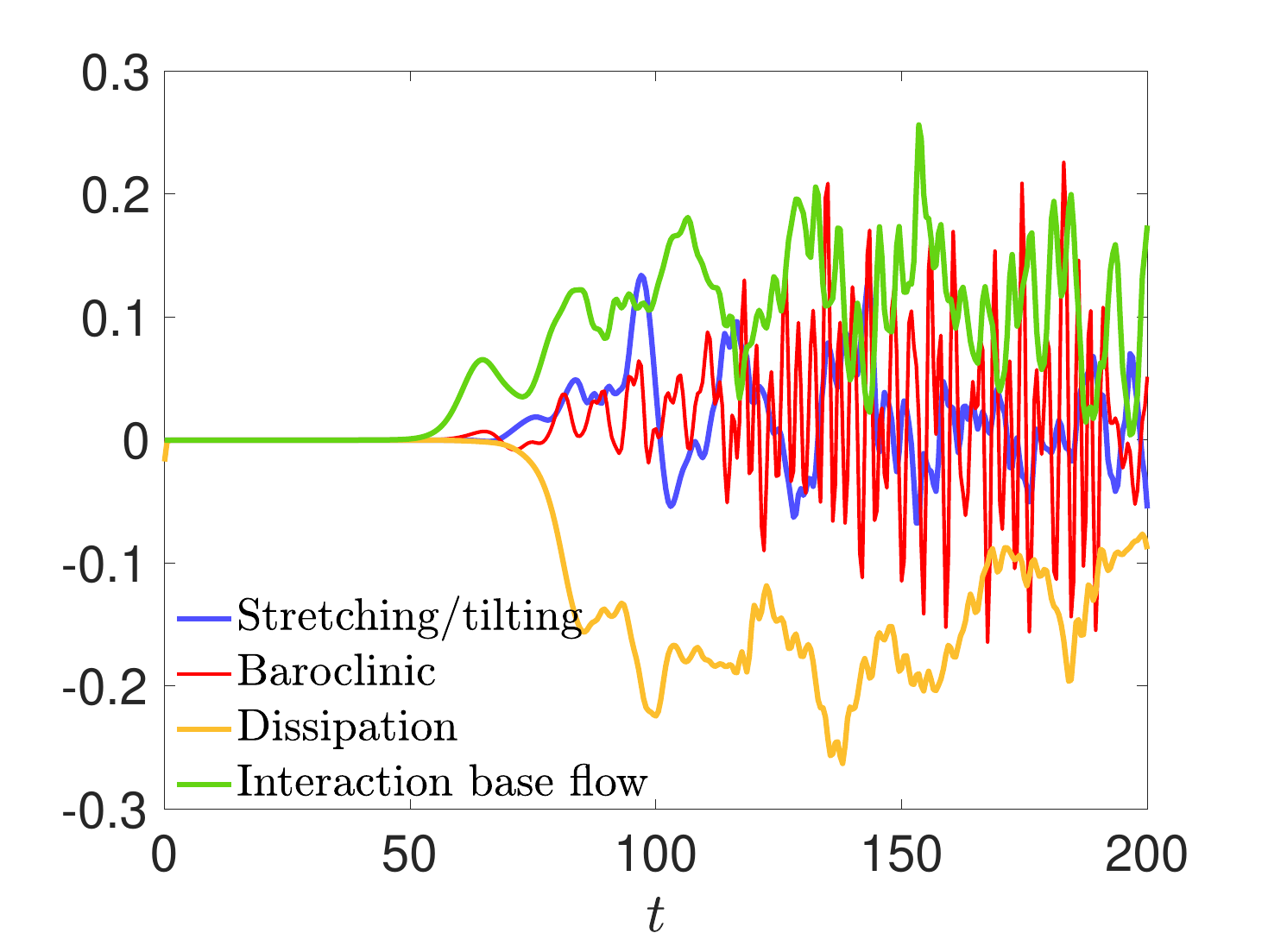}
        \caption{}
        \label{Bilan_enstro_N1_ftilde05_c}
        \end{subfigure}
        \begin{subfigure}[b]{0.495\textwidth}
        \includegraphics[width=\linewidth, trim=0cm 0.1cm 0cm 0cm, clip]{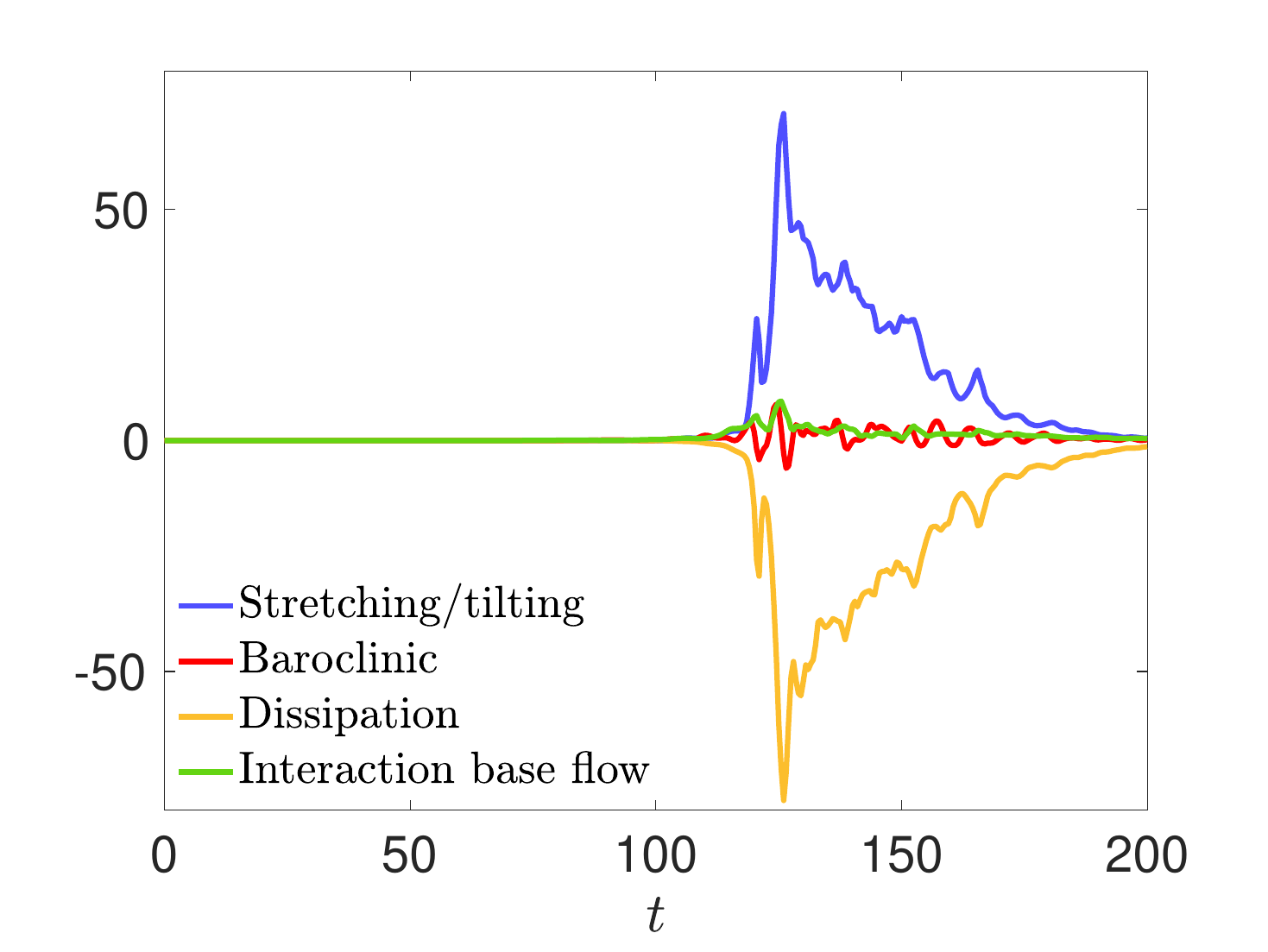}
        \caption{}
        \label{Bilan_enstro_N05_ftilde05_d}
        \end{subfigure}
        \captionsetup{width=1.\linewidth, justification=justified, format=plain}
        \caption{Different terms of the enstrophy budget (\ref{termes_bilan_enstro}) for (a) $(N, \tilde{f})=(5, 0)$, (b) $(N, \tilde{f})=(2, 1.5)$, (c) $(N, \tilde{f})=(1, 0.5)$, (d) $(N, \tilde{f})=(0.5, 0.5)$.}
        \label{Bilan_enstro}
\end{figure}

\noindent Figure \ref{Bilan_enstro_N5_ftilde0_a} shows these four terms in the traditional case. Only the interaction with the base flow and the dissipation contribute to the enstrophy budget as expected. In the traditional-like case for $(N, \tilde{f})=(2, 1.5)$, a similar evolution is observed (figure \ref{Bilan_enstro_N5_ftilde1_b}) except that the baroclinic term is of the same order as the interaction with the base flow. The stretching/tilting term is no longer exactly zero but remains small. For the mixed case for $(N, \tilde{f})=(1, 0.5)$ (figure \ref{Bilan_enstro_N1_ftilde05_c}), the interaction, baroclinic and dissipation terms are still dominant but the stretching/tilting is almost of the same order. It can be noticed also that the dissipation term is one order of magnitude larger than in the previous cases. Finally, for the non-traditional case for $(N, \tilde{f})=(0.5, 0.5)$, figure \ref{Bilan_enstro_N05_ftilde05_d} shows that the considerable increase of $\hat{Z}$ (figure \ref{Enst_N05_ftilde05_a}) is due to the vortex stretching/tilting that becomes, by far, the dominant source of enstrophy when secondary instabilities develop at $t\geqslant 120$. Simultaneously, the dissipation (in absolute value) grows to the same level: it is the dissipative anomaly linked to the development of turbulence. These pronounced peaks are indeed observed when secondary instabilities arise and lead to a transition to turbulence. Such enstrophy evolution is characteristic of a three-dimensional flow even if the present configuration is only two-dimensional. This is because the flow has three velocity components, enabling the stretching and tilting of vorticity. In contrast, the interaction with the base flow and the baroclinic terms are much smaller.

\subsection{Origin of secondary instabilities}\label{sect_secondary_inst}

In this section, we analyse the origin of secondary instabilities observed in the ``mixed'' simulation for $(N,\tilde{f})=(1, 0.5)$ and in the ``non-traditional'' simulation for $(N,\tilde{f})=(0.5, 0.5)$ (see figure \ref{espace_parametres}). In the first case, the secondary instabilities develop in the periphery of the vortex core but remain localised (figure \ref{fig_localSI_c} and \ref{fig_localSI_d}). In the second case $(N,\tilde{f})=(0.5, 0.5)$, the secondary instabilities arise in the same region but quickly contaminate the whole flow and lead to a transition to small-scale turbulence (figure \ref{fig_stringSI_c}, \ref{fig_stringSI_d}).\\

\begin{figure}
\centering
        \begin{subfigure}[b]{0.52\textwidth}
        \includegraphics[width=\linewidth, trim=0cm 0.8cm 0.8cm 1.2cm, clip]{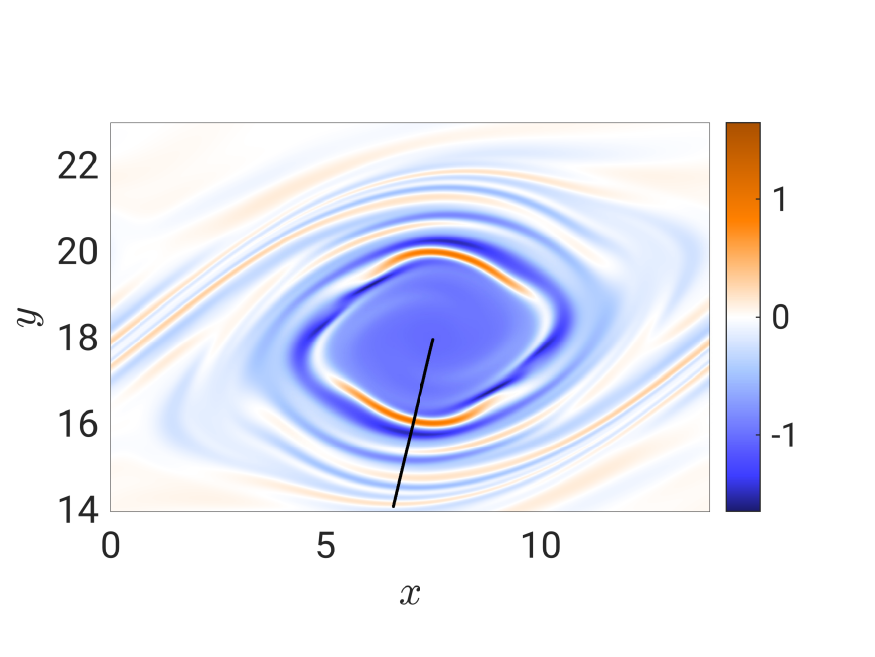}
        \caption{}
        \label{secondary_inst_c}
        \end{subfigure}
        \begin{subfigure}[b]{0.455\textwidth}
        \includegraphics[width=\linewidth, trim=0cm 0cm 0.8cm 1.2cm, clip]{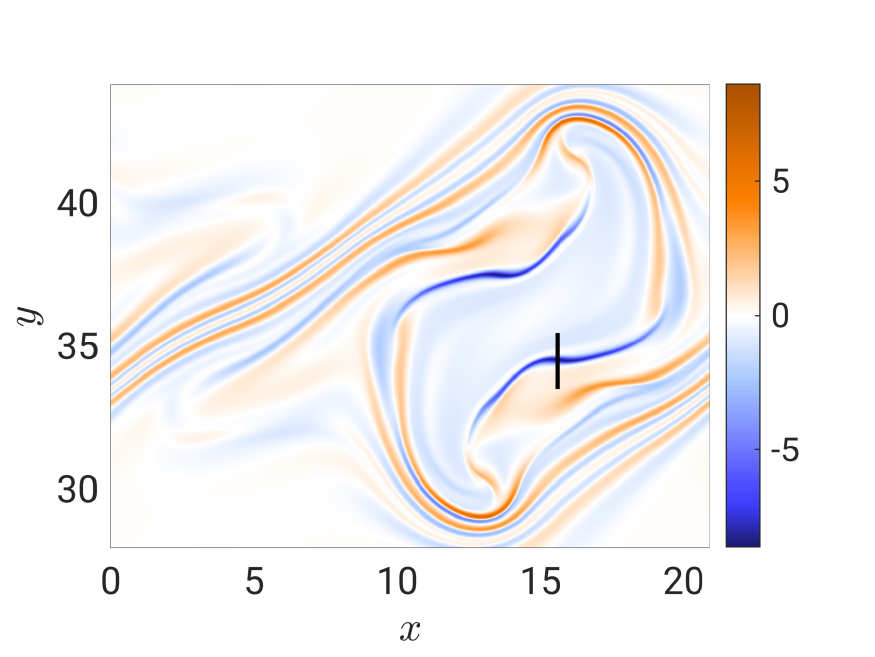}
        \caption{}
        \label{secondary_inst_d}
        \end{subfigure}
         \begin{subfigure}[b]{0.52\textwidth}
        \includegraphics[width=\linewidth, trim=0cm 0.8cm 0.8cm 1.2cm, clip]{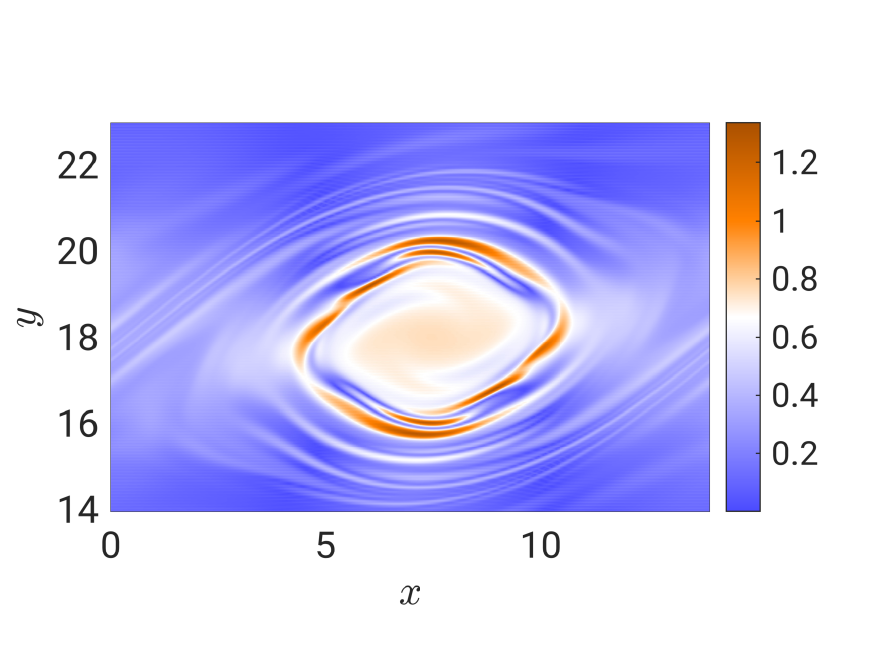}
        \caption{}
        \label{secondary_inst_a}
        \end{subfigure}
        \begin{subfigure}[b]{0.455\textwidth}
        \includegraphics[width=\linewidth, trim=0cm 0cm 0.8cm 1.2cm, clip]{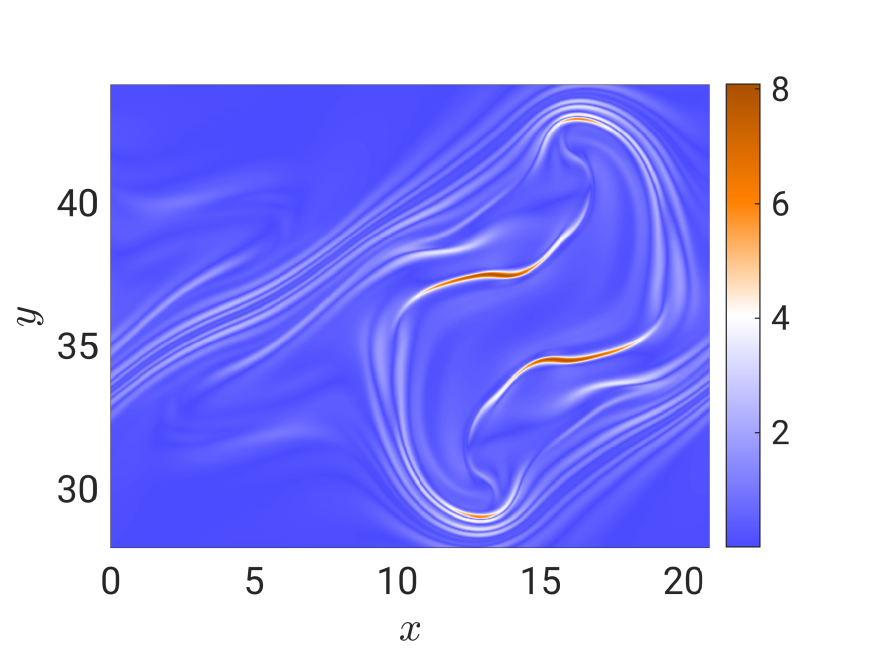}
        \caption{}
        \label{secondary_inst_b}
        \end{subfigure}
        \begin{subfigure}[b]{0.48\textwidth}
        \centering
        \includegraphics[width=0.98\linewidth, trim=0cm 0cm 0.8cm 0.5cm, clip]{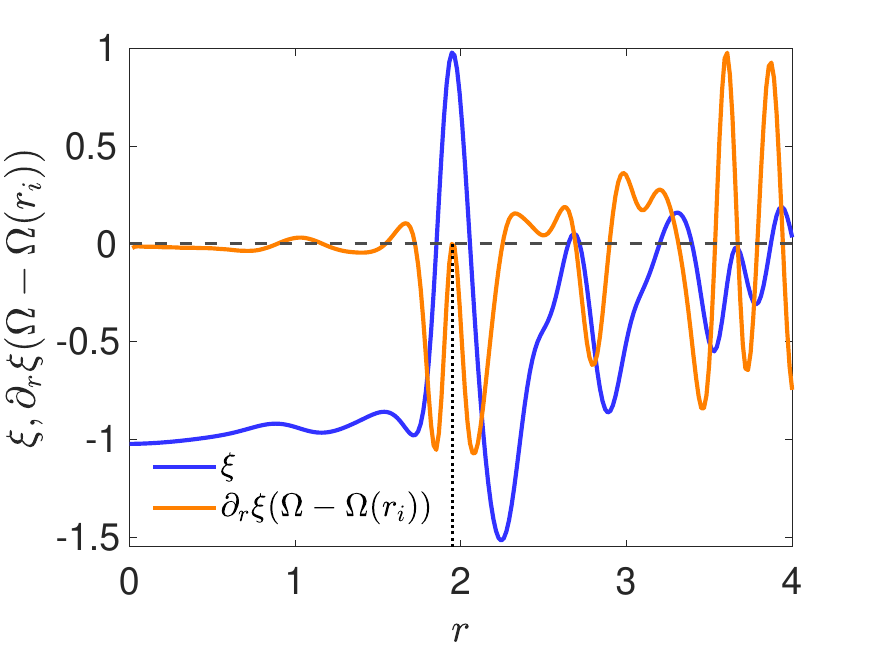}
        \caption{}
        \label{secondary_inst_e}
        \end{subfigure}
        \begin{subfigure}[b]{0.48\textwidth}
        \centering
        \includegraphics[width=0.98\linewidth, trim=0cm 0cm 0.8cm 0.5cm, clip]{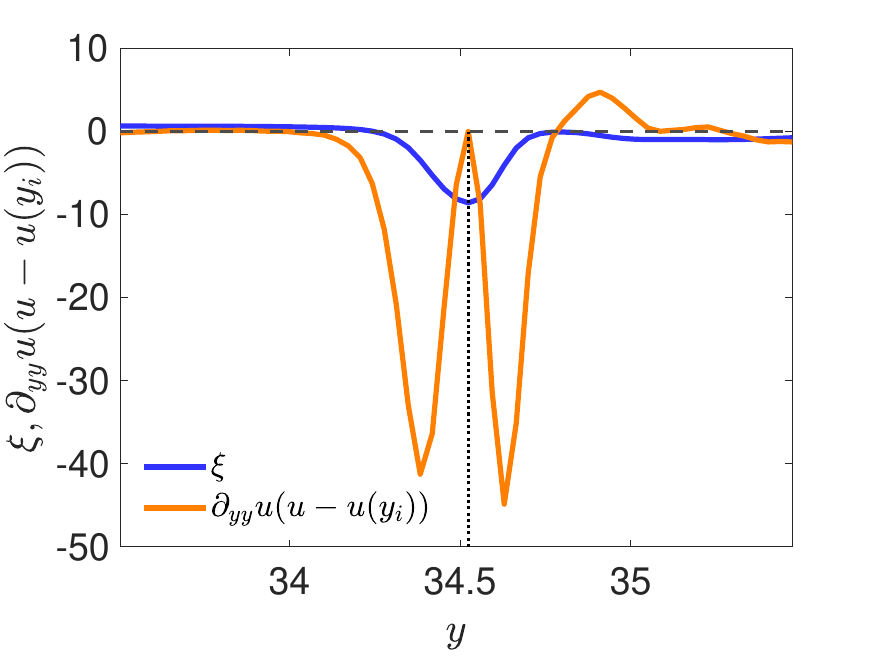}
        \caption{}
        \label{secondary_inst_f}
        \end{subfigure}
        \captionsetup{width=1.\linewidth, justification=justified, format=plain}
        \caption{Shear $S$ at (a) $t=112$ for $(N,\tilde{f})=(1, 0.5)$ and at (b) $t=117.5$ for $(N,\tilde{f})=(0.5, 0.5)$. (c, d) Total vertical vorticity $\xi$ corresponding to (a, b). The black solid lines indicate the lines along which Fj\o rtoft criteria are computed. (e) Vorticity (blue) and cylindrical Fj\o rtoft criterion (\ref{critere_cylindrique}) (orange), computed along the black line in (c). The center of the vortex corresponds to $r=0$. (f) Vorticity (blue) and cartesian Fj\o rtoft criterion (\ref{critere_cartesien}) (orange) along the black line in (d). The black dotted vertical lines in (e) and (f) indicate the location of $r_i$ and $y_i$, respectively.}
        \label{secondary_inst}
\end{figure}

\noindent In order to interpret the origin of these instabilities, we will check whether or not the Fj\o rtoft necessary condition for instability is satisfied \citep[][]{Fjortoft1950}. It is a more stringent criterion than the inflection point condition of Rayleigh. For an inviscid parallel shear flow, it reads:
\begin{equation}
	\partial_{yy}u(y)(u(y)-u(y_{i}))<0,
\label{critere_cartesien}
\end{equation}
where $u(y_{i})$ is the velocity at the inflection point $y_{i}$. There exists also a version of this criterion for an inviscid axisymmetric vortex \citep[][]{Fjortoft1950,SmythCarpenter2019}:
\begin{equation}
	\partial_{r}\xi(r)(\Omega(r)-\Omega(r_{i}))<0,
\label{critere_cylindrique}
\end{equation}
where $(r, \theta, z)$ are cylindrical coordinates in the radial, azimuthal and vertical directions, respectively. $\xi$ is the axial vorticity, $\Omega$ the angular velocity and $r_{i}$ the inflection point where $\partial_{r}\xi(r_{i})=0$.\\

\noindent As shown by \cite{SmythCarpenter2019}, the Fj\o rtoft criterion in Cartesian coordinates is equivalent to require that the inflection point is a maximum of shear:
\begin{equation}
	S=\sqrt{(\partial_{y}u)^{2}+(\partial_{x}v)^{2}}.
\label{absolute_shear}
\end{equation}
\noindent Although these criteria are strictly valid for an axisymmetric vortex and for a parallel shear flow, we will apply them to the flows observed prior to the secondary instabilities onset in the ``mixed case'' and in the ``non-traditional'' case, respectively. Indeed, as seen in figure \ref{secondary_inst_c}, the vortex in the mixed case is approximately axisymmetric whereas, in the non-traditional case (figure \ref{secondary_inst_d}), the vorticity filaments are strongly elongated and can be reasonably approximated as locally parallel.\\

\noindent In the mixed case $(N,\tilde{f})=(1, 0.5)$, figure \ref{secondary_inst_a} shows that $S$ is maximum in the region where secondary instabilities will develop a few time units later. In addition, the criterion (\ref{critere_cylindrique}) is computed along the solid black line in figure \ref{secondary_inst_c} and the results are presented in figure \ref{secondary_inst_e}. It is indeed satisfied around the inflection point which is at $r=1.96$. Furthermore, the associated vorticity profile exhibits also a minimum at $r=2.2$, i.e. there is a second inflection point. Therefore, the configuration is similar to that of a jet, with contiguous maximum and minimum of vorticity. The instability (figure \ref{fig_localSI_c}) indeed resembles the most unstable mode of jet instability i.e. the sinuous mode.\\

\noindent For the non-traditional case $(N,\tilde{f})=(0.5, 0.5)$, the maximum of shear (figure \ref{secondary_inst_b}) is also located in the region where the secondary instabilities will emerge, but it is much more intense than in the previous case. Figure \ref{secondary_inst_f} shows the cartesian Fj\o rtoft criterion (\ref{critere_cartesien}) along the straight black line drawn in figure \ref{secondary_inst_d} across the corresponding vorticity filament. In this case, there is a single inflection point located at $y_{i}=34.52$ (figure \ref{secondary_inst_f}) like for a shear layer. The cartesian Fj\o rtoft criterion for instability (\ref{critere_cartesien}) is satisfied in its neighbourhood.\\

\noindent Therefore, in both mixed and non-traditional simulations, the secondary instabilities can be interpreted as new shear instabilities developing on the vertical vorticity anomalies generated by non-traditional effects.

\subsection{Analysis for strong stratification}\label{strong_strat}
In this section, we will show that the governing equations (\ref{systeme_adim_equations}) can be greatly simplified when the stratification is strong. This will enable us to derive scaling laws for the kinetic and potential energies in terms of $N$ and $\tilde{f}$ when $N\gg1$. We will consider the inviscid limit for simplicity but the following analysis is expected to be valid as long as the Reynolds number is large.\\

\noindent By taking the derivative $D/Dt=\partial_{t}+U\partial_{x}+\boldsymbol{\hat{u}}\cdot\boldsymbol{\nabla}$ of (\ref{systeme_adim_equations_c}), by using (\ref{systeme_adim_equations_d}) and (\ref{systeme_adim_equations_a}), the buoyancy can be eliminated, giving:
\begin{equation}
    \frac{D^2\hat{w}}{Dt^2}+(\tilde{f}^2+N^2)\hat{w}=-\tilde{f}\partial_x\hat{p}.
    \label{eq_calcul_scaling1}
\end{equation}
Even if the linear stability analysis shows that the maximum growth rate $\sigma_{\textrm{m}}$ and most amplified wave number $k_{x\textrm{m}}$ vary somewhat with $N$ and $\tilde{f}$, we will assume a priori here that $D/Dt\sim O(\sigma_{\textrm{m}})\sim O(k_{x\textrm{m}})\sim O(1)$. Therefore, when $N^{2}+\tilde{f}^{2}\gg1$, (\ref{eq_calcul_scaling1}) simplifies to:
\begin{equation}
    \hat{w}\simeq\frac{-1}{\tilde{f}\left(1+\left(\frac{N}{\tilde{f}}\right)^2\right)}\partial_x\hat{p}.
    \label{equation_vitesse_verticale_pression}
\end{equation}
The approximation (\ref{equation_vitesse_verticale_pression}) seems to be valid as soon as $N^{2}+\tilde{f}^{2}\gg1$, i.e. if either $N\gg1$ or/and $\tilde{f}\gg1$. However, we will see a posteriori that the final results are self-consistent with the approximation (\ref{equation_vitesse_verticale_pression}) only if $N\gg1$.\\

\noindent Inserting (\ref{equation_vitesse_verticale_pression}) in the horizontal momentum equations (\ref{systeme_adim_equations_a}), (\ref{systeme_adim_equations_b}) and (\ref{systeme_adim_equations_e}) leads to a closed system of equations for $(\hat{u}, \hat{v}, \hat{p})$:
\begin{subequations}
	\begin{align}
	\displaystyle \partial_t \hat{u}+U\partial_x\hat{u}+\hat{v}\partial_y U+\hat{u}\partial_{x}\hat{u}+\hat{v}\partial_{y}\hat{u}&=-\alpha\partial_x 	\hat{p}, \label{horizontal_system_avec_alpha_a}\\
	\displaystyle \partial_t \hat{v}+U\partial_x\hat{v}+\hat{u}\partial_{x}\hat{v}+\hat{v}\partial_{y}\hat{v}&=-\partial_y \hat{p}, \label{horizontal_system_avec_alpha_b}\\
	\partial_x \hat{u}+\partial_y \hat{v}&=0, \label{horizontal_system_avec_alpha_c}
	\end{align}
	\label{horizontal_system_avec_alpha}
\end{subequations}
which are very similar to the equations in the traditional limit except for the presence of the parameter $\alpha$ in (\ref{horizontal_system_avec_alpha_a}):
\begin{equation}
\alpha=\frac{N^{2}}{\tilde{f}^{2}+N^{2}}.
\label{expression_alpha}
\end{equation}
\noindent This parameter is a measure of the non-traditional Coriolis parameter compared to the Brunt-V\"ais\"al\"a frequency: it is equal to unity in the traditional limit $\tilde{f}=0$, and it decreases as $\tilde{f}/N$ increases. When $\tilde{f}$ is non-zero, (\ref{horizontal_system_avec_alpha}) can be however transformed into the traditional equations by applying the following change of variables:
\begin{equation}
\hat{u}=\hat{u}', \quad \hat{v}=\frac{1}{\sqrt{\alpha}}\hat{v}', \quad \hat{p}=\frac{1}{\alpha}\hat{p}', 
\label{scaling1}
\end{equation}
\begin{equation}
\partial_{x}=\frac{1}{\sqrt{\alpha}}\partial_{x}', \quad \partial_{y}=\partial_{y}', \quad \partial_{t}=\frac{1}{\sqrt{\alpha}}\partial_{t}'.
\label{scaling2}
\end{equation}
Indeed, the equations for the primed variables are strictly two-dimensional:
\begin{subequations}
\begin{align}
\displaystyle \partial_t' \hat{u}'+U\partial_x'\hat{u}'+\hat{v}'\partial_y' U+\hat{u}'\partial_{x}'\hat{u}'+\hat{v}'\partial_{y}'\hat{u}'&=-\partial_x' \hat{p}', \label{horizontal_system_sans_alpha_a} \\
\displaystyle \partial_t' \hat{v}'+U\partial_x'\hat{v}'+\hat{u}'\partial_{x}'\hat{v}'+\hat{v}'\partial_{y}'\hat{v}'&=-\partial_y' \hat{p}', \label{horizontal_system_sans_alpha_b} \\
\partial_x'\hat{u}'+\partial_y'\hat{v}'&=0. \label{horizontal_system_sans_alpha_c}
\end{align}
\label{horizontal_system_sans_alpha}
\end{subequations}
Before discussing some properties associated to (\ref{horizontal_system_sans_alpha}) and (\ref{scaling1}-\ref{scaling2}), we first check a posteriori whether or not the assumption $N^{2}+\tilde{f}^{2}\gg d^{2}/dt^{2}$ used to derive (\ref{horizontal_system_sans_alpha}) is satisfied. The change of variable (\ref{scaling2}) implies that $D/Dt\sim\partial_{t}\sim\partial_{x}=O(1/\sqrt{\alpha})$ since $\partial_{t}'\sim O(1)$ and $\partial_{x}'\sim O(1)$. Hence, we should have:
\begin{equation}
	N^{2}+\tilde{f}^{2}\gg1/\alpha.
	\label{inegalite_scaling_}
\end{equation}
From (\ref{expression_alpha}), we see that this is the case only if $N^{2}\gg1$ and not if $\tilde{f}\gg1$ with $N\leqslant O(1)$ contrary to what (\ref{eq_calcul_scaling1}) suggests. Hence, the reduced system (\ref{horizontal_system_avec_alpha}) is expected to be valid only if the stratification is strong, i.e. $N\gg1$.\\

\noindent The transformation (\ref{scaling1}-\ref{scaling2}) has several interesting properties. First, concerning the linear stability of the base flow (section \ref{subsect3}), it implies that the dispersion relation for any $\tilde{f}$ and $N\gg1$ is $\sigma=\mathcal{F}(k_{x}\sqrt{\alpha})/\sqrt{\alpha}$, where $\sigma=\mathcal{F}(k_{x})$ is the dispersion relation for $\tilde{f}=0$. Figure \ref{scaled_Park_et_al} indeed shows that the rescaled growth rate $\sigma\sqrt{\alpha}$ represented as a function of $k_{x}\sqrt{\alpha}$ collapses toward the growth rate curve for $\tilde{f}=0$ when $N\gg1$ and $\tilde{f}$ is non-zero.\\

\noindent Another important property of $(\ref{horizontal_system_sans_alpha})$ is that the total vertical vorticity $\xi '=\partial_{x}'\hat{v}'-\partial_{y}'\hat{u}'-\partial_{y}'U$ is conserved following the motion since these equations are purely two-dimensional. When rewritten in terms of the unprimed variables, the total vertical vorticity becomes $\xi '=\alpha\partial_{x}\hat{v}-\partial_{y}\hat{u}-\partial_{y}U$ and its governing equation remains:
\begin{equation}
	\partial_{t}\xi'+U\partial_{x}\xi'+\boldsymbol{\hat{u}}\cdot\boldsymbol{\nabla}\xi'=0.
	\label{conservation_vorticite_prime}
\end{equation}
It can be easily shown that (\ref{conservation_vorticite_prime}) derives from the conservation of potential vorticity $\Pi=(\boldsymbol{\omega}+\tilde{f}\boldsymbol{e_{y}})\cdot(N^{2}\boldsymbol{e_{z}}+\boldsymbol{\nabla}b)$.\\

\noindent In the following sections, we further show that the levels of saturation of the different energy components for any $\tilde{f}$ and sufficiently large $N$ can be deduced from those for $\tilde{f}=0$ thanks to the transformation (\ref{scaling1}-\ref{scaling2}).
\subsubsection{Scaling for the horizontal kinetic energy of the perturbation}
The transformation (\ref{scaling1}-\ref{scaling2}) implies that the horizontal kinetic energy can be written:
\begin{equation}
\hat{K}_h=\frac{1}{2L_{x}}\int_0^{L_x}\int_0^{L_y}(\hat{u}^2+\hat{v}^{2})\,dy dx=\frac{1}{2L_{x}'}\int_0^{L_x'}\int_0^{L_y'}\left(\hat{u}'^2+\frac{\hat{v}'^2}{\alpha}\right)\,dy' dx',
\label{Kh} 
\end{equation}
where $L_{x}'=L_{x}\sqrt{\alpha}$ and $L_{y}'=L_{y}$.\\

\begin{figure}
\centering
        \includegraphics[width=0.5\linewidth, trim=0cm 0cm 0cm 0cm, clip]{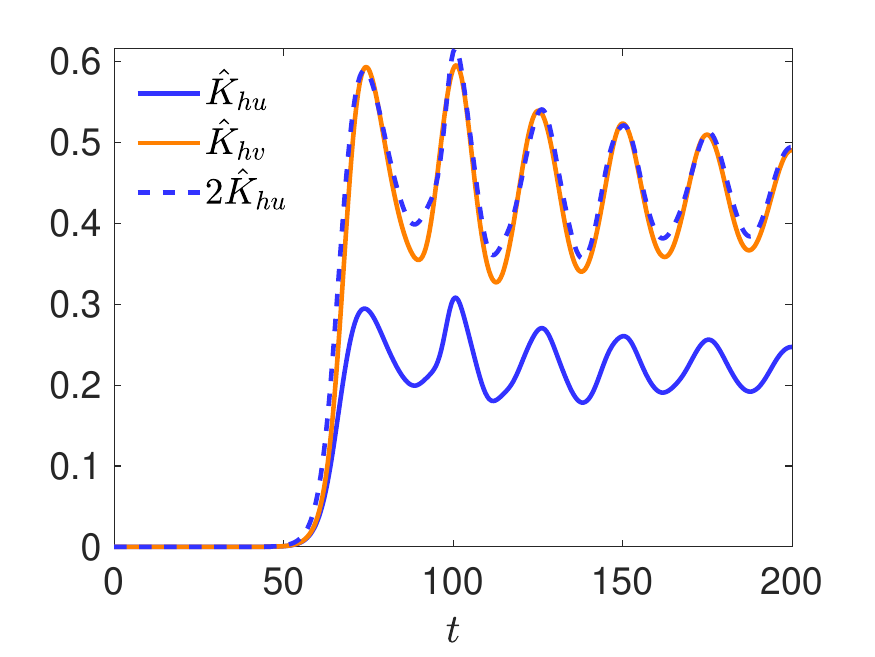}
        \captionsetup{width=1.\linewidth, justification=justified, format=plain}
        \caption{Evolution of $\displaystyle \hat{K}_{hu}=\frac{1}{2L_{x}}\int_0^{L_x}\int_0^{L_y}\hat{u}^2\,dy dx$ and $\displaystyle \hat{K}_{hv}=\frac{1}{2L_{x}}\int_0^{L_x}\int_0^{L_y}\hat{v}^2\,dy dx$ for $\tilde{f}=0$, $Re=2000$ and $Sc=1$. The dashed line, which represents $2\hat{K}_{hu}$ shows that $\hat{K}_{hv}\simeq2\hat{K}_{hu}$ as assumed in (\ref{empiric}).}
        \label{fig_relation_empirique}
\end{figure}

\noindent The numerical simulations show that, at saturation, we have the empiric approximate relation (see figure \ref{fig_relation_empirique}):
\begin{equation}
	 \int_0^{L_x'}\int_0^{L_y'}\hat{v}'^2\,dy' dx' \simeq 2 \int_0^{L_x'}\int_0^{L_y'}\hat{u}'^2\,dy' dx'
	\label{empiric}
\end{equation}
\noindent Therefore, the horizontal kinetic energy for any $\tilde{f}$ and $N\gg1$ is approximately given by:
\begin{equation}
	\hat{K}_h\simeq\frac{1}{3}\left(3+2\left(\frac{\tilde{f}}{N}\right)^{2}\right)\hat{K}_h',
	\label{relation_Kh_scaled_Kh}
\end{equation}
\noindent where $\hat{K}_h'$ is the horizontal kinetic energy for $\tilde{f}=0$. Thus, the horizontal kinetic energy of the perturbations depends only on the ratio $\tilde{f}/N$. It increases with $\tilde{f}$ and decreases with $N$ in agreement with the trends observed in figure \ref{Ec_Ep_ftilde_et_N_fixe}.
\begin{figure}
\centering
\begin{subfigure}[b]{0.49\textwidth}
        \includegraphics[width=\linewidth, trim=0cm 0cm 0cm 0cm, clip]{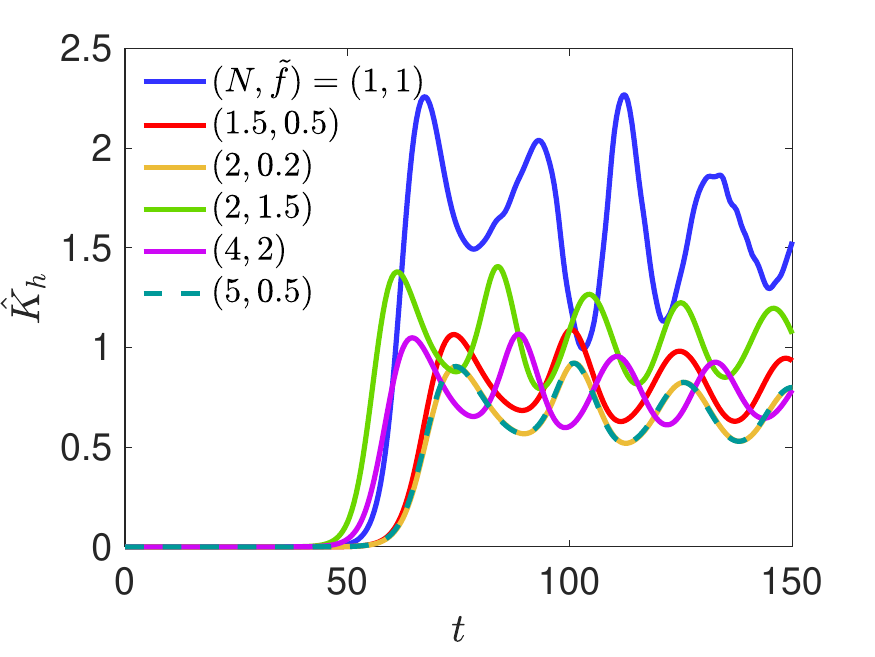}
        \caption{}
        \label{Ech_scaling_N_ftilde_quelconques_et_ratio_1_et_10_a}
        \end{subfigure}
        \begin{subfigure}[b]{0.49\textwidth}
        \includegraphics[width=\linewidth, trim=0cm 0cm 0cm 0cm, clip]{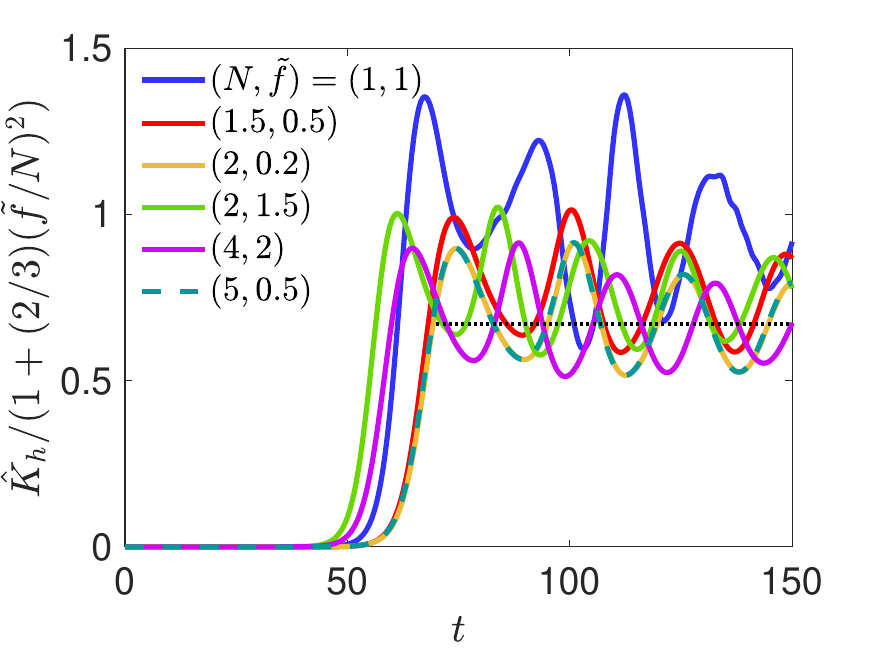}
        \caption{}
        \label{Ech_scaling_N_ftilde_quelconques_et_ratio_1_et_10_b}
        \end{subfigure}
        \captionsetup{width=1.\linewidth, justification=justified, format=plain}
        \caption{(a) Horizontal kinetic energy of the perturbations for different parameters $N$ and $\tilde{f}$. (b) Horizontal kinetic energy of the perturbations rescaled by $(1+(2/3)(\tilde{f}/N)^{2})$ for the same parameters as in (a). The black dotted line in (b) shows the mean value towards which the rescaled curves collapse.}
        \label{Ech_scaling_N_ftilde_quelconques_et_ratio_1_et_10}
\end{figure}
The relation (\ref{relation_Kh_scaled_Kh}) is further tested in figure \ref{Ech_scaling_N_ftilde_quelconques_et_ratio_1_et_10}. Figure \ref{Ech_scaling_N_ftilde_quelconques_et_ratio_1_et_10_a} displays the evolutions of $\hat{K}_h$ for various sets of parameters $(N, \tilde{f})$ such that $N\geqslant 1$. Figure \ref{Ech_scaling_N_ftilde_quelconques_et_ratio_1_et_10_b} shows $\hat{K}_h$ rescaled by $(1+2/3(\tilde{f}/N)^{2})$ for the same parameters. All the horizontal kinetic energies in the saturated regime then collapse approximately around the same mean value. This value is indicated in figure \ref{Ech_scaling_N_ftilde_quelconques_et_ratio_1_et_10_b} by a horizontal black dotted line corresponding to the mean horizontal kinetic energy at saturation of the simulation for $(N,\tilde{f})=(5, 0.5)$ which is closest to the traditional limit. We can notice that even the curve for $(N, \tilde{f})=(1, 1)$ collapses relatively well on the others, even if $N$ is not much larger than unity.
\subsubsection{Scaling for the vertical kinetic energy}
Using (\ref{equation_vitesse_verticale_pression}) and the transformation (\ref{scaling1}-\ref{scaling2}), the vertical velocity can be written:
\begin{equation}
	\hat{w}\simeq-\frac{\tilde{f}}{N^{2}+\tilde{f}^{2}}\frac{\partial_{x}'\hat{p}'}{\alpha^{3/2}}.
	\label{relation_calcul_Kv}
\end{equation} 
\noindent Hence, the vertical kinetic energy can be expressed as:
\begin{equation}
\hat{K}_v=\frac{1}{2L_{x}}\int_0^{L_x}\int_0^{L_y}\hat{w}^2\,dy dx\simeq\left(\frac{\tilde{f}}{N}\right)^{2}\left(\frac{N^{2}+\tilde{f}^{2}}{N^{4}}\right)\left(\frac{1}{2L_{x}'}\int_0^{L_x'}\int_0^{L_y'}(\partial_{x}'\hat{p}')\,dy' dx'\right).
\label{Kv} 
\end{equation}
\begin{figure}
\centering
\begin{subfigure}[b]{0.49\textwidth}
        \includegraphics[width=\linewidth, trim=0cm 0cm 0cm 0cm, clip]{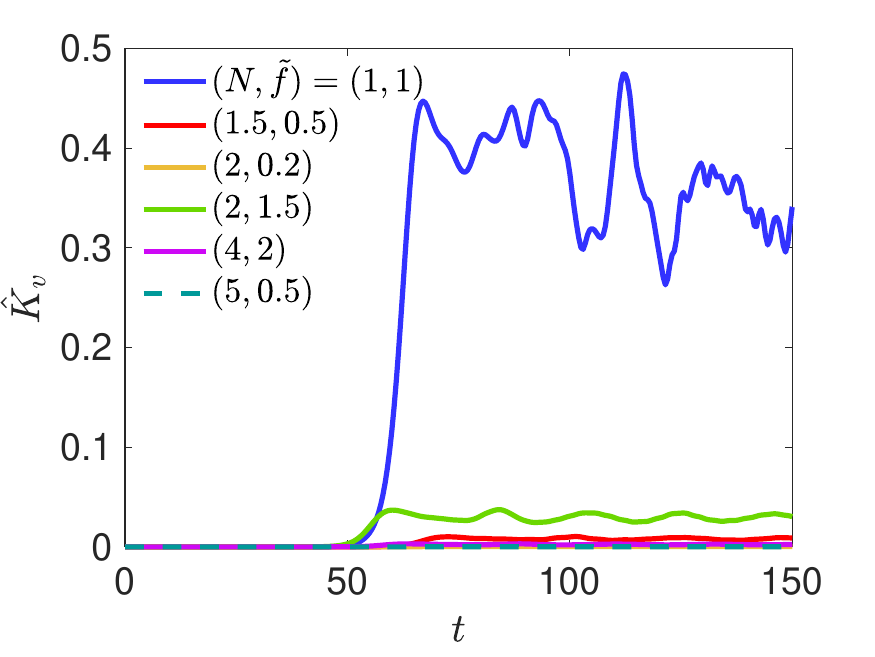}
        \caption{}
        \label{Ec_v_scaling_N_ftilde_quelconques_a}
        \end{subfigure}
        \begin{subfigure}[b]{0.49\textwidth}
        \includegraphics[width=\linewidth, trim=0cm 0cm 0cm 0cm, clip]{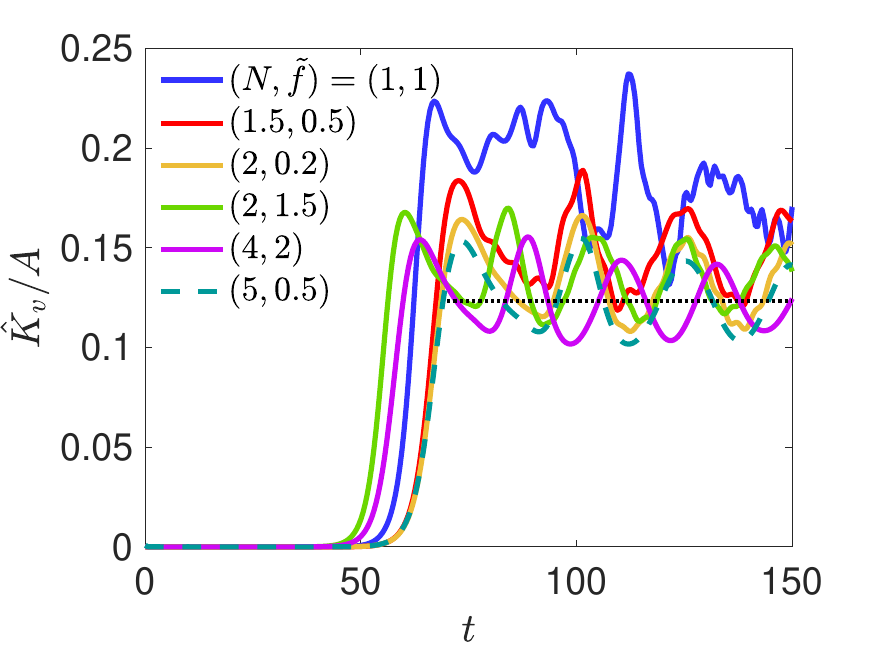}
        \caption{}
        \label{Ec_v_scaling_N_ftilde_quelconques_b}
        \end{subfigure}
        \captionsetup{width=1.\linewidth, justification=justified, format=plain}
        \caption{(a) Vertical kinetic energy of the perturbations for different parameters $N$ and $\tilde{f}$. (b) Vertical kinetic energy of the perturbations rescaled by $(\tilde{f}/N)^{2}(\tilde{f}^{2}+N^{2})/N^{4}$ for the same parameters as in (a). The black dotted line in (b) shows the mean value towards which the rescaled curves collapse.}
        \label{Ec_v_scaling_N_ftilde_quelconques}
\end{figure}
In other words, $\hat{K}_v$ should be proportional to the factor $A=(\tilde{f}/N)^{2}(1+(\tilde{f}/N)^{2})/N^{2}$. As seen in figure \ref{Ec_v_scaling_N_ftilde_quelconques_a}, the vertical kinetic energy saturates at very different levels when $\tilde{f}$ and $N$ are varied (keeping $N\geqslant1$) but when rescaled by $A$ (figure \ref{Ec_v_scaling_N_ftilde_quelconques_b}), all the curves collapse around the same mean level of saturation (shown by a black dotted line). The rescaled vertical kinetic energy for $(N, \tilde{f})=(1, 1)$ is 30\% away from the others since the assumption $N\gg1$ is less well verified than for the other sets of parameters.
\subsubsection{Scaling for the potential energy}
Similarly, the vertical momentum equation (\ref{systeme_adim_equations_c}) in the inviscid limit, i.e.:
\begin{equation}
	\frac{D\hat{w}}{Dt}-\tilde{f}\hat{u}=\hat{b}
	\label{frac_dwdt_}
\end{equation}
can be used to obtain an expression for $\hat{b}$ in terms of primed quantities, thanks to (\ref{scaling1}-\ref{scaling2}) and (\ref{relation_calcul_Kv}):
\begin{equation}
	\hat{b}=-\tilde{f}\hat{u}'+\left(\frac{\tilde{f}}{N}\right)^{2}\left(1+\left(\frac{\tilde{f}}{N}\right)^{2}\right)\frac{D}{Dt'}(\partial_{x}'\hat{p}').
\label{relation_calcul_P}
\end{equation} 
It shows that for strong stratification $N\gg1$, the buoyancy can be approximated by $\hat{b}=-\tilde{f}\hat{u}'$, which corresponds to the quasi-hydrostatic balance, i.e. the hydrostatic equilibrium in presence of the non-traditional Coriolis force \citep[][]{Gerkemaetal2008} and in the absence of the vertical pressure gradient.\\

\noindent Therefore, the potential energy of the perturbations can be approximated by:
\begin{equation}
\hat{P}=\frac{1}{2L_{x}N^{2}}\int_0^{L_x}\int_0^{L_y}\hat{b}^2\,dy dx\simeq\left(\frac{\tilde{f}}{N}\right)^{2}\frac{1}{2L_{x}'}\int_0^{L_x'}\int_0^{L_y'}\hat{u}'^{2}\,dy' dx'.
\label{P} 
\end{equation}
\begin{figure}
\centering
        \begin{subfigure}[b]{0.49\textwidth}
        \includegraphics[width=\linewidth, trim=0cm 0cm 0cm 0cm, clip]{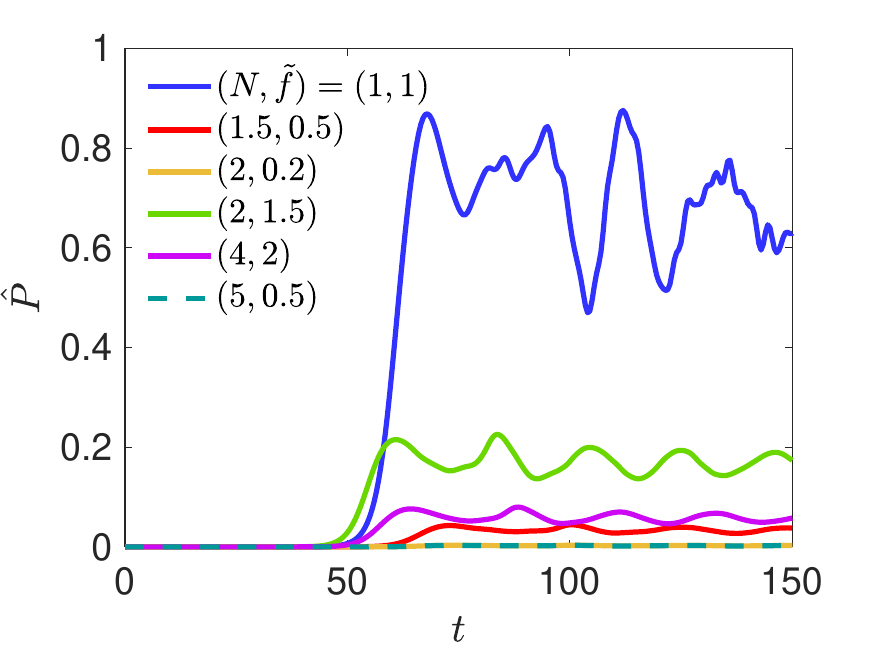}
        \caption{}
        \label{Ep_scaling_N_ftilde_quelconques_et_ratio_1_et_10_a}
        \end{subfigure}
        \begin{subfigure}[b]{0.49\textwidth}
        \includegraphics[width=\linewidth, trim=0cm 0cm 0cm 0cm, clip]{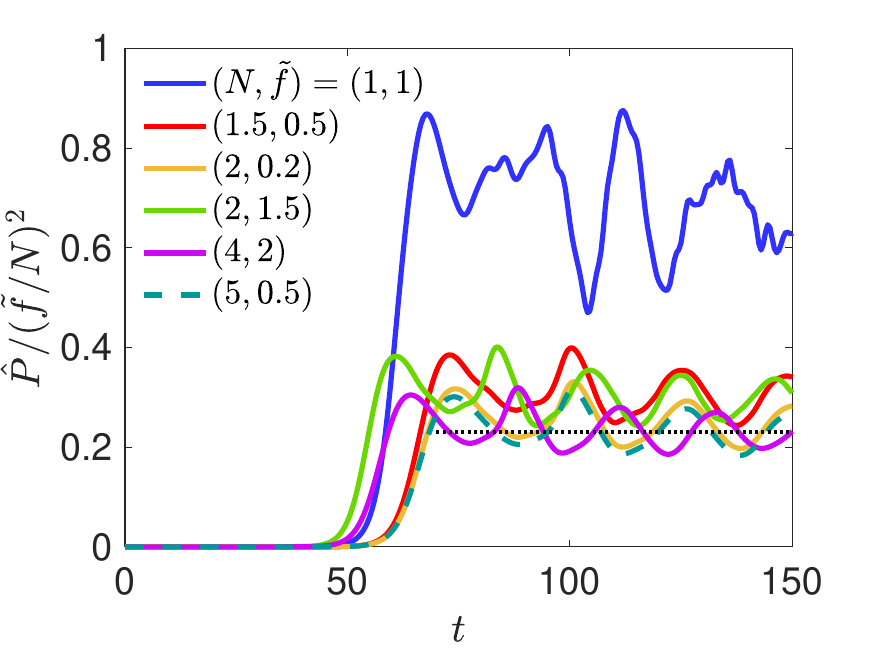}
        \caption{}
        \label{Ep_scaling_N_ftilde_quelconques_et_ratio_1_et_10_b}
        \end{subfigure}
        \captionsetup{width=1.\linewidth, justification=justified, format=plain}
        \caption{(a) Potential energy of the perturbations for different parameters $N$ and $\tilde{f}$. (b) Potential energy of the perturbations rescaled by $(\tilde{f}/N)^{2}$ for the same parameters as in (a). The black dotted line in (b) shows the mean value towards which the rescaled curves collapse.}
        \label{Ep_scaling_N_ftilde_quelconques_et_ratio_1_et_10}
\end{figure}
Figures \ref{Ep_scaling_N_ftilde_quelconques_et_ratio_1_et_10_a} and \ref{Ep_scaling_N_ftilde_quelconques_et_ratio_1_et_10_b} show the bare potential energy and the potential energy rescaled by $(\tilde{f}/N)^2$ respectively, for the same values of $\tilde{f}$ and stratification as in figure \ref{Ech_scaling_N_ftilde_quelconques_et_ratio_1_et_10} and \ref{Ec_v_scaling_N_ftilde_quelconques}. While the levels of saturation of the potential energy are widely spread in figure \ref{Ep_scaling_N_ftilde_quelconques_et_ratio_1_et_10_a}, they gather approximately around the same mean value when rescaled (black dotted line on figure \ref{Ep_scaling_N_ftilde_quelconques_et_ratio_1_et_10_b}). However, the rescaling is less satisfactory for $(N, \tilde{f})=(1, 1)$ because the assumption of a strong stratification is not well satisfied and because the relative amplitude of the neglected term in the quasi-hydrostatic balance is proportional to $(1/N^{2})(1+(\tilde{f}/N)^{2})$, i.e. unity, as seen in (\ref{relation_calcul_P}).
\subsubsection{Summary of the scaling of the energies}
\begin{figure}
\centering
        \begin{subfigure}[b]{0.48\textwidth}
        \includegraphics[width=\linewidth, trim=0cm 0cm 0.8cm 0cm, clip]{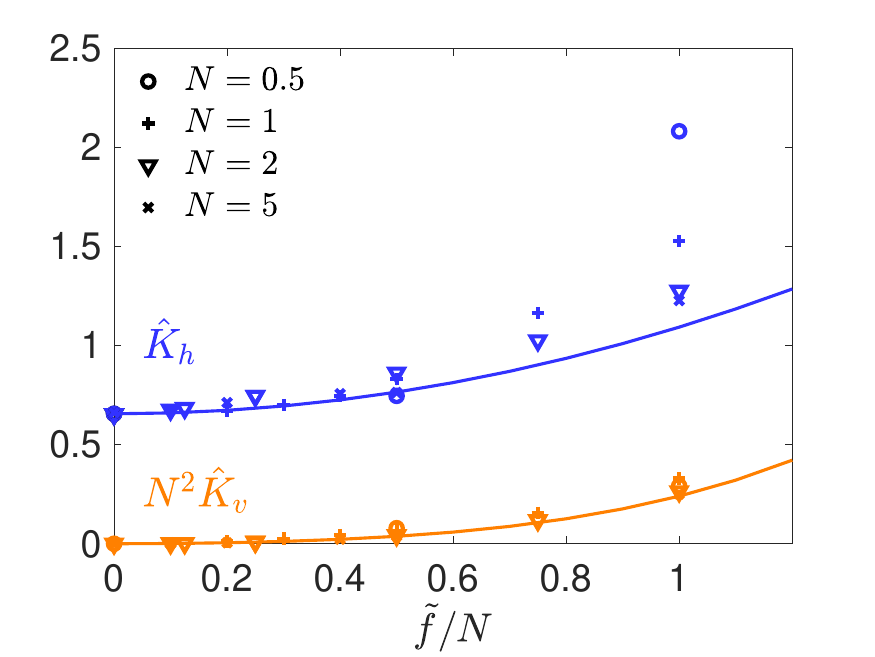}
        \caption{}
        \label{Scaling_Kh_Kv_P_a}
        \end{subfigure}
        \begin{subfigure}[b]{0.48\textwidth}
        \includegraphics[width=\linewidth, trim=0cm 0cm 0.8cm 0cm, clip]{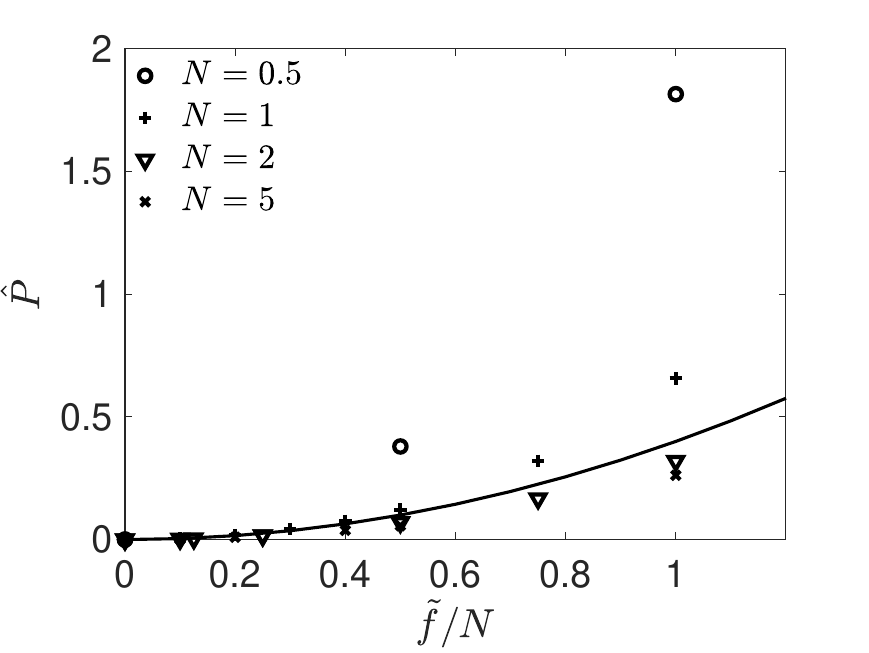}
        \caption{}
        \label{Scaling_Kh_Kv_P_b}
        \end{subfigure}
        \captionsetup{width=1.\linewidth, justification=justified, format=plain}
        \caption{(a) Mean horizontal kinetic energy (blue) and vertical kinetic energy multiplied by $N^{2}$ (orange) at saturation as a function of the ratio $\tilde{f}/N$. (b) Mean potential kinetic energy (black). The different symbols represent simulations at different $N$. Solid lines show the relations $\hat{K}_h\simeq(1/3)(3+2(\tilde{f}/N)^{2})\hat{K}_h'$ (blue), $N^{2}\hat{K}_v~=~0.12(\tilde{f}/N)^{2}(1+(\tilde{f}/N)^{2})$ (orange) and $P=0.4(\tilde{f}/N)^{2}$ (black).}
        \label{Scaling_Kh_Kv_P}
\end{figure}

The scalings for the three energies are summarized in figure \ref{Scaling_Kh_Kv_P}. The symbols in this figure show the mean level at saturation of $\hat{K}_{h}$, $N^{2}\hat{K}_{v}$ and $\hat{P}$ averaged over a range of $100$ time units after non-linear saturation, as a function of $\tilde{f}/N$ for four values of $N$. The solid lines represent the scaling relations derived in the strongly stratified limit for the three energies. The relation (\ref{relation_Kh_scaled_Kh}) for the horizontal kinetic energy is entirely given by the kinetic energy in the traditional limit $\hat{K}_{h}'\simeq 0.65$ and the ratio $\tilde{f}/N$. In contrast, the multiplying factors in front of the relations (\ref{Kv}) and (\ref{P}) for the vertical kinetic energy and potential energy have been determined by a fit. These scaling relations are in good agreement with the simulations when $\tilde{f}/N$ is small and $N$ is large: $\tilde{f}/N\leqslant0.5$ and $N\geqslant2$. On the other hand, some slight and significant departures exist for $N\geqslant2$ and when $N\leqslant1$, respectively, for $\tilde{f}/N\geqslant0.5$, especially for $\hat{K}_{h}$ and $\hat{P}$. This is expected because the assumption of a strongly stratified fluid used to compute these scaling laws is satisfied only when $N\gg1$.

\section{Conclusion and discussion} \label{Sect_conclusion}
In this paper, we have investigated, by means of direct numerical simulations, the non-linear dynamics of a horizontal parallel shear flow in the presence of the full Coriolis acceleration and of stable stratification. Building upon the linear stability analysis of \cite{Parketal2021}, we have studied the effects of the non-dimensional non-traditional Coriolis parameter $\tilde{f}$ and non-dimensional stratification $N$  for constant Reynolds and Schmidt numbers $Re=2000$, $Sc=1$, in the case of the hyperbolic tangent velocity profile. The study has been restricted to two-dimensional perturbations but with three velocity components in order to focus only on the shear instability associated with the inflection point.\\

We have identified three different types of non-linear evolutions for the shear instability. For a sufficiently strong stratification $N>1$ and $\tilde{f}<N$, the evolution is similar to the traditional case leading to the formation of a stable and coherent vortex. In contrast, for weak stratification $N\lesssim 0.5$ and for sufficiently large $\tilde{f}$, secondary instabilities develop vigorously and lead to small-scale turbulence in a large part of the flow. In between, i.e., for moderate stratification, secondary instabilities develop but remain confined within the Kelvin-Helmholtz billows.\\
A key quantity at the origin of these different evolutions is the vertical velocity field. Indeed, the horizontal dynamics generates vertical velocity and buoyancy perturbations through the vertical non-traditional Coriolis acceleration and the advection of the base buoyancy field. In turn, the horizontal non-traditional Coriolis acceleration due to the vertical velocity modifies the horizontal dynamics, i.e. creates vertical vorticity anomalies. The typical amplitude of the vertical velocity increases as the non-traditional Coriolis parameter increases for a given stratification, or as the stratification decreases for a given non-traditional Coriolis parameter. Accordingly, the analysis of the evolution of the kinetic and potential energies have shown that an increase of the stratification tends to render the dynamics closer to the traditional one whereas an increase of the non-traditional Coriolis parameter leads to the opposite effect.\\
The study of the enstrophy evolution shows that the enstrophy in the traditional-like cases remains close to the one in the traditional case. Nevertheless, it may increase slightly in contrast to the traditional limit due to the active vortex stretching and tilting. In the ``mixed'' and ``non-traditional'' cases, the enstrophy increases moderately and dramatically, respectively. Hence, the enstrophy dynamics resembles the one of fully three-dimensional flows, especially in the ``non-traditional'' case, even if the flow is two-dimensional. \\
The nature of the secondary instabilities have also been investigated. We have shown that inflection points satisfying locally the Fj\o rtoft necessary condition for instability exist prior to the onset of secondary instabilities. This suggests that they are shear instabilities due to the vorticity generated by tilting of the horizontal component of background rotation.\\
Finally, the strongly stratified regime $N\gg 1$ has been further studied in the inviscid limit. It has been shown that the traditional equations can be recovered by means of a transformation only depending on $N$ and $\tilde{f}$. This implies that the linear stability characteristics for any $\tilde{f}$ for $N\gg 1$ can be deduced from those of the traditional limit. Similarly, the dependencies on $\tilde{f}$ and $N$ of the horizontal and vertical kinetic energies as well as the potential energy at saturation have been predicted thanks to this transformation. A good agreement with the simulations has been observed when $N\gg 1$. Although all these results pertain to the specific case of an hyperbolic tangent velocity profile, we expect that similar results would be obtained for other base velocity profiles.\\

In summary, the non-linear dynamics of the horizontal shear instability under the complete Coriolis acceleration can be widely different from the classical scenario under the traditional approximation. In particular, the emergence of small-scale turbulence can render transport and mixing processes much more efficient than expected from studies and parametrizations assuming the traditional approximation. In the future, our study should help refine transport and mixing processes parametrizations for stellar interiors, planetary atmospheres and oceans.\\
Strikingly, such transition to turbulence occurs for weakly stratified fluid for non-dimensional buoyancy frequency $N\lesssim 0.5$ and even if the non-dimensional non-traditional Coriolis parameter is small $\tilde{f}\gtrsim 0.025$. Hence, the ratio $\tilde{f}/N$ should be above $0.05$. Interestingly, this range corresponds to the typical values in the atmosphere, oceans and solar-like stars for which $\tilde{f}/N\sim0.01$, $\tilde{f}/N\sim0.05-0.1$ and $\tilde{f}/N\sim0.01-0.1$, respectively. Hence, the transition to turbulence could potentially happen in many regions sufficiently close to the equator in stars, the atmosphere and oceans provided that the effect of the stratification is sufficiently weak.\\

In the present paper, the Reynolds number and Schmidt number were fixed to $Re=2000$ and $Sc=1$. In the future, we plan to investigate the effects of varying these parameters. This should allow us to determine what controls the size of small-scale motions and the maximum level of the enstrophy when turbulence arises. In addition, we will study low values of the Schmidt number such as $Sc=10^{-6}$ pertaining to stars but not the atmosphere and oceans, where $Sc$ is of order unity for temperature or very large $Sc\sim 700$ for salt.\\
In the linear case, \cite{Parketal2021} have already shown that the destabilizing effect of the non-traditional Coriolis acceleration on the shear instability is reduced as the Schmidt number is decreased. We will determine if this is also the case in the non-linear regime.\\
It would also be interesting to study the full three-dimensional dynamics for the present configuration. In particular, the shear associated to the vertical velocity field might trigger secondary shear instabilities in the vertical plane as observed by \cite{ToghraeiBillant2025} for a stratified vortex under the complete Coriolis force or by \cite{Boulangeretal2007, Boulangeretal2008} for a tilted vortex in a stratified fluid. In addition, other orientations of the shear could be considered, especially vertical shear \citep[][]{ParkMathis2025} which is also ubiquitous in geophysical and astrophysical flows. 

\textbf{Acknowledgements: }{The authors thank the referees for their detailed ad constructive reports, which helped them to improve the article. We also thank F. Gallaire for fruitful discussions and V. Toai for technical assistance.\\
A CC-BY public copyright license has been applied by the authors to the present document and will be applied to all subsequent versions up to the Author Accepted Manuscript arising from this submission, in accordance with the grant’s open access conditions.}

\textbf{Funding: }{C.M. and S.M. acknowledge support from the European Research Council (ERC) under the Horizon Europe program (Synergy Grant agreement 101071505: 4D-STAR), from the CNES SOHO-GOLF and PLATO grants at CEA-DAp, and from PNP and PNPS (CNRS/INSU). While partially funded by the European Union, views and opinions expressed are however those of the authors only and do not necessarily reflect those of the European Union or the European Research Council. Neither the European Union nor the granting authority can be held responsible for them. A part of this work was performed using HPC resources from GENCI-IDRIS (Grant 2023- AD012A07419R2).}

\textbf{Declaration of interests: }{The authors report no conflict of interest.}

\appendix
\section{Effect of the viscous and mass diffusions of the base flow}\label{appA}
The numerical simulations presented in the paper are based upon (\ref{systeme_brut_equations}) where the flow is decomposed into a steady base state and a perturbation. The advantage is that the problem can be simulated in a doubly-periodic domain, even if the base state is not periodic in the $y$ direction, because the perturbation tends to zero for large $y$. However, the viscous and diffusive dissipations of the base state $(U,B)$ are omitted in (\ref{systeme_brut_equations}).
In this appendix, we check the effect of the latter approximation by replacing the base velocity (\ref{profil_vitesse_base}) by a double hyperbolic tangent profile:
\begin{equation}
	U(y)=U_0 \left[ \tanh\left(\frac{y}{L_{0}}-\frac{L_{y}}{4} \right) - \tanh\left(\frac{y}{L_{0}}-\frac{3\,L_{y}}{4} \right) - 1 \right],
    \label{double_tanh}
\end{equation}

\noindent as done, for example, by \cite{Lecoanetetal2016}. Then, the base state $(U,B)$ tends to zero for large $y$ and the simulations can be performed directly using (\ref{systeme_vectoriel}) in a doubly-periodic domain, i.e. without the decomposition (\ref{base_state_perturbation_decomposition}). Nevertheless, to have the same confinement and resolution in the $y$ direction as in the simulations with a single hyperbolic tangent profile, the size $L_y$ and the number of points $N_y$ need to be doubled. As such, the computational cost of these simulations is twice higher.
\begin{figure}
\centering
        \begin{subfigure}[b]{0.48\textwidth}
        \includegraphics[width=\linewidth, trim=0cm 0cm 0.8cm 0cm, clip]{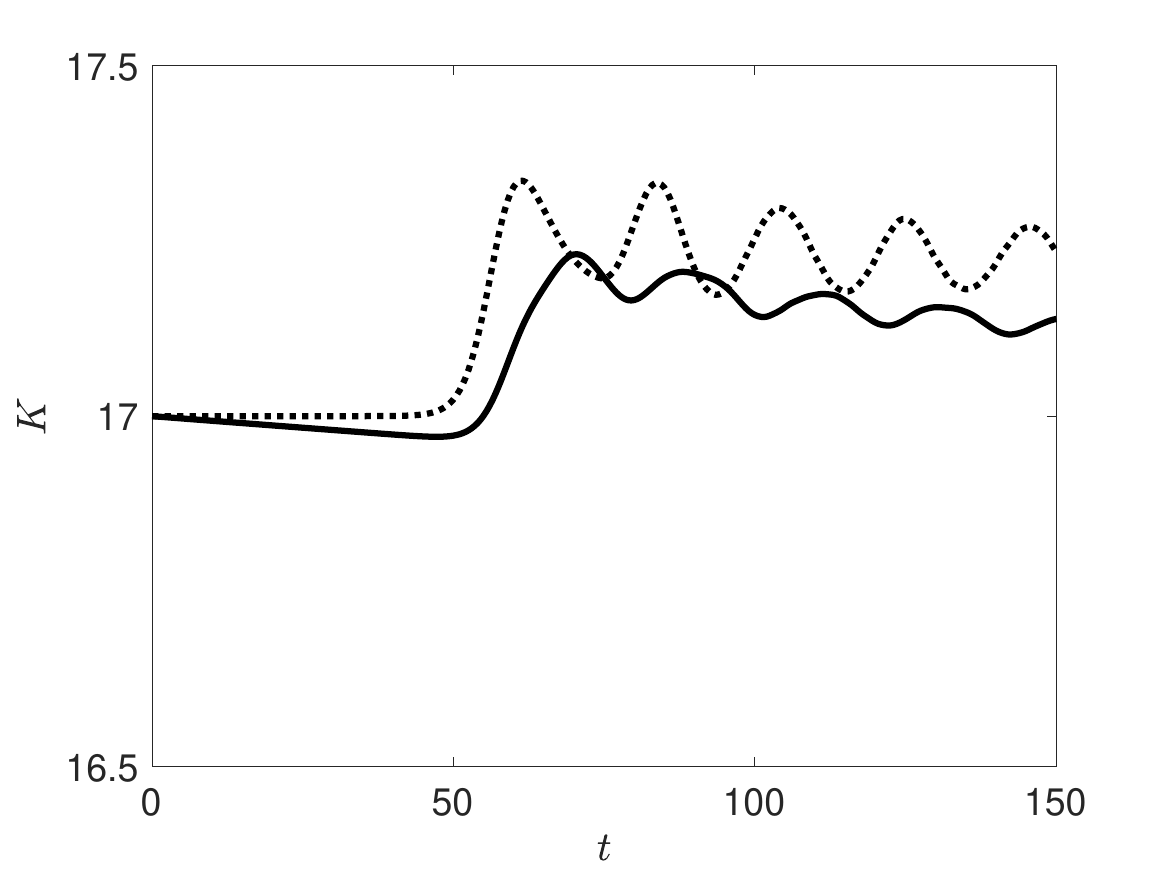}
        \caption{}
        \label{Appendix1}
        \end{subfigure}
        \begin{subfigure}[b]{0.48\textwidth}
        \includegraphics[width=\linewidth, trim=0cm 0cm 0.8cm 0cm, clip]{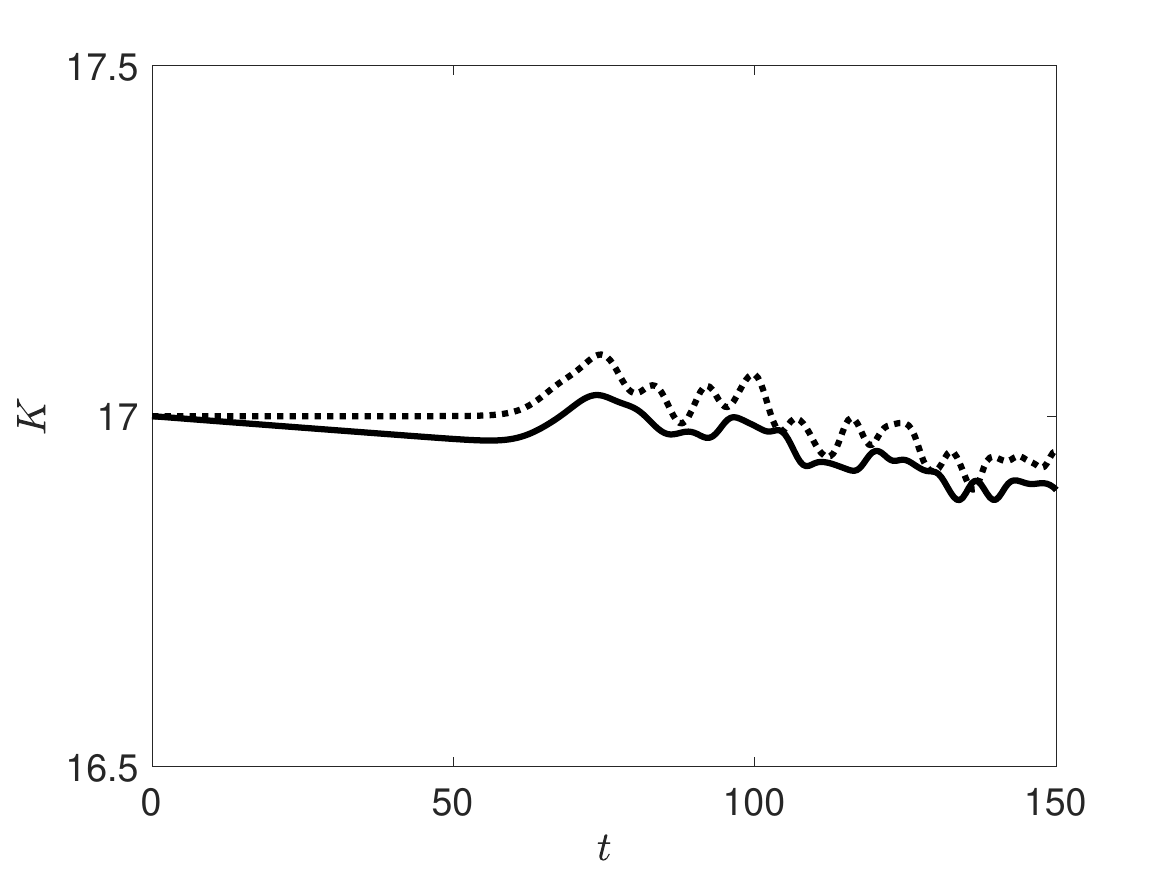}
        \caption{}
        \label{Appendix2}
        \end{subfigure}
        \captionsetup{width=1.\linewidth, justification=justified, format=plain}
        \caption{(a) Total kinetic energy for $(N, \tilde{f})=(2,1.5)$ and $Re=2000$, $Sc=1$ when the dissipation of the base state is omitted (dashed line) and when it is taken into account (solid line). (b) Same as (a) but for $(N, \tilde{f})=(1,0.5)$.}
        \label{fig_appendix}
\end{figure}
\noindent The dotted lines in figure \ref{fig_appendix} show the evolutions of the total kinetical energy
\begin{equation}
    K=\frac{1}{2L_{x}}\int_0^{L_x}\int_0^{L_y}(u^2+v^2+w^2)dy dx, 
    \label{Maintained_K}
\end{equation}
\noindent for the simulations for $(N,\tilde f)=(2,1.5)$ (traditional-like, figure \ref{Appendix1}) and $(N,\tilde f)=(1,0.5)$ (mixed, figure \ref{Appendix2}) listed in table \ref{tab_ref_simu}. The solid lines represent half the total kinetic energy of the same simulations but with the double hyperbolic tangent profile (\ref{double_tanh}). In the latter case, the total kinetic energy slightly decays because of the viscous diffusion of the base flow (\ref{profil_vitesse_base}) before the instability develops instead of being constant. However, the evolutions are globally similar. In particular, the instability develops at the same time for each case. Oscillations at saturation are also visible in both simulations for $(N,\tilde f)=(2,1.5)$ (figure \ref{Appendix1}) even if they are less pronounced when the dissipation of the base state is taken into account. For $(N,\tilde f)=(1,0.5)$, the evolutions of the total kinetic energy are also very close and the secondary instabilities appear around the same time.

\section{Simulation of two shear instability wavelengths}\label{appB}
In the paper, the box size in the streamwise direction $L_{x}$ is set to one wavelength of the shear instability in order to focus on the non-linear evolution of the Kelvin-Helmholtz instability. In this appendix, we present a simulation of the typical ``non-traditional'' evolution for $(N,\tilde{f})=(0.5,0.5)$ in which the length $L_{x}$ of the domain is doubled, $L_{x}=41.88$, so that two Kelvin-Helmholtz vortices can develop and the paring instability can occur. To keep the same resolution, $N_{x}$ is also doubled $N_{x}=2048$. Consequently, the computational cost of these simulations is twice higher. All the other parameters are the same as in table \ref{tab_ref_simu}.\\

\begin{figure}
\centering
        \begin{subfigure}[b]{0.495\textwidth}
        \includegraphics[width=\linewidth, trim=0cm 0.5cm 0.5cm 1.5cm, clip]{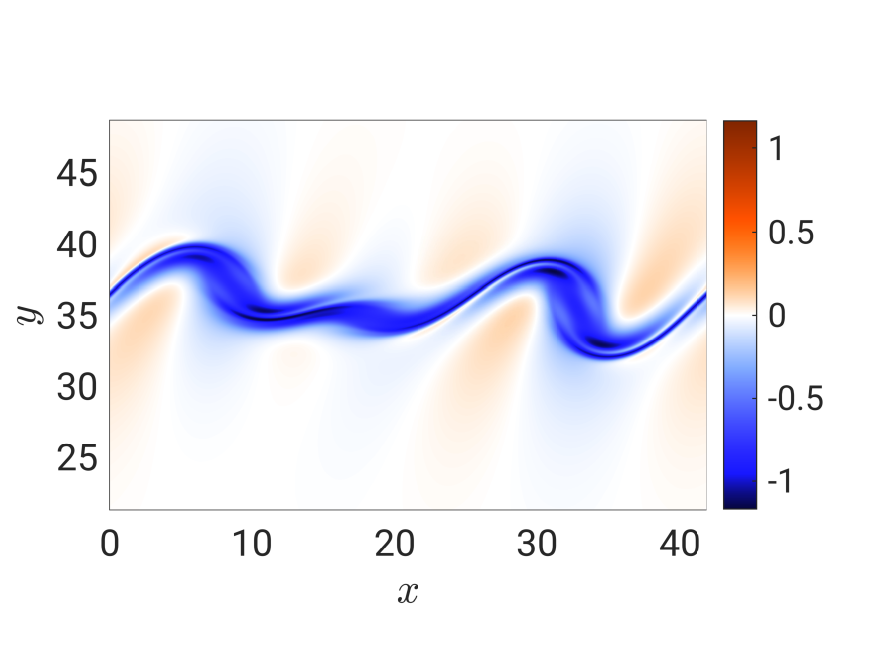}
        \caption{ }
        \label{fig_2Lx_a}
        \end{subfigure}
        \begin{subfigure}[b]{0.495\textwidth}
        \includegraphics[width=\linewidth, trim=0cm 0.5cm 0.5cm 1.5cm, clip]{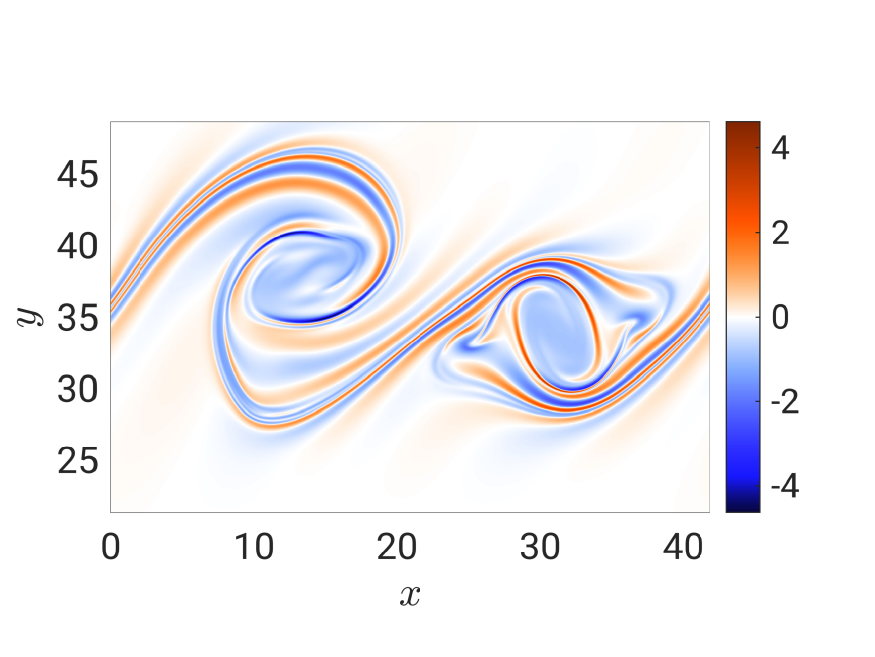}
        \caption{ }
        \label{fig_2Lx_b}
        \end{subfigure}
        \begin{subfigure}[b]{0.495\textwidth}
        \includegraphics[width=\linewidth, trim=0cm 0.5cm 0.5cm 1.5cm, clip]{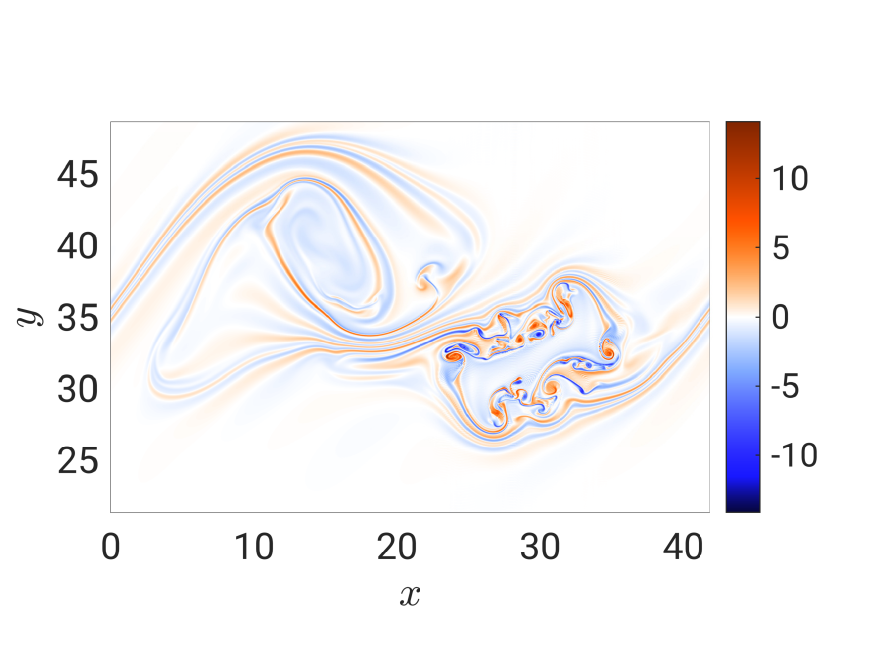}
        \caption{ }
        \label{fig_2Lx_c}
        \end{subfigure}
        \begin{subfigure}[b]{0.495\textwidth}
        \includegraphics[width=\linewidth, trim=0cm 0.5cm 0.5cm 1.5cm, clip]{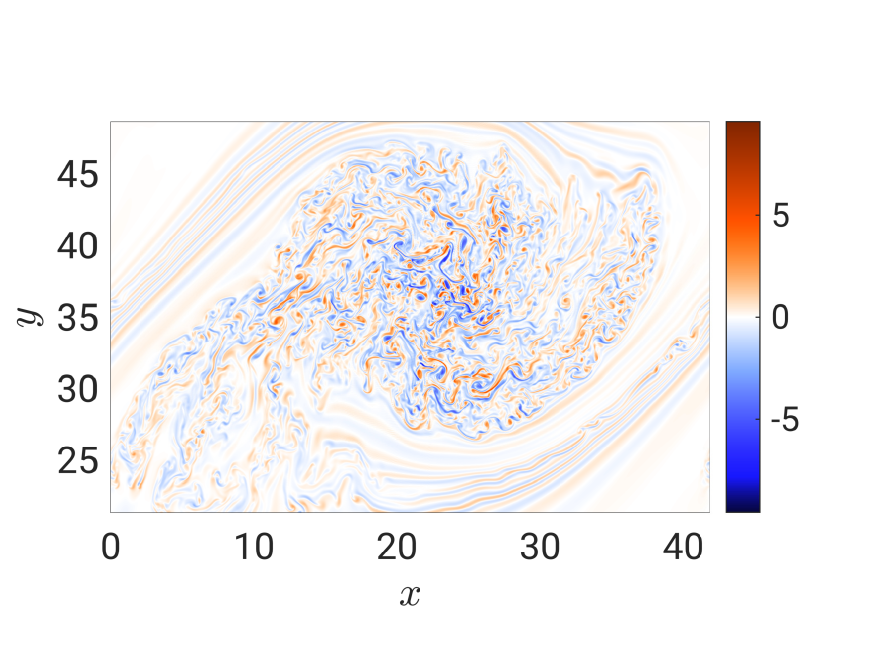}
        \caption{ }
        \label{fig_2Lx_d}
        \end{subfigure}
        \captionsetup{width=1.\linewidth, justification=justified, format=plain}
        \caption{Total vertical vorticity $(\xi=\partial_{x}v-\partial_{y}u)$, at (a), $t=87$, (b), $t=108$, (c), $t=117$, (d), $t=160$ for $N=0.5$, $\tilde{f}=0.5$ (non-traditional evolution), $Re=2000$ and $Sc=1$. The box size is $L_{x}=41.88$ instead of $L_{x}=20.94$ in figure \ref{Quatre_instants_Cas_N_petit}. The $y$-axis has been cropped compared to the original computational domain.}
        \label{Quatre_instants_Cas_2Lx}
\end{figure}

\noindent Figure \ref{Quatre_instants_Cas_2Lx} shows the total vertical vorticity at several instances. At first, the shear layer begins to roll at two distinct locations (figure \ref{fig_2Lx_a}). This creates two vortices (figure \ref{fig_2Lx_b}) that are slightly asymmetric due to the subharmonic instability, i.e. the pairing instability. They exhibit thin layers of intense positive and negative vorticity as observed in figure \ref{fig_stringSI_b}. This leads to the formation of secondary instabilities (figure \ref{fig_2Lx_c}). They appear first in the right vortex and then slightly later in the left vortex. Simultaneously, the two vortices merge, creating a large vortex colonised by small-scale turbulence (figure \ref{fig_2Lx_d}) as observed for a single wavelength (figure \ref{fig_stringSI_d}). Overall, the dynamics is therefore similar when two Kelvin-Helmholtz vortices can develop except that the pairing instability occurs concomitantly with the secondary instability. Similar observations have been made for the other typical sets of parameters of table \ref{tab_ref_simu} when the size $L_{x}$ is doubled.

\bibliographystyle{jfm}
\bibliography{jfm}

\begin{thebibliography}{56}
\expandafter\ifx\csname natexlab\endcsname\relax\def\natexlab#1{#1}\fi
\def\au#1{#1} \def\ed#1{#1} \def\yr#1{#1}\def\at#1{#1}\def\jt#1{\textit{#1}}
  \def\bt#1{#1}\def\bvol#1{\textbf{#1}} \def\vol#1{#1} \def\pg#1{#1}
  \def\publ#1{#1}\def\arxiv#1{#1}\def\org#1{#1}\def\st#1{\textit{#1}}

\bibitem[{Aerts} {\em et~al.\/}(2019){Aerts}, {Mathis} \&
  {Rogers}]{AertsMathisRogers2019}
{\sc \au{{Aerts}, C.}, \au{{Mathis}, S.} \& \au{{Rogers}, T.~M.}} \yr{2019}
  \at{{Angular Momentum Transport in Stellar Interiors}}.  \jt{Annu. Rev. A\&A}
   \bvol{57},  \pg{35--78}.

\bibitem[{Arobone} \& {Sarkar}(2012)]{AroboneSarkar2012}
{\sc \au{{Arobone}, E.} \& \au{{Sarkar}, S.}} \yr{2012}  \at{{Evolution of a
  stratified rotating shear layer with horizontal shear. Part I. Linear
  stability}}.  \jt{J. Fluid Mech.}  \bvol{703},  \pg{29--48}.

\bibitem[{Arobone} \& {Sarkar}(2013)]{AroboneSarkar2013}
{\sc \au{{Arobone}, E.} \& \au{{Sarkar}, S.}} \yr{2013}  \at{{Evolution of a
  stratified rotating shear layer with horizontal shear. Part 2. Nonlinear
  evolution}}.  \jt{J. Fluid Mech.}  \bvol{732},  \pg{373--400}.

\bibitem[{Barker} {\em et~al.\/}(2019){Barker}, {Jones} \&
  {Tobias}]{Barkeretal2019}
{\sc \au{{Barker}, A.~J.}, \au{{Jones}, C.~A.} \& \au{{Tobias}, S.~M.}}
  \yr{2019}  \at{{Angular momentum transport by the GSF instability: non-linear
  simulations at the equator}}.  \jt{Mon. Not. R. Astron. Soc.}
  \bvol{487}~(2),  \pg{1777--1794}.

\bibitem[{Barker} {\em et~al.\/}(2020){Barker}, {Jones} \&
  {Tobias}]{Barkeretal2020}
{\sc \au{{Barker}, A.~J.}, \au{{Jones}, C.~A.} \& \au{{Tobias}, S.~M.}}
  \yr{2020}  \at{{Angular momentum transport, layering, and zonal jet formation
  by the GSF instability: non-linear simulations at a general latitude}}.
  \jt{Mon. Not. R. Astron. Soc.}  \bvol{495}~(1),  \pg{1468--1490}.

\bibitem[{Basak} \& {Sarkar}(2006)]{BasakSarkar2006}
{\sc \au{{Basak}, S.} \& \au{{Sarkar}, S.}} \yr{2006}  \at{{Dynamics of a
  stratified shear layer with horizontal shear}}.  \jt{J. Fluid Mech.}
  \bvol{568},  \pg{19--54}.

\bibitem[{Boulanger} {\em et~al.\/}(2007){Boulanger}, {Meunier} \& {Le
  Diz{\`e}s}]{Boulangeretal2007}
{\sc \au{{Boulanger}, N.}, \au{{Meunier}, P.} \& \au{{Le Diz{\`e}s}, S.}}
  \yr{2007}  \at{{Structure of a stratified tilted vortex}}.  \jt{J. Fluid
  Mech.}  \bvol{583},  \pg{443}.

\bibitem[{Boulanger} {\em et~al.\/}(2008){Boulanger}, {Meunier} \& {Le
  Diz{\`e}s}]{Boulangeretal2008}
{\sc \au{{Boulanger}, N.}, \au{{Meunier}, P.} \& \au{{Le Diz{\`e}s}, S.}}
  \yr{2008}  \at{{Tilt-induced instability of a stratified vortex}}.  \jt{J.
  Fluid Mech.}  \bvol{596},  \pg{1--20}.

\bibitem[de~Bruyn~Kops \& Riley(1998)]{deBruynKopsRiley1998}
{\sc \au{de~Bruyn~Kops, S.~M.} \& \au{Riley, J.~J.}} \yr{1998}  \at{Direct
  numerical simulation of laboratory experiments in isotropic turbulence}.
  \jt{Phys. Fluids}  \bvol{10}~(9),  \pg{2125--2127}.

\bibitem[{Carpenter} {\em et~al.\/}(2012){Carpenter}, {Tedford}, {Heifetz} \&
  {Lawrence}]{Carpenteretal2012}
{\sc \au{{Carpenter}, J.~R.}, \au{{Tedford}, E.~W.}, \au{{Heifetz}, E.} \&
  \au{{Lawrence}, G.~A.}} \yr{2012}  \at{{Instability in Stratified Shear Flow:
  Review of a Physical Interpretation Based on Interacting Waves}}.
  \jt{Applied Mech. Rev.}  \bvol{64}~(6),  \pg{061001}.

\bibitem[{Caulfield}(2021)]{Caulfield2021}
{\sc \au{{Caulfield}, C.~P.}} \yr{2021}  \at{{Layering, Instabilities, and
  Mixing in Turbulent Stratified Flows}}.  \jt{Annu. Rev. Fluid Mech.}
  \bvol{53},  \pg{113--145}.

\bibitem[{Chaboyer} \& {Zahn}(1992)]{ChaboyerZahn1992}
{\sc \au{{Chaboyer}, B.} \& \au{{Zahn}, J.~P.}} \yr{1992}  \at{{Effect of
  horizontal turbulent diffusion on transport by meridional circulation.}}
  \jt{A\&A}  \bvol{253},  \pg{173--177}.

\bibitem[{Cope} {\em et~al.\/}(2020){Cope}, {Garaud} \&
  {Caulfield}]{Copeetal2020}
{\sc \au{{Cope}, L.}, \au{{Garaud}, P.} \& \au{{Caulfield}, C.~P.}} \yr{2020}
  \at{{The dynamics of stratified horizontal shear flows at low P{\'e}clet
  number}}.  \jt{J. Fluid Mech.}  \bvol{903},  \pg{A1}.

\bibitem[{Corcos} \& {Sherman}(1976)]{CorcosSherman1976}
{\sc \au{{Corcos}, G.~M.} \& \au{{Sherman}, F.~S.}} \yr{1976}  \at{{Vorticity
  concentration and the dynamics of unstable free shear layers}}.  \jt{J. of
  Fluid Mech.}  \bvol{73},  \pg{241--264}.

\bibitem[{Deloncle}(2014)]{Deloncleetal2014}
{\sc \au{{Deloncle}, A.}} \yr{2014}  \at{{NS3D v2.14: User's manual}}.
  \jt{Ecole Polytechnique} .

\bibitem[{Deloncle} {\em et~al.\/}(2008){Deloncle}, {Billant} \&
  {Chomaz}]{Deloncleetal2008}
{\sc \au{{Deloncle}, A.}, \au{{Billant}, P.} \& \au{{Chomaz}, J.}} \yr{2008}
  \at{{Nonlinear evolution of the zigzag instability in stratified fluids: a
  shortcut on the route to dissipation}}.  \jt{J. Fluid Mech.}  \bvol{599},
  \pg{229--239}.

\bibitem[{Deloncle} {\em et~al.\/}(2007){Deloncle}, {Chomaz} \&
  {Billant}]{Deloncleetal2007}
{\sc \au{{Deloncle}, A.}, \au{{Chomaz}, J.-M.} \& \au{{Billant}, P.}} \yr{2007}
   \at{{Three-dimensional stability of a horizontally sheared flow in a stably
  stratified fluid}}.  \jt{J. Fluid Mech.}  \bvol{570},  \pg{297--305}.

\bibitem[{Dymott} {\em et~al.\/}(2023){Dymott}, {Barker}, {Jones} \&
  {Tobias}]{Dymottetal2023}
{\sc \au{{Dymott}, R.~W.}, \au{{Barker}, A.~J.}, \au{{Jones}, C.~A.} \&
  \au{{Tobias}, S.~M.}} \yr{2023}  \at{{Linear and non-linear properties of the
  Goldreich-Schubert-Fricke instability in stellar interiors with arbitrary
  local radial and latitudinal differential rotation}}.  \jt{Mon. Not. R.
  Astron. Soc.}  \bvol{524}~(2),  \pg{2857--2882}.

\bibitem[{Fj\o rtoft}(1950)]{Fjortoft1950}
{\sc \au{{Fj\o rtoft}, R.}} \yr{1950}  \at{{Application of integral theorems in
  deriving criteria of stability for laminar flows and for the baroclinic
  circular vortex}}.  \jt{Geofys. Publ.}  \bvol{17}~(6),  \pg{1--52}.

\bibitem[{Flament} {\em et~al.\/}(2001){Flament}, {Lumpkin}, {Tournadre} \&
  {Armi}]{Flamentetal2001}
{\sc \au{{Flament}, P.}, \au{{Lumpkin}, R.}, \au{{Tournadre}, J.} \&
  \au{{Armi}, L.}} \yr{2001}  \at{{Vortex pairing in an unstable anticyclonic
  shear flow: discrete subharmonics of one pendulum day}}.  \jt{J. Fluid Mech.}
   \bvol{440}~(1),  \pg{401--409}.

\bibitem[{Fontane} \& {Joly}(2008)]{FontaneJoly2008}
{\sc \au{{Fontane}, J.} \& \au{{Joly}, L.}} \yr{2008}  \at{{The stability of
  the variable-density Kelvin-Helmholtz billow}}.  \jt{J. Fluid Mech.}
  \bvol{612},  \pg{237--260}.

\bibitem[{Gerkema} \& {Shrira}(2005)]{GerkemaShrira2005}
{\sc \au{{Gerkema}, T.} \& \au{{Shrira}, V.~I.}} \yr{2005}  \at{{Near-inertial
  waves in the ocean: beyond the ``traditional approximation''}}.  \jt{J. Fluid
  Mech.}  \bvol{529},  \pg{195--219}.

\bibitem[{Gerkema} {\em et~al.\/}(2008){Gerkema}, {Zimmerman}, {Maas} \& {van
  Haren}]{Gerkemaetal2008}
{\sc \au{{Gerkema}, T.}, \au{{Zimmerman}, J.~T.~F.}, \au{{Maas}, L.~R.~M.} \&
  \au{{van Haren}, H.}} \yr{2008}  \at{{Geophysical and astrophysical fluid
  dynamics beyond the traditional approximation}}.  \jt{Rev. Geophysics}
  \bvol{46}~(2),  \pg{RG2004}.

\bibitem[{Ho} \& {Huerre}(1984)]{HoHuerre1984}
{\sc \au{{Ho}, C.~M.} \& \au{{Huerre}, P.}} \yr{1984}  \at{{Perturbed free
  shear layers}}.  \jt{Annu. Rev. Fluid Mech.}  \bvol{16},  \pg{365--424}.

\bibitem[{Johannessen} {\em et~al.\/}(1996){Johannessen}, {Shuchman},
  {Digranes}, {Lyzenga}, |{Wackerman}, {Johannessen} \&
  {Vachon}]{Johannessen1996}
{\sc \au{{Johannessen}, J.~A.}, \au{{Shuchman}, R.~A.}, \au{{Digranes}, G.},
  \au{{Lyzenga}, D.~R.}, \au{|{Wackerman}, C.}, \au{{Johannessen}, O.~M.} \&
  \au{{Vachon}, P.~W.}} \yr{1996}  \at{{Coastal ocean fronts and eddies imaged
  with ERS 1 synthetic aperture radar}}.  \jt{J. Geophys. Res.}
  \bvol{101}~(C3),  \pg{6651--6667}.

\bibitem[{Kippenhahn} {\em et~al.\/}(2013){Kippenhahn}, {Weigert} \&
  {Weiss}]{Kippenhahn2013}
{\sc \au{{Kippenhahn}, R.}, \au{{Weigert}, A.} \& \au{{Weiss}, A.}} \yr{2013}
  {\em {Stellar Structure and Evolution}\/}.  \publ{Springer-Verlag, Berlin}.

\bibitem[{Klaassen} \& {Peltier}(1985{\natexlab{{\em
  a\/}}})]{KlaassenPeltier1985a}
{\sc \au{{Klaassen}, G.~P.} \& \au{{Peltier}, W.~R.}} \yr{1985{\natexlab{{\em
  a\/}}}}  \at{{Evolution of finite amplitude Kelvin-Helmholtz billows in two
  spatial dimensions}}.  \jt{J. Atmos. Sciences}  \bvol{42},  \pg{1321--1339}.

\bibitem[{Klaassen} \& {Peltier}(1985{\natexlab{{\em
  b\/}}})]{KlaassenPeltier1985}
{\sc \au{{Klaassen}, G.~P.} \& \au{{Peltier}, W.~R.}} \yr{1985{\natexlab{{\em
  b\/}}}}  \at{{The onset of turbulence in finite-amplitude Kelvin-Helmholtz
  billows}}.  \jt{J. Fluid Mech.}  \bvol{155},  \pg{1--35}.

\bibitem[{Klaassen} \& {Peltier}(1989)]{KlaassenPeltier1989}
{\sc \au{{Klaassen}, G.~P.} \& \au{{Peltier}, W.~R.}} \yr{1989}  \at{{The role
  of transverse secondary instabilities in the evolution of free shear
  layers}}.  \jt{J. Fluid Mech.}  \bvol{202},  \pg{367--402}.

\bibitem[{Klaassen} \& {Peltier}(1991)]{KlaassenPeltier1991}
{\sc \au{{Klaassen}, G.~P.} \& \au{{Peltier}, W.~R.}} \yr{1991}  \at{{The
  influence of stratification on secondary instability in free shear layers}}.
  \jt{J. Fluid Mech.}  \bvol{227},  \pg{71--106}.

\bibitem[{Lecoanet} {\em et~al.\/}(2016){Lecoanet}, {McCourt}, {Quataert},
  {Burns}, {Vasil}, {Oishi}, {Brown}, {Stone} \& {O'Leary}]{Lecoanetetal2016}
{\sc \au{{Lecoanet}, D.}, \au{{McCourt}, M.}, \au{{Quataert}, E.}, \au{{Burns},
  K.~J.}, \au{{Vasil}, G.~M.}, \au{{Oishi}, J.~S.}, \au{{Brown}, B.~P.},
  \au{{Stone}, J.~M.} \& \au{{O'Leary}, R.~M.}} \yr{2016}  \at{{A validated
  non-linear Kelvin-Helmholtz benchmark for numerical hydrodynamics}}.
  \jt{Mon. Not. R. Astron. Soc.}  \bvol{455}~(4),  \pg{4274--4288}.

\bibitem[{Mashayek} \& {Peltier}(2011)]{MashayekPeltier2011}
{\sc \au{{Mashayek}, A.} \& \au{{Peltier}, W.~R.}} \yr{2011}
  \at{{Three-dimensionalization of the stratified mixing layer at high Reynolds
  number}}.  \jt{Phys. Fluids}  \bvol{23}~(11),  \pg{111701--111701--4}.

\bibitem[{Mashayek} \& {Peltier}(2012{\natexlab{{\em
  a\/}}})]{MashayekPeltier2012}
{\sc \au{{Mashayek}, A.} \& \au{{Peltier}, W.~R.}} \yr{2012{\natexlab{{\em
  a\/}}}}  \at{{The `zoo' of secondary instabilities precursory to stratified
  shear flow transition. Part 1 Shear aligned convection, pairing, and braid
  instabilities}}.  \jt{J. Fluid Mech.}  \bvol{708},  \pg{5--44}.

\bibitem[{Mashayek} \& {Peltier}(2012{\natexlab{{\em
  b\/}}})]{MashayekPeltier2012b}
{\sc \au{{Mashayek}, A.} \& \au{{Peltier}, W.~R.}} \yr{2012{\natexlab{{\em
  b\/}}}}  \at{{The `zoo' of secondary instabilities precursory to stratified
  shear flow transition. Part 2 The influence of stratification}}.  \jt{J.
  Fluid Mech.}  \bvol{708},  \pg{45--70}.

\bibitem[{Mathis} {\em et~al.\/}(2014){Mathis}, {Neiner} \& {Tran
  Minh}]{MathisNeinerTranMihn2014}
{\sc \au{{Mathis}, S.}, \au{{Neiner}, C.} \& \au{{Tran Minh}, N.}} \yr{2014}
  \at{{Impact of rotation on stochastic excitation of gravity and
  gravito-inertial waves in stars}}.  \jt{A\&A}  \bvol{565},  \pg{A47}.

\bibitem[{Mathis} {\em et~al.\/}(2018){Mathis}, {Prat}, {Amard}, {Charbonnel},
  {Palacios}, {Lagarde} \& {Eggenberger}]{Mathisetal2018}
{\sc \au{{Mathis}, S.}, \au{{Prat}, V.}, \au{{Amard}, L.}, \au{{Charbonnel},
  C.}, \au{{Palacios}, A.}, \au{{Lagarde}, N.} \& \au{{Eggenberger}, P.}}
  \yr{2018}  \at{{Anisotropic turbulent transport in stably stratified rotating
  stellar radiation zones}}.  \jt{A\&A}  \bvol{620},  \pg{A22}.

\bibitem[{Mathis} \& {Zahn}(2004)]{MathisZahn2004}
{\sc \au{{Mathis}, S.} \& \au{{Zahn}, J.~P.}} \yr{2004}  \at{{Transport and
  mixing in the radiation zones of rotating stars. I. Hydrodynamical
  processes}}.  \jt{A\&A}  \bvol{425},  \pg{229--242}.

\bibitem[{Michalke}(1964)]{Michalke1964}
{\sc \au{{Michalke}, A.}} \yr{1964}  \at{{On the inviscid instability of the
  hyperbolic-tangent velocity profile}}.  \jt{J. Fluid Mech.}  \bvol{19},
  \pg{543--556}.

\bibitem[{Park} \& {Mathis}(2025)]{ParkMathis2025}
{\sc \au{{Park}, J.} \& \au{{Mathis}, S.}} \yr{2025}  \at{{Vertical shear
  instabilities in rotating stellar radiation zones: effects of the full
  Coriolis acceleration and thermal diffusion}}.  \jt{Mon. Not. R. Astron.
  Soc.}  \bvol{540},  \pg{298--318}.

\bibitem[{Park} {\em et~al.\/}(2021){Park}, {Prat}, {Mathis} \&
  {Bugnet}]{Parketal2021}
{\sc \au{{Park}, J.}, \au{{Prat}, V.}, \au{{Mathis}, S.} \& \au{{Bugnet}, L.}}
  \yr{2021}  \at{{Horizontal shear instabilities in rotating stellar radiation
  zones. II. Effects of the full Coriolis acceleration}}.  \jt{A\&A}
  \bvol{646},  \pg{A64}.

\bibitem[{Patnaik} {\em et~al.\/}(1976){Patnaik}, {Sherman} \&
  {Corcos}]{Patnaiketal1976}
{\sc \au{{Patnaik}, P.~C.}, \au{{Sherman}, F.~S.} \& \au{{Corcos}, G.~M.}}
  \yr{1976}  \at{{A numerical simulation of Kelvin-Helmholtz waves of finite
  amplitude}}.  \jt{J. Fluid Mech.}  \bvol{73},  \pg{215--240}.

\bibitem[{Pedley}(1969)]{Pedley1969}
{\sc \au{{Pedley}, T.~J.}} \yr{1969}  \at{{On the instability of viscous flow
  in a rapidly rotating pipe}}.  \jt{J. Fluid Mech.}  \bvol{35},  \pg{97--115}.

\bibitem[{Peltier} \& {Caulfield}(2003)]{Peltier2003}
{\sc \au{{Peltier}, W.~R.} \& \au{{Caulfield}, C.~P.}} \yr{2003}  \at{{Mixing
  Efficiency in Stratified Shear Flows}}.  \jt{Annu. Rev. Fluid Mech.}
  \bvol{35}~(35),  \pg{135--167}.

\bibitem[Pierrehumbert \& Widnall(1982)]{PierrehumbertWidnall1982}
{\sc \au{Pierrehumbert, R.~T.} \& \au{Widnall, S.~E.}} \yr{1982}  \at{The two-
  and three-dimensional instabilities of a spatially periodic shear layer}.
  \jt{J. Fluid Mech.}  \bvol{114},  \pg{59–82}.

\bibitem[{Reinaud} {\em et~al.\/}(2000){Reinaud}, {Joly} \&
  {Chassaing}]{ReinaudJolyChassaing2000}
{\sc \au{{Reinaud}, J.}, \au{{Joly}, L.} \& \au{{Chassaing}, P.}} \yr{2000}
  \at{{The baroclinic secondary instability of the two-dimensional shear
  layer}}.  \jt{Phys. Fluids}  \bvol{12}~(10),  \pg{2489--2505}.

\bibitem[{Rieutord}(2006)]{Rieutord2006}
{\sc \au{{Rieutord}, M.}} \yr{2006} {On the dynamics of radiative zones in
  rotating stars}.  \bt{In {\em EAS Publications Series\/} (ed. \ed{Michel
  {Rieutord} \& Berengere {Dubrulle}})},  \st{EAS Publications Series},
  \vol{vol.~21},  \pg{pp. 275--295}.

\bibitem[{Smyth} \& {Carpenter}(2019)]{SmythCarpenter2019}
{\sc \au{{Smyth}, W.~D.} \& \au{{Carpenter}, J.~R.}} \yr{2019} {\em Instability
  in Geophysical Flows\/}.  \publ{Cambridge University Press}.

\bibitem[{Smyth} \& {Peltier}(1994)]{SmythPeltier1994}
{\sc \au{{Smyth}, W.~D.} \& \au{{Peltier}, W.~R.}} \yr{1994}
  \at{{Three-dimensionalization of barotropic vortices on the f-plane}}.
  \jt{J. Fluid Mech.}  \bvol{265},  \pg{25--64}.

\bibitem[{Staquet}(1995)]{Staquet1995}
{\sc \au{{Staquet}, C.}} \yr{1995}  \at{{Two-dimensional secondary
  instabilities in a strongly stratified shear layer}}.  \jt{J. Fluid Mech.}
  \bvol{296},  \pg{73--126}.

\bibitem[{Toghraei} \& {Billant}(2022)]{ToghraeiBillant2022}
{\sc \au{{Toghraei}, I.} \& \au{{Billant}, P.}} \yr{2022}  \at{{Dynamics of a
  stratified vortex under the complete Coriolis force: two-dimensional
  three-components evolution}}.  \jt{J. Fluid Mech.}  \bvol{950},  \pg{A29}.

\bibitem[{Toghraei} \& {Billant}(2025)]{ToghraeiBillant2025}
{\sc \au{{Toghraei}, I.} \& \au{{Billant}, P.}} \yr{2025}  \at{Dynamics of a
  stratified vortex under the complete coriolis force: three-dimensional
  evolution}.  \jt{J. Fluid Mech.}  \bvol{1009},  \pg{A11}.

\bibitem[{Vallis}(2006)]{Vallis2006}
{\sc \au{{Vallis}, G.~K.}} \yr{2006} {\em {Atmospheric and Oceanic Fluid
  Dynamics}\/}.

\bibitem[{Yanase} {\em et~al.\/}(1993){Yanase}, {Flores}, {M{\'e}tais} \&
  {Riley}]{Yanaseetal1993}
{\sc \au{{Yanase}, S.}, \au{{Flores}, C.}, \au{{M{\'e}tais}, O.} \&
  \au{{Riley}, J.~J.}} \yr{1993}  \at{{Rotating free-shear flows. I. Linear
  stability analysis}}.  \jt{Phys. Fluids A}  \bvol{5}~(11),  \pg{2725--2737}.

\bibitem[{Zahn}(1984)]{Zahn1984}
{\sc \au{{Zahn}, J.~P.}} \yr{1984} {Introductory report : Stability of Rotation
  Laws}.  \bt{In {\em Liege International Astrophysical Colloquia\/}},
  \st{Liege International Astrophysical Colloquia},  \vol{vol.~25},  \pg{pp.
  407--418}.

\bibitem[{Zahn}(1992)]{Zahn1992}
{\sc \au{{Zahn}, J.~P.}} \yr{1992}  \at{{Circulation and turbulence in rotating
  stars.}}  \jt{A\&A}  \bvol{265},  \pg{115--132}.

\bibitem[{Zeitlin}(2018)]{Zeitlin2018}
{\sc \au{{Zeitlin}, V.}} \yr{2018}  \at{{Symmetric instability drastically
  changes upon inclusion of the full Coriolis force}}.  \jt{Phys. Fluids}
  \bvol{30}~(6),  \pg{061701}.

\end{thebibliography}

\end{document}